\documentclass[12pt]{article}
\usepackage[mathscr]{eucal}
\usepackage{epsfig,amsfonts}
\usepackage[fleqn]{amsmath}
\usepackage{amsthm,amssymb}
\usepackage{graphicx}
\usepackage{hhline}
\usepackage{cite}

\makeatletter
\@addtoreset{equation}{section}
\makeatother

\def\be{\begin{equation}}
\def\ee{\end{equation}}
\def\bdm{\begin{displaymath}}
\def\edm{\end{displaymath}}
\def\bea{\begin{eqnarray}}
\def\eea{\end{eqnarray}}

\def\m{\mu}

\def\ri{{\rm i}}
\def\Xint#1{\mathchoice
    {\XXint\displaystyle\textstyle{#1}}%
    {\XXint\textstyle\scriptstyle{#1}}%
    {\XXint\scriptstyle\scriptscriptstyle{#1}}%
    {\XXint\scriptscriptstyle\scriptscriptstyle{#1}}%
    \!\int}
\def\XXint#1#2#3{{\setbox0=\hbox{$#1{#2#3}{\int}$}
    \vcenter{\hbox{$#2#3$}}\kern-.5\wd0}}

\def\dashint{\Xint-}

\newcommand{\rd}{\mbox{d}}
\newcommand{\re}{\mbox{e}}

\begin{document}

\begin{titlepage}
\begin{flushright}
RUNHETC-06-01\\
\end{flushright}

\vspace{2cm}

\begin{center}
\begin{LARGE}

{\bf Notes on parafermionic QFT's}

\vspace{0.3cm}
{\bf  with boundary interaction}

\end{LARGE}
\vspace{1.3cm}

\begin{large}

{\bf  Sergei  L. Lukyanov}$^{1,2}$

\end{large}

\vspace{1.cm}

${}^{1}$NHETC, Department of Physics and Astronomy\\
     Rutgers University\\
     Piscataway, NJ 08855-0849, USA\\
\vspace{.2cm}
and\\
\vspace{.2cm}
${}^{2}$L.D. Landau Institute for Theoretical Physics\\
  Chernogolovka, 142432, Russia\\
\vspace{1.0cm}

\end{center}

\begin{center}

\centerline{\bf Abstract} \vspace{.8cm}
\parbox{11cm}{
The main result  of these notes  is
an analytical expression for the
partition function of the circular brane model~\cite{SLAZ}
for  arbitrary values of  the topological angle.
The  model has important applications in condensed
matter physics. It is related to
the dissipative rotator (Ambegaokar-Eckern-Sch\"on) model~\cite{AES}
and
describes a
``weakly blocked'' quantum dot with an infinite number of
tunneling channels
under a finite gate voltage bias.
A  numerical check of the analytical solution by means of Monte Carlo simulations has been  performed recently in
Ref.~\cite{SWar}.
To derive the main result we
study the so-called boundary parafermionic sine-Gordon model.
The latter is of certain interest to condensed matter
applications, namely as a toy model for a  point junction
in the multichannel quantum wire~\cite{saleur}.
 }
\end{center}
\vspace{.6cm}
\begin{flushleft}
\rule{4.1 in}{.007 in}\\
{June  2006}
\end{flushleft}
\vfill
\end{titlepage}
\newpage

\tableofcontents

\section{Introduction}
\label{intro}

Since the seminal work  by  Fateev and Zamolodchikov~\cite{ZamFat}
CFT models with  parafermionic symmetry  were extensively studied.
In the original formulation the  parafermionic current algebra
appears  as an extended conformal symmetry of self-dual
multicritical points of ${\mathbb Z}_k$ symmetric statistical
systems \cite{Wu, Elit, Frad, FZaue}. Later  this algebra  has
been employed  in string theory~\cite{Gepner}. Much effort has
been devoted to the parafermionic  models with a boundary\
\cite{Carda, SalB, Card,AOSR, Moor}.

\begin{figure}[ht]
\centering
\includegraphics[width=7cm]{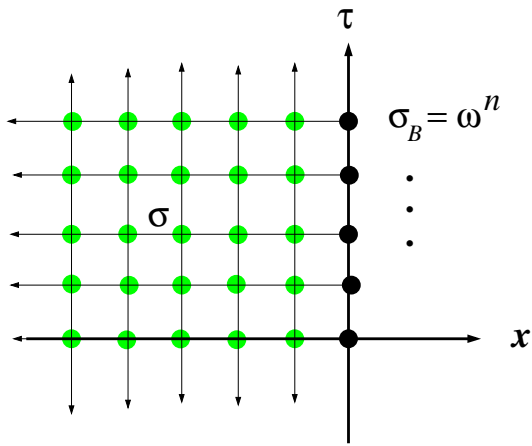}
\caption{ ${\mathbb Z}_k$ symmetric statistical system with fixed boundary condition.
The spins $\sigma$\
\eqref{lissl}\ are located at the vertices of the half-infinite  square lattice.
For an explicit form of  self-dual  Boltzmann weights see Ref.~\cite{FZaue}.  }
\label{fig-lat}
\end{figure}

In conformal field theory there is a special class of Conformally
invariant Boundary Conditions (CBC's) \cite{Carda}. In the case of
parafermionic models some of CBC's   are easily visualized in
terms of the original ${\mathbb Z}_k$ symmetric statistical
systems~\cite{Carda, SalB, Card}. The fluctuating  variables
(``spins'') in  such  systems can be thought of as a set of  $k$
special  points on the unit circle, \bea\label{lissl}
\sigma\in\big\{\, \omega^{n}\,\big\}_{n=0}^{k-1} \ \ \ \ {\rm
with}\ \ \ \omega=\re^{2\pi{\rm i}\over k}\ . \eea The simplest
microscopic boundary conditions are those for which all  spin boundary
values are the same, say $\sigma_B=\omega^{n}$. The scaling limit
of the  self-dual  multicritical ${\mathbb Z}_k$  system depicted
in   Fig.~\ref{fig-lat}, is described  in terms of the minimal
parafermionic  model~\cite{ZamFat}   on the half line $x< 0$  with
the so-called  fixed CBC imposed at $x=0$. With some abuse of
notation we  refer  to these CBC's, as well as the corresponding
RG fixed points, as  ${\cal B}_{n,n}$ $(n=0,\,1,\ldots k-1)$.
\bigskip

\bigskip

Yet another simple  type of CBC is  the ``free'' one. In this
case, the microscopic  boundary  spins are free to take any value
from the set~\eqref{lissl}. In the present work, we denote such CBC
and the associated   RG fixed point by ${\cal B}_{\rm free}$. The
universal ratio~\cite{SalB, Card,AOSR, Moor}, \bea\label{aslss}
{{\rm g}_{\rm free}\over {\rm g}_{\rm fixed}}=\sqrt{k}\ , \eea of
the corresponding boundary degeneracies~\cite{AfLud}
is greater than one. Thus
a   unitary
boundary RG flow ${\cal R}^{(k)}$
from  ${\cal B}_{\rm free}$ to ${\cal B}_{n,n}$ seems to exist, or at least
does not contradict the ${\rm g}$-theorem~\cite{AfLud, FRied}.
Indeed, ${\cal R}^{(2)}$ (the Ising model case)
is a  textbook example  of
boundary flow~\cite{CZ}.
For an arbitrary  $k$  the   existence of ${\cal R}^{(k)}$  can be
advocated within the   general approach~\cite{Schom}.
This boundary flow has  been also  argued    in Refs.~\cite{AOS, AOSR}.

In this paper we  study the QFT model  underlying the boundary
flow ${\cal R}^{(k)}$. In Section~\ref{firstssec}  we 
introduce its  Hamiltonian ${\boldsymbol H}^{(k)}_{\theta}$. It
is   crucial   that ${\boldsymbol  H}^{(k)}_{\theta}$ depends
on an additional dimensionless angular  parameter,  $\theta\equiv
\theta+2\pi k$. The Hamiltonian ${\boldsymbol  H}^{(k)}_{\theta}$
suffers from a specific ultraviolet divergence, which is very
similar to the  ``small instanton'' divergence 
in the 2D $O(3)$ nonlinear
$\sigma$-model~\cite{Frolov,Luscher}. In Section~\ref{secsinG} we consider
a certain non-perturbative regularization of ${\boldsymbol
H}^{(k)}_{\theta}$ involving an additional Bose  field which does not suffer 
from  this problem. The regularized model  is referred to as the
Boundary Parafermionic (BP) sinh-Gordon model. The sinh-Gordon
parameter $b$ plays the  role of a regularization parameter and the
divergence shows up when $b\to 0$. In Section~\ref{secexash} we
propose an analytical expression~\eqref{ksssk} for the partition
function of the BP sinh-Gordon model in terms of solutions of a
certain  second order linear differential equation. Our motivation
behind Eq.~\eqref{ksssk} follows closely along the lines of Al.B.
Zamolodchikov's  unpublished notes~\cite{AlZ}. Namely, in a view
of Appendices~A and  B, the proposed   formula is  a
straightforward generalization of    Zamolodchikov's result for
the conventional  (without  parafermions)  boundary sinh-Gordon
model. Unfortunately, even though  Eq.~\eqref{ksssk} can be easily
guessed, its  derivation  from first principles  is still lacking.
For this reason we  perform some consistency checks
for~\eqref{ksssk} in Appendix~C.

The main quantity we are interested in, is the partition function $Z^{(k)}_\theta$
corresponding to the Hamiltonian ${\boldsymbol  H}^{(k)}_{\theta}$. It
can be extracted   from the  limiting $b^2\to+0$ behavior
of the BP sinh-Gordon  partition function.
An analytical expression for $Z^{(k)}_\theta$
is given    in Section~\ref{secondssec}.
Using this   result,  we describe  a qualitative picture of the boundary flow governed
by ${\boldsymbol  H}^{(k)}_{\theta}$ in Section~\ref{thirdssec}.
It turns out that for each $\theta$ from the open segment
$(2 n-1)\,\pi <\theta< (2 n+1)\,\pi$,
the boundary flow is terminated at the  fixed point ${\cal B}_{n,n}$.
At the same time the RG  trajectories associated with
$\theta=\pi (2 n+1)$ with $n=1,\ldots k-1$ posses nontrivial infrared fixed points whose boundary
degeneracies  are all the same and  equal to $2\,  \cos({\pi\over k+2})\  {\rm g}_{\rm fixed}$.
Hence, the infrared physics of the boundary flow ${\cal R}^{(k)}$
essentially depends on the parameter $\theta$. For this reason
we  shall include an additional  symbol  in the boundary flow  notation: ${\cal R}^{(k)}_\theta$.

In the rest of the main body of the paper we  focus on the large-$k$ limit.
In Section~\ref{twosec} we
identify $Z_\theta=\lim_{k\to\infty}Z^{(k)}_{\theta}$  with the
partition function of the circular brane
model\ \cite{SLAZ}, and  the parameter $\theta$ is  understood  as the topological angle.
In Section~\ref{threesec} a particular ultraviolet regularization of the
circular brane model is considered. The regularized theory is known as the
dissipative quantum  rotator model~\cite{AES} and commonly used in 
describing the Coulomb charging in quantum dots. 
The applicability of our results to
the quantum dot in  weakly blockaded regime is
briefly discussed in Conclusion.

Numerous technical details   of the work are relegated  to  appendices.
In Appendices~A and B we study
the model which,   from a very formal point of view,
is the  BP sinh-Gordon model
considered at purely imaginary values of the
sinh-Gordon parameter $b$.
In essence, these appendices constitute 
work done in 1999 and which  was never published before. It
follows closely   the approach developed in the series of
papers~\cite{BLZ,blz2,blz3,blz5,blzz}. In particular, some of the
key statements involved have been already proven. In those cases
we omit the  proofs  and refer the reader for details
to~\cite{blz3,blz5,blzz}. Among  results obtained in Appendices~A
and B  is   equation~\eqref{lssajssy}, which is interesting in its own right. 
It generalizes a conjecture of P. Fendley and
H. Saleur~\cite{saleur} for the  DC conductance in the multichannel
quantum wire to the case of finite temperature. It should be
emphasized that, although the style  of our presentation is
somewhat sketchy, it seems possible to transform  the derivation
of~\eqref{lssajssy} into a rigorous mathematical proof.

In Appendix~C
we  perform some consistency checks of the
proposed exact analytical expression  for the partition
function of the BP sinh-Gordon model.
Finally,  for  reference purposes, we
collect  some formulas concerning the  low-temperature
expansion of  the partition function $Z^{(k)}_{\theta}$ in Appendix~D.


\section{\label{onesec} Boundary flow ${\cal R}^{(k)}_{\theta}$}

\subsection{\label{firstssec} Hamiltonian ${\boldsymbol  H}^{(k)}_{\theta}$}

Let us consider
the minimal parafermionic model   on a
half lane, $x\leq 0$, constrained
by the free CBC.
By employing  the  magic of modular transformation~\cite{Card} to the
parafermionic characters\ \cite{Kac,Gepner},
one can  observe  operators of the scaling dimensions
\bea\label{lsjs}
\Delta_n={n(k-n)\over k}\ \ \ \ \ \ \ \ \ \ (n=0,\ldots,k-1)
\eea
among  boundary fields of the theory.
Below they   are denoted  as
\bea\label{saskisys}
\big\{{\boldsymbol \Psi}_n(\tau) \big\}_{n=0}^{k-1}\ .
\eea The variable
$\tau$ labels points along  the boundary $x=0$ (see Fig.~\ref{fig-lat}).
Notice that ${\boldsymbol \Psi}_0$ coincides with the unit operator ${\boldsymbol I}$.
We also reserve    special notations for the two relevant
boundary fields of  the  lowest nontrivial scaling
dimension $\Delta_1=\Delta_{k-1}=1-{1\over k}$, namely
\bea\label{sjsahks}
{\boldsymbol \Psi}_+\equiv{\boldsymbol \Psi}_1\, ,\ \ \ \ \
{\boldsymbol \Psi}_-\equiv{\boldsymbol \Psi}_{k-1}\ .
\eea
The boundary fields ${\boldsymbol \Psi}_n(\tau)$  can be thought of as
Fateev-Zamolodchikov chiral parafermionic  currents brought to the boundary
point $\tau$.\footnote{A precise meaning of this statement is discussed in
Section~\ref{nshgsgt}.} Thus,
their  boundary Operator Product Expansions (OPE's)
have a form similar to those  of   the chiral parafermionic algebra:
\bea\label{KJaskjs}
{\boldsymbol \Psi}_j(\tau_1)\, {\boldsymbol \Psi}_m(\tau_2)=
{C_{jm}\over |\tau_1-\tau_2|^{\Delta_{j+m}-\Delta_j-\Delta_m}}\ \
{\boldsymbol \Psi}_{j+m }(\tau_2)+\ldots\ .
\eea
Here
the sum $j+m$ should be  understood modulo $k$ and the structure constants
are the same as in\ \cite{ZamFat}.
With a properly normalized    ${\boldsymbol \Psi}_j$,
the structure constants
read explicitly as
\bea\label{alssjlsa}
C_{jm}=\sqrt{{(j+m)!(k-j)!(k-m)!\over j!\,m!\,k!\,(k-m-j)!}}\ .
\eea

\begin{figure}[ht]
\centering
\includegraphics[width=6cm]{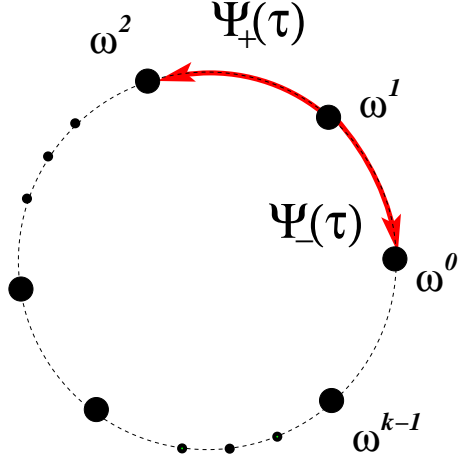}
\caption{Discontinuity diagram  in the fixed boundary conditions corresponding to the
insertion of the boundary parafermions.}
\label{fig-nodes}
\end{figure}

Let ${\cal H}^{(k)}_{\rm free}$ and ${\cal H}^{(k)}_n$  be the spaces of states
of  minimal parafermionic models  on the half line
constrained, respectively,
by the free $({\cal B}_{\rm free})$   and  fixed    $({\cal B}_{n,n})$ CBC's.
Obviously, all the  spaces ${\cal H}^{(k)}_n$ are formally
isomorphic for different integers $n$:
\bea\label{slsjhsl}
{\cal H}^{(k)}_0\simeq {\cal H}^{(k)}_1\simeq\ldots \simeq {\cal H}^{(k)}_{k-1}\ .
\eea
By virtue of  the simple microscopic nature of the
fixed and free CBC's  (see Introduction),
it is expected that
\bea\label{assasa}
{\cal H}^{(k)}_{\rm free}\subset\oplus_{n=0}^{k-1}\,{\cal H}^{(k)}_n\ .
\eea
Then,  the fields ${\boldsymbol \Psi}_\pm(\tau)$, being considered as  operators
acting  in ${\cal H}^{(k)}_{\rm free}$,
intertwine
the linear subspaces
${\cal H}^{(k)}_{n}\bigcap {\cal H}^{(k)}_{\rm free} $ and
 ${\cal H}^{(k)}_{n\pm 1 ({\rm mod} k)} \bigcap {\cal H}^{(k)}_{\rm free}$.
Heuristically, one can think about
${\boldsymbol \Psi}_\pm (\tau)$ as
operators  creating discontinuities  in the  fixed boundary condition
${\cal B}_{n, n}\to {\cal B}_{n\pm 1, n\pm 1}$ at the
boundary  point $\tau$   (see heuristic Fig.\ref{fig-nodes}).

In this work   we  study   the model described by   the
Hamiltonian \bea\label{ssakjiuy} {\boldsymbol H}^{(k)}_\theta=
{\boldsymbol H}_{\rm free}^{(k)}-\mu\, \big[\, \re^{{\rm
i}\theta\over k}\ {\boldsymbol \Psi}_++\re^{-{{\rm i}\theta\over
k}}\ {\boldsymbol \Psi}_-\, \big]\ , \eea where ${\boldsymbol
H}_{\rm free}^{(k)}$ is the  Hamiltonian of the minimal
parafermionic model subject to the free CBC, ${\boldsymbol
\Psi}_\pm\equiv{\boldsymbol \Psi}_\pm(0)$, and, hence,
\bea\label{sasskj} {\boldsymbol H}_{\theta}^{(k)}\ :\ \ {\cal
H}^{(k)}_{\rm free}\to{\cal H}^{(k)}_{\rm free}\ . \eea The
parameter $\mu$ in \eqref{ssakjiuy} carries the dimension
$[\,length\,]^{-{1\over k}}$, so that $\mu^{k}$  sets the
physical  energy scale (the Kondo temperature) in the theory. It
is useful to keep in mind that the OPE's\
\eqref{KJaskjs},\,\eqref{alssjlsa} specify  the normalization of
the boundary fields\ \eqref{saskisys} modulo global ${\mathbb
Z}_k$ transformations, \bea\label{lissa} {\boldsymbol \Psi}_j\to
\omega^{aj}\ {\boldsymbol \Psi}_j\ \ \ (a=1,\ldots k-1)\ . \eea
For this reason the  angular parameter $\theta\equiv
\theta+2\pi\,k$  in Eq.\eqref{ssakjiuy} is not completely
unambiguous; it is defined  modulo transformations
\bea\label{uyslksa} \theta\to \theta -2\pi\, a\, \ \ \ (a=1,\ldots
k-1)\ . \eea The partition function \bea\label{sjksks}
Z^{(k)}_{\theta}= {\rm Tr}_{{\cal H}^{(k)}_{\rm free}}\Big[\,
\exp\big(-{\boldsymbol H}^{(k)}_\theta/T\big)\, \Big] \eea is an
even $2\pi$-periodic function of $\theta$. Therefore the
uncertainty   in   $\theta$ does not  affect  the  thermodynamics
of the model, and 
we  may  consider $\theta$ in the domain $0\leq\theta\leq \pi$ only.


\subsection{\label{secsinG} The BP  sinh-Gordon model}

The key idea behind our   calculation of \eqref{sjksks} is that
${\boldsymbol H}_\theta^{(k)}$ appears in a certain limit of a more
general  Hamiltonian, involving additional bosonic degrees of
freedom. The generalized model is referred to below as the Boundary
Parafermionic (BP) sinh-Gordon model.

Let  ${\boldsymbol \Phi}(x)$ and
${\boldsymbol \Pi}(x)$ be Bose  field operators obeying  canonical commutation
relations
\bea\label{aslsajs}
[\, {\boldsymbol \Pi}(x)\,,\, {\boldsymbol\Phi}(x')\,]=-2\pi{\ri }\ \delta(x-x')\ .
\eea
The Hamiltonian of the   BP  sinh-Gordon model
is described as follows:
\bea\label{uyywtrsahsa}
{\boldsymbol H}_{\rm bshg}^{(k)}=
{\boldsymbol   H}_{\rm free}^{(k)}+{\boldsymbol   H}_{\rm free}^{(\Phi)}
+ h\, {\boldsymbol \Phi}_B
-\mu\, \Big(\,  {\boldsymbol \Psi}_+\ \re^{{b\over\sqrt{k}} {\boldsymbol \Phi}_B}
  +{\boldsymbol \Psi}_-\ \re^{-{b\over\sqrt{k}} {\boldsymbol \Phi}_B}\, \Big)
\ ,
\eea
where
\bea\label{slsajsl}
{\boldsymbol   H}_{\rm free}^{(\Phi)}={1\over 4\pi}\ \int_{-\infty}^0\rd x\ \big(\,
{\boldsymbol  \Pi}^2+(\partial_x {\boldsymbol   \Phi})^2\, \big)\ .
\eea
We do not impose any
constraint on the boundary values
of the canonical fields\ \eqref{aslsajs}. Hence
the Hamiltonian\ \eqref{uyywtrsahsa} acts  in the space
\bea\label{llsajl}
{\cal H}=
{\cal H}_{\rm free}^{(k)}\otimes{\cal H}_{\rm free}^{(\Phi)}\ ,
\eea
where
${\cal H}^{(k)}_{\rm free}$ is the same as in Eq.\eqref{assasa} and ${\cal H}_{\rm
 free}^{(\Phi)}$
is the space of states of the one component boson ${\boldsymbol \Phi}(x)$ with no constraint at
the boundary $x=0$.
The parameter $h$ in \eqref{uyywtrsahsa}
plays the role of an  external field  coupled to the
boundary values ${\boldsymbol \Phi}_B\equiv{\boldsymbol \Phi}(x)|_{x=0}$.
To give an unambiguous definition of the coupling constant  $\mu$,
one should specify the normalization of the boundary fields
\bea\label{kausy}
{\boldsymbol V}_\pm=
 {\boldsymbol \Psi}_\pm\ \re^{\pm{b\over\sqrt{k}} {\boldsymbol \Phi}_B}\ .
\eea
In what follows  these composite fields are assumed to be canonically
normalized with respect their short-distance behavior, i.e.
\bea\label{uyttkalhj}
{\boldsymbol V}_\pm(\tau_1)\,{\boldsymbol V}_\mp(\tau_2)=
|\tau_1-\tau_2|^{-2d}\times {\boldsymbol I}+\ldots\   ,
\eea
and
\bea\label{adfdrlsjlsa}
{\boldsymbol V}_\pm(\tau_1){\boldsymbol V}_\pm(\tau_2)
\cdots {\boldsymbol V}_\pm(\tau_k)
={k!\over k^{k\over 2}}\prod_{i<j}|\tau_i-\tau_j|^{2(1-d)
}\times   \re^{\pm b\sqrt{k} {\boldsymbol \Phi}_B}+\ldots\ ,
\eea
while
\bea\label{alaalaj}
\re^{\pm b\sqrt{k} {\boldsymbol \Phi}_B}(\tau_1)\,
\re^{\pm b\sqrt{k}{\boldsymbol \Phi}_B}(\tau_2)=
|\tau_1-\tau_2|^{2k b^2 }\times {\boldsymbol I}+\ldots\ .
\eea
Here
\bea\label{slsasl}
d=1-{1\over k}-{b^2\over k}
\eea
is the scaling dimension of the boundary fields ${\boldsymbol V}_\pm$\ \eqref{kausy}.
Also notice that the   factor  $k!/k^{k/2}$ in \eqref{adfdrlsjlsa} is  a
product $\prod_{j=1}^{k-1}C_{1j}$ of the structure constants\ \eqref{alssjlsa}.

Formally, when  the dimensionless  parameter  $b^2$ in
\eqref{uyywtrsahsa} is equal to  zero, the  Hamiltonian ${\boldsymbol
H}_{\rm bshg}^{(k)}$ becomes the  sum of a free Bose Hamiltonian
and ${\boldsymbol H}_\theta^{(k)}$. In fact, because of the
particular divergence generated by the interaction term in\
\eqref{uyywtrsahsa} at $b^2\to 0$, the limiting   procedure  is
not  a trivial matter. The divergence appears due to the leading
term in the OPE\ \eqref{adfdrlsjlsa}. Keeping this in mind, we
rewrite the Hamiltonian \eqref{uyywtrsahsa} as a sum of two
operators \bea\label{trslsisjl} {\boldsymbol H}_{\rm bshg}^{(k)}=
{\boldsymbol H}_{1}[{\boldsymbol \Phi}]+ {\boldsymbol H}_{2
}[{\boldsymbol \Psi},{\boldsymbol \Phi}_B]\ , \eea with
\bea\label{jshsghsg} {\boldsymbol H}_{1}[{\boldsymbol
\Phi}]={\boldsymbol   H}_{\rm free}^{(\Phi)} +h\, {\boldsymbol
\Phi}_B+{\textstyle{2 \over kb^2}} \, \big(\, A\,
\cosh\big(b\sqrt{k} {\boldsymbol \Phi}_B\big)-B\, {\boldsymbol
I}\, \big) \eea and \bea\label{ajlsajlsa} {\boldsymbol
H}_{2}[{\boldsymbol \Psi},{\boldsymbol \Phi}_B] =
{\boldsymbol   H}_{\rm free}^{(k)}
-\mu\, \Big(  {\boldsymbol \Psi}_+ \re^{{b\over\sqrt{k}} {\boldsymbol \Phi}_B}
  +{\boldsymbol \Psi}_- \re^{-{b\over\sqrt{k}} {\boldsymbol \Phi}_B} \Big)-
{\textstyle{2\over kb^2}}\,
 \big( A \cosh\big(b\sqrt{k} {\boldsymbol \Phi}_B\big)-B {\boldsymbol I}\big)\,
 .
\eea Our basic assumption is that {\it the coefficients $A$ and
$B$ for the local boundary counterterms in \eqref{ajlsajlsa} can
be adjusted in  such a way that the limit  of the operator
${\boldsymbol   H}_{2}$ does exist when  $b^2\to +0$}. In other
words, we expect that   the singular behavior at $b^2\to +0$ is
fully controlled by  the operator ${\boldsymbol   H}_{1}$
\eqref{jshsghsg}, which is just  a Hamiltonian of the
conventional boundary  sinh-Gordon model.

The  $b^2\to +0$ limit is the classical limit for ${\boldsymbol
H}_{1}$. Indeed, after the field redefinition
${\boldsymbol\varphi}=b\sqrt{k}\ {\boldsymbol \Phi}$ this becomes
particularly striking. Since quantum fluctuations of
${\boldsymbol\varphi}$ are suppressed as $b^2\to 0$ we may apply
the saddle point approximation to account for  their contribution.
The bulk Hamiltonian for ${\boldsymbol\varphi}$ is free and
massless, and,  consequently, the saddle point is achieved for
some constant classical field configuration which we  denote  as
$\ri\theta$. Now, since all matrix elements of ${\boldsymbol
H}_{2}$ are expected to be    finite at  $b^2=0$, we can safely
replace ${\boldsymbol \Phi}_B$ in \eqref{ajlsajlsa} by the
$c$-number ${\ri\theta\over b\sqrt{k}}$. Obviously, the limit
\bea\label{lssksdjuuyq} \big[{\boldsymbol
H}_{\theta}^{(k)}\big]_{\rm reg}= \lim_{b^2\to +0}{\boldsymbol
H}_2\big[{\boldsymbol \Psi}, {\textstyle{\ri \theta\over
b\sqrt{k}}}\big] \eea should be treated as the regularized version
of the Hamiltonian ${\boldsymbol   H}_{\theta}^{(k)}$.

The saddle point configuration ${\boldsymbol\Phi}={\ri\theta\over b\sqrt{k}}$ corresponds to a
minimum of the boundary potential ${\boldsymbol U}[{\boldsymbol \Phi}_B]$,
which is a sum of the last two terms   in the boundary  sinh-Gordon
Hamiltonian\ \eqref{jshsghsg}. The boundary potential
is minimized at
\bea\label{kjss}
\sin(\theta)=\ri\  b \sqrt {k}\ \  {h\over 2 A}\ .
\eea
We should also   take into  account  the effect of  Gaussian
fluctuations around the classical saddle-point configuration.
Let us
write ${\boldsymbol \Phi}(x)=\ri\, {\theta\over b \sqrt{k}}+\delta{\boldsymbol \Phi}(x)$.
Then,
expanding the boundary  potential
near its minima,
one gets
\bea\label{akjlashs}
{\boldsymbol U}[{\boldsymbol \Phi}_B]={\textstyle {2\over k b^2}}\ \big(\, A\,
{ S}(\theta)-B\, \big)\ {\boldsymbol I}+
A\, \cos(\theta)\
(\delta{\boldsymbol \Phi}_B)^2+O\big(\,(\delta {\boldsymbol \Phi}_B)^3\,\big)\ ,
\eea
with
\bea\label{uyjshss}
{ S}(\theta)= \theta\, \sin(\theta)+\cos(\theta)\ .
\eea

Let  $Z_{\rm Gauss}$ be  the partition function of the  Gaussian
theory with the quadratic boundary potential, ${\bar
Z}^{(k)}_\theta(\kappa)$ be the partition function corresponding
to the regularized Hamiltonian\ \eqref{lssksdjuuyq}, and $Z_{\rm
bshg}^{(k)}(h)$ be   the partition function of the BP sinh-Gordon
model: \bea\label{slssjshy} Z_{\rm bshg}^{(k)}(h)= {\rm Tr}_{{\cal
H}}\Big[\, \exp\big(-{\boldsymbol H}^{(k)}_{\rm bshg}/T\big)\,
\Big]\ , \eea where the trace is taken over the space of states\
\eqref{llsajl}. Then, the above consideration implies the
following relation between these three partition  functions as
$b^2\to+0$: \bea\label{sjsaljsasa} Z_{\rm bshg}^{(k)}\big(\,
{\textstyle {2 A\over \ri b\sqrt{k}}} \, \sin\theta\,
\big)\big|_{b^2\to+0}\to\re^{-{2\over  kb^2}\ ({A\over T}\, {
S}(\theta)-{B\over T})}\ Z_{\rm Gauss}\  {\bar Z}^{(k)}_\theta\ .
\eea Finally, according to Ref.\cite{Witten}, \bea\label{lkssluy}
Z_{\rm Gauss}={\rm g}_D\ {\Gamma(1+ 2\,{A\over T}\,
\cos\theta)\over \sqrt{4\pi\, {A\over T} \,\cos(\theta)}}\
\Big({C\over T}\Big)^{-2\,{A\over T}\,\cos(\theta)}\ \ . \eea Here
${\rm g}_D=2^{-{1\over 4}}$ is the boundary degeneracy
associated with the Dirichlet CBC  of the uncompactified  Bose field
and $C$ is  some nonuniversal dimensionful constant.

\subsection{\label{secexash} Exact expression for the partition function $Z_{\rm bshg}^{(k)}(h)$}

Appendixes~A-C  are meant to  demonstrate   that the partition
function \eqref{slssjshy} can be expressed in terms of the
Wronskian $W[\, \Theta_+\,,\, \Theta_-\,]=\Theta_+\partial_u
\Theta_--\Theta_-\partial_u \Theta_+$ of two particular solutions of the
Schr${\ddot {\rm o}}$dinger equation
\bea\label{OsDEra} \Big[\,
-\partial_u^2+ \kappa^2\,\big(\, \re^{-{b u\over Q} }+\re^{{u\over
bQ} } \, \big)^k-\xi^2 \, \Big]\, \Theta(u) = 0\, , \eea where
\bea\label{sddslaskslk} Q=b+b^{-1}\ . \eea More precisely,  let
$\Theta_+$  and $\Theta_-$ be regular solutions of \eqref{OsDEra}  as
$u\to+\infty$ and $u\to-\infty$, respectively. They are specified
unambiguously via their own  WKB asymptotics: \bea
\label{psiassminus} \Theta_{\pm}(u) &\to&\ (2\kappa)^{-{1\over 2}}\
\exp\big(\, F(  b^{\pm 1}\, |\, \pm u)\,  \big) \ \ \ \ \ \   \
{\rm as}\ \ \ \ \ \ \  u\to \pm \infty\ , \eea where $F$ is
expressed in terms of the conventional hypergeometric function,
\bea\label{nahggt} F( b\,|\, u)= -{\textstyle{k \over 4bQ}}\, u
-{\textstyle {2 bQ\kappa \over k}}\ \re^{{k u\over 2bQ}}\
{}_2F_{1}\big(\, -{\textstyle{k\over 2bQ}}, -{\textstyle{k\over
2}}\,;\, 1-{\textstyle{k\over 2bQ}}\, |\,-\re^{-u}\big)\ . \eea
Then, the partition function   \eqref{slssjshy} is given by\footnote{
Eq.\eqref{ksssk} is a generalization of the original proposal from
\cite{AlZ} to   the case $k>1$. For $k=1$  Al.B. Zamolodchikov 
presented a set of  convincing arguments in support of
\eqref{ksssk}. In particular,   the Wronskian $W[
\, \Theta_+, \Theta_-]$ was expressed   in terms of the  solution  of the
massless thermodynamic Bethe ansatz equation associated with the
boundary sinh-Gordon model\ \eqref{jshsghsg}.}
\bea\label{ksssk} Z_{\rm bshg}^{(k)}(h)={\rm g}_D\ {\rm g}_{\rm
fixed}\  W[\, \Theta_+\,,\, \Theta_-\,]\ . \eea 
The parameters of
the Hamiltonian \eqref{uyywtrsahsa} are related  to $\xi$ and
$\kappa$ in \eqref{OsDEra} as follows: \bea\label{liuyytkasnlas}
&&\xi=\ri\ {\sqrt{k}\over 2Q}\ {h\over T}\, ,\\
&&\kappa={1\over 2\pi T}\ {k\over bQ}\ \bigg[{2\pi\mu\over
\sqrt{k}\Gamma(1-{bQ\over k})}\bigg]^{k\over bQ}\ .\label{lkasnlas}
\eea

Notice that the Wronskian $W[\, \Theta_+\,,\, \Theta_-\,]$ can be viewed as a
spectral determinant of the  Schr${\ddot {\rm o}}$dinger operator
\eqref{OsDEra}. Indeed the spectrum of \eqref{OsDEra} is  simple,
discrete and positive: \bea\label{ksjlasj} {\rm Spect}=\big\{\,
\xi^2_n\, \big\}_{n=1}^{\infty}\ \ \ \ (\xi^2_n>\xi^2_m>0\ \ \
{\rm if}\ \ \ n>m)\ , \eea 
and, as follows from
the  Bohr-Sommerfeld
quantization condition,
\bea\label{sskasa} \xi_n= {\pi k\over Q^2}\ {n\over \log n}\
\Big[\, 1+O\big( {\textstyle{\log(\log n)\over \log n}}\,\big)\,
\Big]\ \  \ \ \ \ {\rm as }\ \ \ \ \ \ n\to \infty\ . \eea 
Therefore  the spectral
determinant can be defined through the equation
\bea\label{ssjljsa} D(\kappa,\xi)=D_0(\kappa)\ \prod_{n=1}^\infty
\Big(1-{{\xi^2\over \xi_n^2}}\,\Big)\ , \eea where $D_0(\kappa)$
is some  overall $\xi$-independent factor. 
At the same time, 
$W[\,\Theta_{+}\,,\, \Theta_{-}\,]$ is an entire function of the
variable $\xi^2$ which shares the same set of zeros as
\eqref{ssjljsa}.\footnote{ Let  ${\tilde \xi}^2$  be a zero of
$W[\,\Theta_{+}\,,\, \Theta_{-}\,]$. In this case the solution $\Theta_+(u)$ and
$\Theta_{-}(u)$   are linearly dependent. It just means that the
Schr${\ddot {\rm o}}$dinger equation possesses a solution which is
both  regular at $x\to+\infty$ and $x\to-\infty$, hence ${\tilde
\xi}^2\in{\rm Spect}$. It is also obvious  that $W[\,\Theta_{+}\,,
\,\Theta_{-}]\big|_{\xi^2=\xi^2_n} =0$.} Using the standard WKB
methods \cite{Landau}, one can show that $\log W[\,\Theta_{+}\,,\,
\Theta_{-}\,]=o(\xi^{2})$ as $\xi^2\to\infty$. Hence, as  follows
from the Liouville theorem,  the ratio $D/W$ does not depend on
$\xi$, and we can always adjust the overall normalization factor
$D_0$ in  \eqref{ssjljsa} to make it equal to  one, i.e.,
\bea\label{qqzsksls} D(\kappa,\xi)=W[\,\Theta_{+}(u)\,,\,\Theta_{-}(u)\,]\
. \eea

\subsection{\label{secondssec} Exact expression for  
the partition function $Z^{(k)}_{\theta}$}

With the exact expression for $Z_{\rm bshg}(h)$, we
can now check the semiclassical    behavior~\eqref{sjsaljsasa}.

Consider  the spectral determinant $D(\kappa,\kappa\sin\theta)$ in the limit $b^2\to+0$.
There is no problem with the limit  for the differential
equation\ \eqref{OsDEra} and for the solution
$\Theta_+$\ \eqref{psiassminus}. Namely,
\bea\label{lksalsasa}
\Theta_+(u)\to\ \exp\Big[\, \big({\textstyle{2\over k b^2}}+  c_+\big)\,
\kappa\,
\sin^2({\textstyle {\pi (k-1)\over 2}})\, \Big]\ \Xi_+(u)\ \ \ \ \
{\rm as}\ \ b^2\to +0\ ,
\eea
where  $\Xi_+$ is a   solution of
\bea\label{OsDErt}
\Big[\, -\partial_u^2+
\kappa^2\,\big(1+
\re^u\big)^k-\kappa^2\sin^2(\theta)
\, \Big]\, \Xi(u) = 0
\eea
defined by  the  asymptotic condition 
\bea\label{kakjs}
\Xi_+(u)\to {\re^{-(u+c_+)\kappa}\over \sqrt{2\kappa}\, (1+\re^{u})^{k\over 4}}
\ \exp\bigg\{ -  \kappa
\int_0^{\re^u}
{\rd z\over z}\, \big(\, (1+z)^{k\over 2}-1\, \big)\, \bigg\}\, \ \ \ 
{\rm as}\ \ u\to+\infty\ .
\eea
The constant $c_+$ in \eqref{lksalsasa} reads explicitly
\bea\label{lsksjk}
c_+=\psi\big(1+{\textstyle{k\over 2}}\big)+\gamma_E\ ,
\eea
where
$\psi(z)={{\rm d} \over {\rm d} z}\log\Gamma(z)$ and $\gamma_E$ is Euler's constant.
Notice that
the argument of the exponential function in \eqref{lksalsasa} vanishes
for  any odd $k$.

The asymptotic condition\ \eqref{psiassminus} for $\Theta_-$ is singular
as $b^2\to+0$.  For this reason
the limiting behavior of this solution  is a slightly  delicate issue.
It can be analyzed  using the
WKB approximation for $\Theta_-$:
\bea\label{shksjs}
\Theta_-^{\rm (wkb)}(u)= {1\over \sqrt{2\kappa {\cal P}(u)}}
\exp\bigg\{\kappa
 \int_{-\infty}^u\rd v\, \big(\, {\cal P}(v)-\re^{-{b k\over 2 Q} v}\, \big)
-
{\textstyle{2Q\kappa\over bk}}\  \re^{-{bk\over 2Q} u}\, \bigg\}\ ,
\eea
where $ {\cal P}(x)$ is given by
\bea\label{ssjaso}
{\cal P}(u)=
\sqrt{\big(\, \re^{-{b u\over Q} }+\re^{{u\over bQ} }
\, \big)^k-\sin^2(\theta)}\ \ \ \ \ \ \ {\rm with}\ \ \
\ \ \ \ \ \sin(\theta)=\xi/\kappa\ .
\eea
Notice that
the subtraction term
$\re^{-{b k\over 2 Q} v}$ in the integrand comes from the asymptotic
conditions\ \eqref{psiassminus}\ and ensures   convergence of the
integral in\ \eqref{shksjs}.
Consider now the argument $\{\, \ldots\, \}$ of the
exponential in \eqref{shksjs} as
$b^2\to +0$. It is easy to see that $\{\, \ldots\, \}
\to F(u)-{2\over k b^2}\ \kappa { S}(\theta)$,
where ${ S}(\theta)$ is given by\ \eqref{uyjshss}, while
$F(u)\to  u\kappa\cos(\theta) $ as $u\to-\infty$.
Then
\bea\label{ssslsl}
\Theta_-(u)\to \Xi_-(u)\
{\re^{-c_-\kappa\cos\theta} \over \sqrt{2\kappa \cos(\theta)}}\
\ \re^{ -{2\over k b^2}\ \kappa { S}(\theta)}\ \ \ \ {\rm as}\ \ \ b^2\to +0\ ,
\eea
where $\Xi_-$ is a solution of the differential equation
\eqref{OsDErt} such that
\bea\label{kakjsuy}
\Xi_-(u)\to \re^{(u+c_-)\kappa\cos\theta}\ \ \ \ \ \ {\rm as}\ \ \ u\to -\infty\ ,
\eea
and  $c_-$  is
an arbitrary constant.
It will be convenient to choose 
\bea\label{slsus}
c_-=\psi\big({\textstyle{k\over 2}}\big)+\gamma_E+2\log 2\, .
\eea
We  emphasize that
$\Xi_-$ can be defined by means of  the  
asymptotic condition\ \eqref{kakjsuy}
for    $0\leq \theta<{\pi\over 2}$ only.
For ${\pi\over 2}< \theta\leq \pi$, 
the solution $\Xi_-$   grows at large negative $u$ and
the asymptotics \eqref{kakjs} does not define $\Xi_-$
unambiguously. Fortunately, 
the function
$\Xi_-/\Gamma(1+2\kappa\cos\theta)$  is an entire function
of the complex variable $\zeta=\cos(\theta)$ for  $u<0$. So
the solution $\Xi_-(u)$    can be
introduced within 
${\pi\over 2}\leq \theta\leq \pi$
through   the analytic continuation with respect  to the variable $\theta$  from the
domain $0\leq \theta<{\pi\over 2}$.

In a view of relations~\eqref{lkasnlas}, 
the singular behavior of $D(\kappa,\kappa\sin\theta)$ as
$b^2\to +0$ matches exactly the structure
\eqref{sjsaljsasa} provided
\bea\label{aasjsajs}
A=T\kappa\, ,\ \ \ \ \ B= T\kappa\ \sin^2({\textstyle {\pi (k-1)\over 2}})\ .
\eea
Furthermore, comparing the pre-exponent  in  \eqref{sjsaljsasa}
with  a similar factor for $D(\kappa,\kappa\sin\theta)$,  we finally obtain
the partition function for the  regularized Hamiltonian \eqref{lssksdjuuyq}
\bea\label{dsgfasta}
{\bar Z}^{(k)}_{\theta}(\kappa)={\rm g}_{\rm fixed}\
\kappa^{2\kappa \cos\theta}\
{{\sqrt{2\pi}}\ W[\Xi_+,\Xi_-]\over \Gamma(1+2\kappa\cos\theta)}\ .
\eea

Two important comments are in order here.
First,
${\bar Z}^{(k)}$  is the function of the dimensionless variable $\kappa$,
which is in fact the inverse temperature measured in the units of the RG invariant  scale $E_*$:
\bea\label{skjssak}
\kappa={E_*\over T}\ .
\eea
As  follows from  Eq.~\eqref{lkasnlas} considered at $b^2=0$
\bea\label{slskas}
E_*={k^{1-{k\over 2}}\over 2\pi}\ \bigg[{2\pi \mu\over \Gamma(1-{1\over k})}\bigg]^{k}\ .
\eea
The parameter 
$\mu$ should be understood now  as
the  coupling of the Hamiltonian\ \eqref{lssksdjuuyq}.

Second, the partition function for the
regularized Hamiltonian \eqref{lssksdjuuyq}
is defined modulo an overall factor  $\re^{c_-\kappa\cos\theta}$,
where $c_-$ is an arbitrary constant.
By fixing   the  value of $c_-$ via  Eq.~\eqref{slsus}
we have enclosed  a particular  normalization condition for
the regularized ground state energy
\bea\label{lisso}
{\bar E}^{(k)}_{\theta}=-\lim_{T\to 0}\big[\, T\log {\bar Z}^{(k)}_{\theta}\, \big]\ .
\eea
Namely,
if $c_-$ is chosen as in \eqref{slsus}, then
${\bar E}^{(k)}_{\theta}$
satisfies    the  condition
\bea\label{slsasal}
{\partial^2\over  \partial\theta^2}
{\bar E}^{(k)}_{\theta}\, \big|_{\theta=0}=0\ .
\eea
In the case of odd $k$ the regularized ground state energy
satisfying \eqref{slsasal} is  unambiguously defined. In particular, 
the dimensionful constant
\bea\label{lskals}
{\bar E}^{(k)}_{\theta}\big|_{\theta=0}=-2\, \big(\, 1-{\textstyle{1\over k}}\, \big)\ E_* 
\eea
can be viewed as
a universal    physical  energy scale  in  the model~\eqref{ssakjiuy}.
Since  the constant $B$ \eqref{aasjsajs} does  not vanish for  even  values of  $k$,
the regularized Hamiltonian \eqref{lssksdjuuyq}
contains an additional    counterterm of  unit operator. In this case
only the difference ${\bar E}^{(k)}_{\theta}-{\bar E}^{(k)}_{\theta=0}$
turns out  to be a  universal quantity.
For even $k$ we 
choose Eq.\eqref{lskals} as an extra normalization condition 
for the regularized ground state energy.

Whereas 
${\bar E}^{(k)}_{\theta}$  satisfying
\eqref{slsasal} and \eqref{lskals} is a universal scaling function,
the ground state energy $E^{(k)}_{\theta}$
for  the original   Hamiltonian\ ${\boldsymbol H}^{(k)}_\theta$ \eqref{ssakjiuy}
does  depend on  details of  the regularization procedure. More precisely, the above
consideration suggests (see Eqs.~\eqref{ajlsajlsa} and \eqref{lssksdjuuyq})
the following general form for $E^{(k)}_{\theta}$
\bea\label{saslhs}
E^{(k)}_{\theta}=-E_*\ \Big[ \, L^{(k)}_1\ \cos(\theta)-L^{(k)}_2\
\sin^2\big({\textstyle{\pi (k-1)\over 2}}\big)\, \Big]
+{\bar E}^{(k)}_{\theta}\ ,
\eea
where $L^{(k)}_{1,2}$  are some dimensionless nonuniversal  constants which are
expected to  absorb all
divergences  in the theory \eqref{ssakjiuy}.\footnote{
A very similar phenomenon was  discussed  in~\cite{FatZam} for
the minimal parafermionic  models  perturbed by  the  parafermionic currents
in the bulk.} Hence, the partition function
\eqref{sjksks} has the form
\bea\label{kaaka}
Z^{(k)}_{\theta}=\exp\Big[\,\kappa\, L^{(k)}_1\, \cos(\theta)
-\kappa\, L^{(k)}_2\,
\sin^2\big({\textstyle{\pi (k-1)\over 2}}\big)\, \Big]\ {\bar Z}^{(k)}_\theta(\kappa)\ ,
\eea
where  $ {\bar Z}^{(k)}_\theta(\kappa)$ is given by Eq.\eqref{dsgfasta}.
For the   regularization described above $L^{(k)}_{1,2}={1\over kb^2}+O(1) $ with $b^2\to+0$.
In    general,
the divergent  constants $L^{(k)}_{1,2}$
are expressed in terms of
some UV cutoff  equipped with   the theory.

One should note that
some particular cases of   $Z^{(k)}_{\theta}$
have been already   discussed in the
literature.
E.g., the  case $k=2$  was studied in  Ref.~\cite{CZ}.
The differential equation \eqref{OsDErt} for $k=2$
can be transformed to  Kummer's equation, so that    ${\bar Z}^{(2)}_{\theta}$ has
the especially simple form,
\bea\label{ksksj}
{\bar Z}^{(2)}_{\theta}(\kappa)={{\rm g}_{\rm fixed}\, \sqrt{2\pi} \over \Gamma({1\over 2}+2\kappa
\cos^2({\theta\over 2}))}\ \ (2\kappa)^{2\kappa \cos^2({\theta\over 2})}\ \re^{-\kappa}\ .
\eea
Eq.~\eqref{ksksj} is in  complete agreement with the result of \cite{CZ}.
It should be also  mentioned that  for $\sin(\theta)=0$
the partition function is described in terms of the
differential equation  originally  introduced
in  \cite{LVZ}  in the context of the so-called paperclip boundary flow.
The difference between $\theta=0$ and $\theta=\pi$
appears in a choice of $\Xi_-$-solution. Namely,
$\Xi_-^{(\theta=0)}$ and  $\Xi_-^{(\theta=\pi)}$
are the
two Bloch-wave solutions of \eqref{OsDErt} with  $\sin(\theta)=0$~\cite{LTZ}.
We refer the reader to those papers for
a wealth of data about the differential equation \eqref{OsDErt} with
$\sin(\theta)=0$. In particular, the
thermodynamic ansatz  equations describing
the partition function $Z^{(k)}_\theta$ \eqref{sjksks} for
$\theta=0$ and $\theta=\pi$ can be found  in Ref.~\cite{LTZ}.

\subsection{\label{thirdssec} Qualitative pattern of the boundary flow ${\cal R}^{(k)}_{\theta}$}

Our prime  interest  here is in 
the high- and low-temperature behaviors of $Z^{(k)}_\theta$.
With \eqref{dsgfasta} it is straight-forward to check that the  boundary entropy
$S_{\theta}^{(k)}=\big(1+T\, {\rd\over \rd T} \big)\, \log Z^{(k)}_\theta$,
has the following   $\theta$-independent limit as $T\to\infty$:
\bea\label{lslassajlas}
\lim_{T\to \infty}S_{\theta}^{(k)}(T)=\log({\rm g}_{\rm free})\ .
\eea
At the same time,
\bea\label{hjdhdkss}
\lim_{T\to 0}S_{\theta}^{(k)}(T)=
\begin{cases}
\log({\rm g}_{\rm fixed})\ \ \ &{\rm as}\ \ \ 0\leq\theta<\pi\\
\log({\rm g}_2) \ \ \ &{\rm as}\ \ \  \theta=\pi
\end{cases}\  .
\eea
Here  the boundary degeneracy ${\rm g}_2$ is given by the formula
\bea\label{lsalsa}
{\rm g}_{s}={\rm g}_{\rm fixed}\ {\sin({\textstyle{s\pi\over k+2}})\over \sin({\textstyle{\pi\over k+2}})}\  ,
\eea
with $s=2$.
Whereas \eqref{lslassajlas} is just a simple  consequence of the fact
that ${\cal R}_{\theta}^{(k)}$ starts from   ${\cal B}_{\rm free}$,
Eq.~\eqref{hjdhdkss} is much less trivial.
It shows that within $0\leq\theta<\pi$ the infrared behavior of the  theory 
is defined  by the fixed CBC, 
but for  $\theta=\pi$ the RG flow is terminated at some nontrivial fixed point.

The CBC associated with the boundary degeneracy ${\rm g}_2$
can be easily identified.
Recall that the minimal parafermionic model
possesses  the Cardy states~\cite{Card}
with    the  additional  conformal symmetry, namely
${\rm W}_k$  symmetry introduced in
Refs.~\cite{SasZ, FATZ, FL}.\footnote{See Section~\ref{nshgsgt}
for the brief review of ${\rm W}_k$ symmetry and ${\rm W}_k$  invariant
CBC's   in the minimal parafermionic  models.}
They
are in one-to-one correspondence with the
$k(k+1)/2$  number  of  ${\rm W}_k$ 
primary fields of the model. 
Let us  denote the CBC's
corresponding to the  ${\rm W}_k$  invariant Cardy states by
${\cal B}_{l,n}$ with  $0\leq n\leq l\leq k-1$.
Their boundary degeneracies  are given
by Eq.~\eqref{lsalsa} with $s=l-n+1$.\footnote{
In string 
language~\cite{Moor},  ${\cal B}_{l,n}$
are the so-called ``A-branes''. The fixed CBC's   ${\cal B}_{n,n}$
are D0-branes
coinciding 
with  the points $\omega^{n}$ on the unit circle.
The CBC's
${\cal B}_{ln}$ with $l>n$  are  interpreted as $D1$-branes,
the chords stretched between the  points
$\omega^n$ and $\omega^{l}$.  The ``lightest  B-brane''
from~\cite{Moor} is identified
with the free CBC ${\cal B}_{\rm free}$.}
In particular the boundary degeneracies
associated with  ${\cal B}_{0,1},\, \ldots,\, {\cal B}_{k-2,k-1}$
and ${\cal B}_{k-1,0}$
are all the same and given by the constant  ${\rm g}_2=2\, \cos({\pi\over k+2})\,
{\rm g}_{\rm fixed}$.

\begin{figure}[ht]
\centering
\includegraphics[width=9cm]{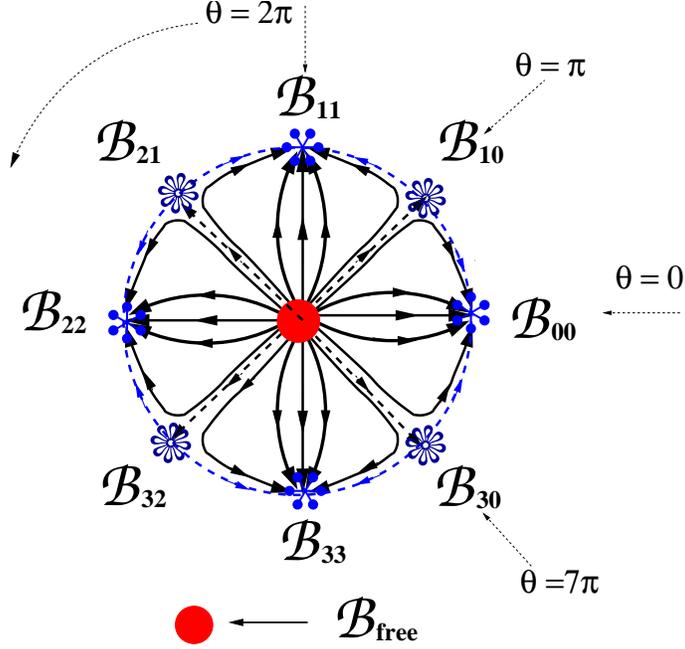}
\caption{
The boundary flow ${\cal R}_{\theta}^{(k=4)}$
in the 2D plane of coupling constants. Polar  coordinates $(\rho,\phi)$
in this plane are some properly chosen ``running''
couplings, $\rho=\rho(T,\theta)$, 
$\phi=\phi(T,\theta)$, such
that $\lim_{T\to\infty}\rho(T,\theta)=
0$ and $\rho_{\max}=\lim_{T\to 0}\rho(T,\theta)$.
The parameter $\theta$ is constant along each trajectory at  $T\not=0$.}
\label{fig-flow}
\end{figure}

Now, in view of the
global ${\mathbb Z}_k$ invariance~\eqref{lissa},\,\eqref{uyslksa}, it is easy to imagine
the whole  qualitative picture  of the boundary RG flow
${\cal R}_{\theta}^{(k)}$ for
$0\leq \theta<2\pi k$. In the case $k=4$ it is depicted  in   Fig.~\ref{fig-flow}.
Among  the flow lines shown in Fig.~\ref{fig-flow} there are trajectories
adjacent  to the big circle  associated with zero   temperature.
In Appendix~D we present an  exact formula~\eqref{wejsre} for the normalized ground state energy
\eqref{lisso},\,\eqref{slsasal}.
It turns out   that
 ${\bar E}^{(k)}_{\theta}$
is a singular function at $\theta=\pi$ (see formula~\eqref{sssslsa}):
\bea\label{kkasaks}
{\bar E}^{(k)}_{\theta}\propto  |\theta-\pi|^{1\over 1-d_{\varepsilon}}\ ,
\eea
where
\bea\label{kakakayu}
d_\varepsilon={2\over k+2}\ .
\eea
It is worth noting that
$d_\varepsilon$ is the scaling dimension of the so-called
first energy boundary  operator, ${\boldsymbol \varepsilon}$~\cite{ZamFat}. Hence,
an  effective Hamiltonian, describing the system in a vicinity of
the nontrivial infrared fixed points ${\cal B}_{n+1,n}\,$,
contains this  relevant  boundary field  along with   irrelevant 
conformal and ${\rm W}_k$  descendents of ${\boldsymbol \varepsilon}$ and ${\boldsymbol I}$.
Notice that 
the coupling constant for ${\boldsymbol \varepsilon}$
vanishes at $\theta=\pi,\,3\pi\ldots$, so 
the leading low-temperature corrections come from   the first
${\boldsymbol W}_3$  descendent
of $\varepsilon$ of the scaling dimension $1+d_{\epsilon}$.
The low-temperature  expansion of \eqref{dsgfasta}
for the radial  trajectories  $\theta=\pi,\,3\pi\ldots$   was studied in Ref.~\cite{LTZ}.
More details about  the  low-temperature expansions of
$Z^{(k)}_\theta$ for $\theta\not=\pi$ are presented  in Appendix~D.

\section{
\label{twosec}
Circular brane partition function}

\subsection{$Z^{(k)}_\theta$ in the limit  $k\to\infty$}

The regularized
partition function~\eqref{dsgfasta}
possesses a finite limit  as $k\to\infty$ provided the physical energy scale $E_*$
is kept fixed.
Indeed, let us shift the variable $u$ in the differential equation\ \eqref{OsDErt}:
\bea\label{kqkqkh}
u=y-\log(k)\ .
\eea
Then,    the large-$k$ limit   brings \eqref{OsDErt}  to the form,
\bea\label{Osrt}
\Big[\, -\partial_y^2+
\kappa^2\,\exp\big(\,\re^{y}\,\big)
-\kappa^2\sin^2(\theta)
\, \Big]\, \Xi(y) = 0\, .
\eea
The solutions $\Xi_+$ and $\Xi_-$ remain finite
for any fixed $y$ when
$k\to \infty$.
Thus
\bea\label{asksskaa}
{\bar Z}_\theta(\kappa)=\lim_{k\to\infty}{\bar Z}_\theta^{(k)}(\kappa)
\eea
can be calculated using the same Eq.\eqref{dsgfasta} with
$W[\,\Xi_+\, ,\, \Xi_-\,]$ being the Wronskian of two 
solutions of the differential equation\ \eqref{Osrt}
which are fixed by the asymptotic conditions
\bea\label{llisa}
\Xi_-(y)\to \big(2\re^{\gamma_E}\big)^{\kappa\cos\theta}\
\  \re^{y\kappa\cos\theta}\ \ \ \ \ \ {\rm as}\ \ \ \ y\to-\infty\ ,
\eea
and
\bea\label{llauisa}
{ \Xi}_+(y)\to (2\kappa)^{-{1\over 2}}\  \exp\Big[\,-{\textstyle {1\over 4}}\, \re^y-
\kappa\, 
{\rm Ei}\big(\,\textstyle\frac{1}{2}\,\re^y\,\big)\, \Big]\, \ \ \ \ {\rm as}\ \ \ \ y\to+\infty\ ,
\eea
with ${\rm Ei}(z)=\dashint_{-z}^\infty\frac{{\rm d} x}{x}\, \re^{-x}$.
Again, the asymptotic condition\ \eqref{llisa} is
applicable for $0\leq\theta<{\pi\over 2}$ only. Within  the domain
${\pi\over 2}\leq\theta<\pi$ the solution ${ \Xi}_-$
should be  defined through the analytic continuation
with respect to the variable $\theta$.

Let us assume now  that $k$ is odd and consider
the partition function\ \eqref{kaaka} at $k\to\infty$. One has
\bea\label{saka}
\lim_{k\to\infty\atop
k=3,\,5\ldots}Z_{\theta}^{(k)}=
\re^{ \kappa \cos(\theta)\, L}\ \ {\bar Z}_\theta(\kappa)\ ,
\eea
where $L$ is the limiting value of the nonuniversal constant $L^{(k)}_1$.
As  follows from \eqref{kaaka},
the large-$k$ limit taken among the even $k$ may cause  the  result
to differ from  \eqref{saka} by an extra multiplicative factor
$\re^{\kappa\, const}$. Obviously, this  subtlety does  not affect the
physical content  of  the limiting theory and we shall ignore it below.

\subsection{Circular brane model }

Here we identify
the QFT describing the large-$k$ limit of the model \eqref{ssakjiuy}.

For finite $k$ the
minimal parafermionic
model defined
on the semi-infinite
cylinder $\tau\equiv\tau+1/T,\ x\leq 0$,
 contains two   holomorphic  ${\boldsymbol \psi}_{\pm}(\tau,x)$
$=
{\boldsymbol \psi}_{\pm}(\tau- {\ri}x)$
and two antiholomorphic  ${\bar {\boldsymbol \psi}}_{\pm}(\tau, x)=
{\bar {\boldsymbol \psi}}_{\pm}(\tau+ {\ri}x)$  currents
of conformal dimensions $\Delta_{1}={\bar \Delta}_1=1-{1\over k}$.
When  $k\to\infty$, the unlocal parafermionic currents turn into
chiral  spin-1 currents. The latter can be bosonized in terms of a pair of
fields satisfying free massless equations of motion in the bulk $x<0$:
\bea\label{slsksls}
\partial {\bar \partial } {\boldsymbol X}=\partial {\bar \partial } {\boldsymbol Y}\ ,
\eea
with
$\partial={1\over 2}\ (\partial_{\tau}+\ri\, \partial_{x})$
and $
{\bar \partial}={1\over 2}\ (\partial_{\tau}-\ri\, \partial_{x})$.
Namely,
\bea\label{aslkss}
\lim_{k\to\infty}{\boldsymbol \psi}_{\pm}={\partial {\boldsymbol X}\pm \ri \partial {\boldsymbol Y}\over \sqrt{2}}\, ,\ \ \ \
\lim_{k\to\infty}{\bar {\boldsymbol \psi}
}_{\pm}={{\bar \partial} {\boldsymbol X}\pm \ri {\bar \partial} {\boldsymbol Y}\over \sqrt{2}}\ .
\eea
Hence, the large-$k$  limit
of \eqref{ssakjiuy}\ is described by the theory with the bulk Euclidean action
\bea\label{baction}
{\mathscr   A}_{\rm bulk}=
{1\over \pi}\ \int_0^{1/T}\rd\tau\int_{-\infty}^{0}\rd x\
\big(\, \partial  X{\bar \partial}  X+\partial  Y{\bar \partial}  Y\, \big)\ .
\eea
In view of   the  heuristic picture of the boundary interaction
in \eqref{ssakjiuy} (see Fig.\ref{fig-nodes}) one may  expect
the boundary values $X|_{x=0}\equiv X_B$ and $Y|_{x=0}\equiv Y_B$
satisfy some   $O(2)$-invariant boundary condition.
This boundary condition has been  already identified for
$\theta=0,\,\pi$  in  Refs.~\cite{SLAZ,LTZ}.
In both cases
the limit  \eqref{saka} coincides with
the partition function of the
circular brane model.
The bulk action of this two-dimensional model
is indeed given by\ \eqref{baction}, while
the boundary values  of the Bose fields
are subjected to a nonlinear constraint
\bea\label{bconstaint}
 X_B^2+Y_B^2={1\over g_0}\ .
\eea

Let us recall some facts about the circular brane model~\cite{SLAZ}.
Due to the
nonlinear
boundary condition~\eqref{bconstaint}, this theory
needs renormalization. It has to be
equipped with the ultraviolet (UV) cut-off,
and consistent removal of the
UV divergences requires that the bare coupling  $g_0$ be
given a dependence on the cut-off energy scale $\Lambda$, according to
the Renormalization Group (RG) flow equation
\bea\label{flow}
\Lambda\, {{\rd g_0}\over{\rd\Lambda}} = - 2 \, g_{0}^2 - 4
\,g_{0}^3 + \cdots\ .
\eea
The leading two terms of the 
$\beta$-function written down in~\eqref{flow}
were
computed in~\cite{Kosterlitz} and~\cite{Zwerger},
and indeed agree with the more general
calculations in~\cite{Leigh}\ and~\cite{Rychkov}.
Eq.~\eqref{flow} implies that
\bea\label{scale}
E_*\simeq   \Lambda\ g^{-1}_0\ \re^{-{1\over 2g_0}}\
\Big(1+\sum_{m=1}^{\infty}c_m\, g_0^m\, \Big)\ ,
\eea
where the symbol $\simeq$ stands for asymptotic equality.
The coefficients $c_m$ in  the asymptotic series~\eqref{scale}
are not universal, i.e., they
depend on the details of the regularization.

It is  important  that the general definition of the circular brane
model involves  an additional
parameter, the topological angle $\theta$.
The configuration space for the two component  Bose field $X^{\mu}=(X,\,Y)$  consists of
sectors characterized by an integer $w$, the
number of times the boundary value $ (X_B,\, Y_B)$ winds around the
circle \eqref{bconstaint} when one  goes around the boundary at $x=0$.
The contributions from the topological sectors can be weighted with
the factors $\re^{{\rm i}w\theta}$. Thus, in general
\bea\label{topsum}
Z_{\theta}= \sum_{w=-\infty}^{\infty}\re^{{\rm i}
w\theta}
 \ Z^{(w)}\, ,
\eea
where $Z^{(w)}$ is the path  integral
\bea\label{salskljsk}
Z^{(w)}
=
\int_{(w)}\,{ \cal D} X { \cal D} Y\
\re^{-{\mathscr  A}_{\rm bulk}[X,Y]}
\,,
\eea
evaluated
over the
fields from the sector $w$ only.
As it was pointed out  in
\cite{Wang},\footnote{The circular brane model shows many similarities with the $O(3)$
nonlinear
$\sigma$-model (or ${\bf n}$-field)~\cite{APolyakov, Belavin}, where
a very similar effect  of  the small instanton
divergence was known for a  while~\cite{Frolov,Luscher}.}
the functional integral\ \eqref{salskljsk}
has a particular nonperturbative divergence due to the small-size
instantons\ \cite{Korshunov, Nazarov}, which cannot be absorbed into the renormalization of the
coupling constant $g_0$. Due to this effect,
the partition function\ \eqref{topsum}\ is expected to have the form
\eqref{saka},  where
\bea\label{khsks}
L=2\,\log\Big({{\bar \Lambda}\over E_*}\Big)\ ,
\eea
with  ${\bar \Lambda}$ some ultraviolet cut-off regularizing
the
small instanton divergence, and   $\kappa$ the inverse temperature
measured in units of the physical
scale~\eqref{scale}.\footnote{Notice
that  the instanton cut-off ${\bar \Lambda}$ is different from the perturbative
cut-off $\Lambda$ in the RG flow 
equation\ \eqref{flow}, i.e. their ratio ${\bar \Lambda}/\Lambda$ is
some non-universal function of the bare coupling $g_0$.
In Section~\ref{threesec}
we find this  ratio  for some particular UV regularization
of the circular brane model.}
This suggests  to extend the results of Refs.~\cite{SLAZ,LTZ}
and
identify  \eqref{saka}
with the
partition function of the circular brane model $Z_\theta$
where   $\theta$
plays the role of the topological angle:
\bea\label{spsalsli}
Z_\theta=
\exp\Big(\,2 \kappa \cos(\theta)\, \log\big({\textstyle {{\bar \Lambda}
\over E_*}}\big)\,\Big)\ \ \ {\bar Z}_\theta(\kappa)
\ .
\eea

\subsection{High-temperature expansion}

Here we  discuss the
high-temperature  behavior of the partition function  $Z_\theta$.

First and foremost we recall a simple qualitative picture described in Refs.~\cite{SLAZ,LVZ}.
As follows from Eq.~\eqref{flow}, the circular brane model
is a asymptotically
free theory -- at short
distances the effective size of the circle \eqref{bconstaint}\
becomes
large.
For small  but nonzero values of the  ``bare'' coupling constant $g_0$
we  can locally ignore   the curvature effects and
write the Bose  field  $X^{\mu}(\tau,x) =  X^{\mu}_0 + \delta
 X^{\mu}(\tau,x)$. It makes sense to split the fluctuational part, $\delta
 X^{\mu}$, into the components normal and tangent to the circle\ \eqref{bconstaint} at
the point $X^{\mu}_0$. The  components must satisfy the Dirichlet and
the Neumann BC, respectively. Therefore  the ``microscopic''  boundary degeneracy
for  $g_0\ll 1$ does not depend on $\theta$ and
coincides  with the product of the
boundary degeneracies associated with the Dirichlet $({\rm g}_D)$
and the
Neumann  CBC (${\rm g}_N={\rm g}_D\ {C\over 2\pi}$, where
$C=2\pi/\sqrt{g_0}$ is the full length of the circle\ \eqref{bconstaint}).
The above consideration supplemented by the
common wisdom of the renormalization group suggests that  as $E_*\ll T\ll \Lambda$,
the  partition function must  develop the   following leading high-temperature
behavior:
\bea\label{sssl}
Z_\theta= {{\rm g}_{\rm fixed}\over \sqrt{g(T)}}\Big[\, 1+O\big(\, g(T)\,\big)\, \Big]\ .
\eea
In writing the above equation,
we  replace  the bare coupling  $g_0$ in
the ``microscopic''  boundary degeneracy
${\rm g}_D^2/\sqrt{g_0}$   by
the ``running'' coupling constant $g(T)$.
As  follows from the two-loop
$\beta$-function\ \eqref{flow}, $g=g(T)$ 
can be defined by the equation:
\bea\label{lsalsalk}
\kappa\equiv{E_*\over T}=g^{-1}\ \re^{-{1\over 2g}}\ .
\eea
Also,  the factor ${\rm g}_{\rm fixed}$ in \eqref{sssl} should  be understood  as
\bea\label{sakjasa}
{\rm g}_{\rm fixed}={\rm g}_{D}^2\ .
\eea

The exact partition function~\eqref{spsalsli} indeed 
possesses the leading  short distance behavior~\eqref{sssl}.
The systematic high-temperature expansion  for $\theta=0$
was obtained in Ref.\cite{SLAZ}.
That result     can be easily generalized
to  $\theta\not=0$, yielding an expansion of the form 
\bea\label{kusasa}
Z_{\theta}\simeq
 {{\rm g}_{\rm fixed}\over
 \sqrt{g}} \ \Big({{\bar \Lambda}\over T}\Big)^{2\kappa\cos\theta}\
 g^{-\kappa\cos\theta}\ \sum_{n=0}^{\infty}\kappa^{n}\ z_n(g,\theta)\ ,
\eea
where the coefficients $z_n(g,\theta)$ are power series in the running coupling constant\
\eqref{lsalsalk}.
To describe  some coefficients in \eqref{kusasa}  explicitly,  it is convenient 
to represent
$Z_{\theta}$ as
\bea\label{lslsaj}
Z_{\theta}={{\rm g}_{\rm fixed}\over \sqrt{g}}\ \re^{{\chi\over T}\, \cos\theta}\
{\tilde Z}_{\theta}\ .
\eea
Here 
\bea\label{lsljsajs}
\chi=-T\ \partial^2_{\theta}\log(Z_{\theta})\big|_{\theta=0}\ ,
\eea
and therefore  the factor  ${\tilde Z}_{\theta}$
is subject to the
constraint
\bea\label{asskji}
{\partial^2{\tilde Z}_{\theta}
\over \partial \theta^2}\Big|_{\theta=0}=0\ .
\eea
It is possible to show that the  topological 
susceptibility $\chi$ expands explicitly  as
\bea\label{jsjsaj}
\chi\simeq E_*
\Big[ 2\log\big({\textstyle {{\bar \Lambda}\over T}}\big)-\log g+
3\gamma_E+\log 2-2\gamma_E\, g+O(g^2)-
{\textstyle{2\pi^2\over 3}} \kappa \big(1+O(g)\big)+O(\kappa^2)
 \Big]. 
\eea
The  expansion  coefficients
of ${\tilde Z}_{\theta}$~\eqref{lslsaj} are somewhat simpler
than $z_n$ in~\eqref{kusasa}.
For example,
\bea\label{slskjso}
{\tilde Z}_0=1-(1+\gamma_E)\, g+O(g^2,\kappa^2)\ ,
\eea
and
\bea\label{slskjsouy}
{\tilde Z}_\theta/{\tilde Z}_0=1-{\textstyle{4\pi^2\over 3}}\
\big(1+O(g^2)\big)\  \sin^4\big({\textstyle{\theta\over 2}\big)}\ \kappa^2+O(\kappa^3)\ .
\eea
Notice  that the coefficients ${2\pi^2\over 3}$ and ${4\pi^2\over 3}$ in  Eqs.\eqref{jsjsaj},\,\eqref{slskjsouy}
are in agreement with the result of two-instanton calculation from Ref.~\cite{Feigelman}.

\subsection{Low-temperature expansion}

For  $k=\infty$,
the boundary degeneracy ${\rm g}_{\rm fixed}$
in \eqref{hjdhdkss} 
should be understood as in Eq.~\eqref{sakjasa}.
It implies that the circular brane boundary flow is terminated at the
fixed point corresponding to the Dirichlet
CBC $X^{\mu}_B=0$
for any $\theta$ within the domain $0\leq \theta<\pi$. At the same time,
for $\theta=\pi$ the  flow possesses  a nontrivial IR fixed point whose boundary
degeneracy is given by
\bea\label{slslswe}
{\rm g}_{2}=2\ {\rm g}_{D}^2
\eea
(see Eq.~\eqref{lsalsa} with  $s=2$ and $k=\infty$).
The last equation suggests that the boundary values of $ X^{\mu}$ are constrained
to two points.
We refer the reader to Ref.~\cite{LTZ}
for a comprehensive  description  of this nontrivial RG fixed point.

\begin{figure}[ht]
\centering
\includegraphics[width=9cm]{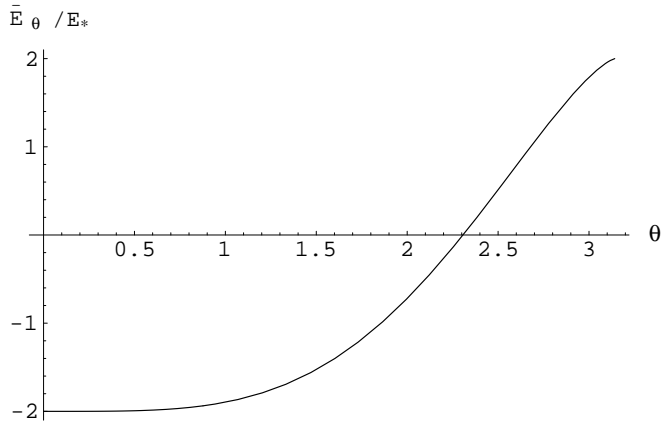}
\caption{The regularized
ground state energy~\eqref{jsiutre}
in the  circular brane model.}
\label{fig-pluiu}
\end{figure}

Here we describe the low-temperature expansion for  $\theta\not=\pi$.
In this case the free energy, $F_\theta=-T\log Z_{\theta}$,
 admits an asymptotic  expansion in terms of $T^2$:
\bea\label{lsals}
F_{\theta}\simeq -2 E_*\cos(\theta)\, \log\big({\textstyle {{\bar \Lambda}\over E_*}}\big)+
{\bar E}_\theta-
T\log({\rm g}_{D}^2)-
\sum_{l=1}^{\infty} F_l(\theta)\, {T^{2l}\over E_*^{2l-1}}\, .
\eea
The regularized ground state energy ${\bar E}_\theta$
is a large-$k$ limit of the function\ \eqref{wejsre}.
It explicitly reads 
\bea\label{jsiutre}
&& {\bar E}_\theta/ E_*=C_0-(\gamma_E+\log 2)\, \cos\theta+
\\ && \ \ \ \ \ \ \
\int_0^1\rd x\
\bigg[\, \sqrt{\pi}\, \cos^2(\theta)\ {\Gamma(x+{1\over 2})\over
\Gamma(x)}\ {}_3F_2\big(x+{\textstyle{1\over 2}}\,,\,
{\textstyle{1\over 2}}\,,\, 1\, ;\, 2\, ,\, {\textstyle{3\over 2}}\, |\, \cos^2(\theta)\, \big)
-\nonumber\\
&& \ \ \ \ \ \ \ \ \ \ \ \ \ \ \ \ \ {\textstyle {2\over 3}}\, \cos^3(\theta)\ x\
 {}_3F_2\big(x+1\,,\, 1\, ,\, 1\, ;\, 2
\,  ,\, {\textstyle{5\over 2}}\, |\, \cos^2(\theta)\, \big)\, \bigg]\ .\nonumber
\eea
Here  ${}_3F_2$ is
the generalized hypergeometric function and
\bea\label{ksjksaoqiu}
C_0=-\sqrt{\pi}\ \dashint_0^1\rd x\ {\Gamma(x-{1\over 2})\over
\Gamma(x)}=-1.44142\ldots\ .
\eea
The function~\eqref{jsiutre} is plotted in  Fig.~\ref{fig-pluiu}.
Notice that
it develops  a non-analytical behavior
at $\theta=\pi$:
\bea\label{saosaj}
{\bar E}_\theta/E_*= 2  +{\pi(\pi-\theta)\over \log (\pi-\theta)}+
O\Big({\pi-\theta\over \log^2 (\pi-\theta)}\Big)\ .
\eea
\bigskip

\bigskip

\noindent
The first two coefficients $F_l$ in the expansion\ \eqref{lsals} explicitly read as follows:
\bea\label{ssksl}
F_1(\theta)={\theta\over 12\sin\theta}\ ,
\eea
and
\bea\label{kkdkj}
F_2
(\theta)={1\over 720\,\sin^3\theta}\ \int_{0}^{\theta}\rd t\
\bigg[\, 6\, \log^2\Big({{\sin\theta\over \sin t}}\Big)-
7\, \log\Big({{\sin\theta\over \sin t}}\Big)-5\, \bigg]\ .
\eea

\section{\label{threesec} The dissipative quantum rotator model}

The circular brane  model has useful interpretation in terms of
Brownian dynamics of a quantum rotator. It was noticed a while ago in
Ref.~\cite{Calan}\ that the
free massless bulk dynamics~\eqref{baction} is equivalent to the
Caldeira-Leggett model of a quantum thermostat~\cite{Legett}.
Upon fixing the boundary
values $X^{\mu}_B(\tau)$
and integrating out the  bulk part of
the field $ X^{\mu}(\tau,x)$,
Eqs.~\eqref{topsum},~\eqref{salskljsk}\
reduce to
\bea\label{partdissip}
Z_\theta ={\rm g}_{\rm fixed}\  \int\,{\cal D}\eta\
\re^{-{\mathscr  A}_{\rm diss}[\,\eta\, ]-{\mathscr  A}_{ \theta}[\,\eta\, ]}\ ,
\eea
where
\bea\label{disact} {\mathscr  A}_{\rm diss}[\,\eta\,] =
{T^2\over{2 g_0 }}\ \int_{0}^{1\over T} \rd\tau \int_{0}^{1\over T}
\rd\tau'\, {{\sin^2\big(\, {{\eta(\tau)-\eta(\tau')}\over 2}\, \big)
}\over\sin^2 (\pi T(\tau-\tau'))
}\ ,
\eea
and
\bea\label{disactte} {\mathscr  A}_{\theta}[\,\eta\,] = -
{\ri\theta\over 2\pi}\, \int_{0}^{1\over T}\rd\tau\,\partial_{\tau}\eta(\tau)\ .
\eea
The  variable $\eta(\tau)$ is the angular
field defined through
\bea\label{sajsakas}
(X_B,Y_B)={1\over \sqrt{g_0}}\ (\cos \eta,\, \sin \eta )\ .
\eea
The path integral \eqref{partdissip}  requires some  UV regularization.
Among a large variety of
regularizations 
there is one of
special interest for applications~\cite{AES,Zaikin}:
\bea\label{cutoff}{\mathscr A}_{\rm DQR}[\,\eta\,]
 = {\mathscr A}_{\rm diss}[\,\eta\,]+{\mathscr  A}_{ \theta}[\,\eta\, ] + {\mathscr A}_{ C}[\,\eta\,]\ ,
\eea
where
\bea\label{kjkjj}
 {\mathscr A}_{C}[\,\eta\,]=  {1\over{4E_C}}\, \int_{0}^{1\over T} \rd\tau\,\eta^2_{\tau}\ .
\eea
The functional \eqref{cutoff}
amounts to
the Caldeira-Leggett action for the dissipative quantum rotator.
In the weak coupling regime  $(g_0\gg 1)$
 the additional term\ \eqref{kjkjj}  just provides an
explicit UV cutoff of the dissipative action\ \eqref{disact}
with the cutoff energy $\Lambda\sim E_C/g_0$.
Now,
once the regularization scheme is chosen,
all coefficient $c_l$
in the asymptotic  expansion\ \eqref{scale} are determined  unambiguously
within the standard perturbation theory, in
particular 
\cite{SLAZ}
\bea\label{lslasaj}
c_1=-{\textstyle{3\over 4}}\ \pi^2\ .
\eea

The  additional  term~\eqref{kjkjj} not only makes the
circular brane model perturbatively well-defined,
but also
regularizes the  small instanton divergence.
For example,   the one-instanton contribution
to the  topological susceptibility\ \eqref{lsljsajs}
in the Gaussian approximation reads  as follows\ \cite{Wang,Larkin}:
\bea\label{lssjlaj}
\chi^{(\rm pert)}={E_C\over 2\pi^2 g_0^2}\,  \re^{-{1\over 2 g_0}}\
\bigg[\, 2\ \int_{a a^*<1} {\rd a\wedge \rd a^*\over
2\pi \ri }\
{\re^{-{\mathscr  A}^{(1)}_{C}}
\over  1-a a^*}+
O(g_0)\, \bigg]\ .
\eea
Here ${\mathscr  A}^{(1)}_{C}$
is the functional\ \eqref{kjkjj} evaluated for  the
one instanton solution.
The  exponential factor $\re^{-{\mathscr  A}^{(1)}_{C}}$
makes the integral over the instanton moduli $|a|<1$ finite.
Indeed, using the   explicit form
of  the one instanton solution  $\eta^{(1)}$\ \cite{Korshunov,Nazarov},
\bea\label{slsj}
\exp\big(\ri \eta^{(1)}(\tau)\big)=
{z-a\over
1-a^*\, z}\, , \ \ \ {\rm where }\ \ \ \ z=\re^{2\pi T \ri\tau}\ ,
\eea
one finds
\bea\label{salksaja}
{\mathscr  A}^{(1)}_C
={\pi^2 T\over  E_C}\ {1+aa^*\over 1-aa^*}\ .
\eea
Now we can integrate over the instanton moduli explicitly:
\bea\label{slksahks}
\chi^{(\rm pert)}={E_C\over 2\pi^2 g_0^2}\   \re^{-{1\over 2 g_0}}\,
\Big[\,  2\,\log\Big({ E_C\re^{-\gamma_E}\over 2\pi^2 T}\Big)+
O\big(g_0,T/E_C\big)\, \Big]\, .
\eea
It is instructive to compare\ \eqref{slksahks}
with the high-temperature expansion\ \eqref{lsljsajs}.
In essence, Eq.\eqref{lsljsajs}
requires
that the   bare coupling constant   expansion
should be of the form
\bea\label{aolaoaias}
\chi^{(\rm pert)}= E_*\, \Big[\, 2\log\big({\textstyle {{\bar \Lambda}\over T}}\big)-\log g_0+
3\gamma_E+\log 2+O(g_0,T/\Lambda )\, \Big]\  .
\eea
Combining  Eqs.\eqref{slksahks} and \eqref{aolaoaias},\,\eqref{scale} yields
the following relations between
the perturbative cut-off ($\Lambda$), the small instanton
cut-off (${\bar \Lambda}$),
and the  energy scale  $E_C$ in  \eqref{kjkjj}:
\bea\label{asksjudsp}
\Lambda={E_C\over 2\pi^2 g_0}\ ,
\eea
and
\bea\label{salksj}
{\bar \Lambda}=E_C\
{ \re^{-{5\gamma_E\over 2}}\over \sqrt{8}\pi^2}\ \sqrt{g_0}\ .
\eea

\section{\label{conclus} Conclusion: Charge fluctuations in  quantum dot}

\begin{figure}[ht]
\centering
\includegraphics[width=7cm]{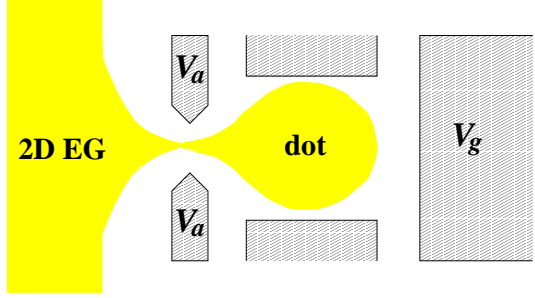}
\caption{ Schematic view of a quantum dot
connected to a bulk 2D electrode. The dot is formed by applying
a negative voltage to the gates. Electrostatic conditions in
the dot are controlled by the gate voltage $V_g$. Voltage $V_a$ applied to
the auxiliary gates controls the transmission through the constriction. }
\label{fig-dot}
\end{figure}

The dissipative rotator  model\ \eqref{cutoff}
was
introduced in Ref.~\cite{AES}
as an effective
field theory describing tunneling of quasiparticles between
superconductors. Later  it has   been involved   in a  surge of activity  in the study of
a low-capacitance metallic island (quantum dot), connected to
an outside lead by a tunnel junction 
\cite{Averin, Matveev, Nazarov, Wang,Zwerger,Feigelman,Larkin}.
A remarkable feature of such a device (see Fig.\ref{fig-dot}, borrowed from
Ref.\cite{Matveev})    is that the essentially many-body phenomena
may be described via a single variable $\eta$ (conjugate to the
number of excess charges on the island).
As was shown in Refs.~\cite{AES, Nazarov},
the fermionic degrees of freedom  can be formally integrated out yielding
the Matsubara
effective action   ${\mathscr A}_{\rm eff}$ for $\eta$.
The Coulomb part of this action
coincides with
${\mathscr A}_{\theta}+{\mathscr A}_{C}$~\eqref{disactte},~\eqref{kjkjj}
provided
the  energy scale $E_C$ is understood now  as
the single electron charging   energy,  i.e.,
\bea\label{salsis}
E_C={e^2\over 2 C}\ ,
\eea
where  $C$ is the island   capacitance.
The parameter
\bea\label{salss}
n_g={\theta\over 2\pi}\
\eea
has also the simple physical meaning of
a  dimensionless  applied  gate voltage $V_g$:
$n_g=C_g\, V_g/e$, where $C_g$ is the gate capacitance (see Fig.~\ref{fig-dot}).

In addition to the Coulomb terms, the effective action contains a highly nontrivial dissipative
piece which  essentially
depends on
details of  the  island-lead interaction.
In fact,
the transmission coefficients $\{\, T_{a}\,\}_{a=1}^{N}$
in the  island-lead interface
are coupling constants of   ${\mathscr A}_{\rm eff}$ 
($N$ is a number  of spinless electron  channels).
The most general   effective action appears to be  too   complicated for
practical purposes. It simplifies drastically in the special   large-$N$ limit
when each individual  transmission coefficient $T_{a}$ vanishes, but
the  dimensionless (measured in units $e^2/h$)  tunneling conductance
\bea\label{laskhjsa}
\alpha={1\over 2\pi^2}\ \sum_{a=1}^NT_a\ ,
\eea
remains finite.
In this limit ${\mathscr A}_{\rm eff}$\
takes   the form of  the dissipative rotator action\ \eqref{cutoff} with
\bea\label{sjlsaj}
g_0={1\over2\pi^2 \alpha }\ ,
\eea
so the basic  thermodynamic quantities
can be extracted from the corresponding
partition function, $Z_{\rm DQR}$.
The  renormalized charging energy ${\tilde E}_C$
\bea\label{jassaj}
{\tilde E}_C=-2\pi^2\, T\ \partial_{\theta}^2\log \big(\,Z_{\rm DQR}\,\big)
\eea
is of special  interest.
As it has been  argued above,
the quantum dissipative rotator model
possesses the  universal scaling behavior provided  $g_0\ll 1$ and
$E_C\gg T$.
This corresponds to the regime of  almost perfect transmission
for the low-capacitance 
metallic island.
In the  scaling regime   $Z_{\rm DQR}$
turns into the circular brane partition function $Z_{\theta}$.
Thus the exact  results obtained in this work give
a  complete description of the  renormalized charging energy in  the scaling limit.
In particular at $T=0$, 
\bea\label{slss}
{{\tilde E}_C\over 2\pi^2 E_*}\Big|_{T=0}\,\longrightarrow\, 2\, \cos(\theta)\
\log\bigg[{ E_C\, \re^{-{5\over 2}\gamma_E}\over 4\pi^3 E_*
\sqrt{\alpha}}\bigg]+
{\partial^2 \over \partial \theta^2}\,
\Big({{\bar E}_\theta\over E_*}\Big)\ \ \ \ \ (\,\alpha,\,E_C\to\infty\,)\ .
\eea
Recall that the function ${\bar E}_\theta/E^*$ is
given by  Eq.~\eqref{jsiutre} and plotted in Fig.~\ref{fig-pluiu}, while
\bea\label{klsajks}
E_*=2\pi^2\ E_C\ \alpha^2\ \big(1-{\textstyle{3\over 8\alpha}}+O(\alpha^{-2})\, \big)\
\re^{-\pi\alpha^2}\ .
\eea
It is worth  noting  that  the quoted  analytical results for $E_C$ with $\theta=0$
have been numerically checked   in Ref.~\cite{SWar} via Monte Carlo simulations
of $Z_{\rm DQR}$.

The quantitative  description of the quantum dot with finite number of
channels remains a challenging problem.
For finite $N$, one  would  still expect some sort of universal scaling behavior  characterized
by the energy  scale~\cite{Nazarov}
\bea\label{salsak}
E_*\propto E_C\ \prod_{a=1}^N\sqrt{1-T_a}\ .
\eea
In the case of  $N=2$ this was explicitly demonstrated in Ref.~\cite{Matveev}.
Matveev argued that the universal behavior  of  the two-channel quantum dot
is described  by the Hamiltonian ${\boldsymbol H}^{(k)}_{\theta}$~\eqref{ssakjiuy} with   $k=2$.
Thus  the thermodynamic properties  of the system can be extracted  from
the Chatergee-Zamolodchikov partition function\ \eqref{ksksj}.
With regard to  $N=2$ and $N=\infty$ ``integrable'' cases,
it would be interesting to explore the possibility of using
${\boldsymbol H}^{(k)}_{\theta}$ \eqref{ssakjiuy} as a  QFT Hamiltonian
underlying  the
universal scaling behavior  of the $N$-channel  quantum dot provided $N=k$.

\section*{Acknowledgments}

I am deeply  indebted  to Alexander B. Zamolodchikov
for teaching  me QFT, previous collaboration and
interest in this work.
I also would like to thank  Alexei B. Zamolodchikov for sharing  insights from
the unpublished work~\cite{AlZ},  and
V.V. Bazhanov, G.Y. Chitov,  V.A. Fateev, 
A. Leclair,  A.M. Tsvelik  and P. Werner
for interesting discussions and important comments.

\bigskip

\noindent This research  is supported in part by DOE grant
$\#$DE-FG02-96 ER 40949.

\appendix

\section{
Appendix : The BP  sine-Gordon model}

In this appendix we  shall study the  BP
sine-Gordon model.
It  provides an  essential step
toward
the main result \eqref{dsgfasta}.

\subsection{Definition of the model}

Let ${\boldsymbol  H}_{\rm free}$ be a Hamiltonian of  the boundary
CFT model
which contains  two noninteracting  sectors,
namely
the  minimal parafermionic CFT   and the  Gaussian theory  of  the  free uncompactified Bose field:
\bea\label{lssaasj}
{\boldsymbol  H}_{\rm free}={\boldsymbol  H}_{\rm free}^{(k)}+
{\boldsymbol  H}_{\rm free}^{(\Phi)}\ .
\eea
Suppose the parameter
$b$ in Eq.\eqref{slsasl}  is purely imaginary,
\bea\label{jssyhstg}
b=\ri\, \beta\ ,
\eea
so that the boundary
fields\ \eqref{kausy}  take the form:
\bea\label{alsksausy}
{\boldsymbol V}_\pm=
 {\boldsymbol \Psi}_\pm\ \re^{\pm\ri {\beta\over\sqrt{k}} {\boldsymbol \Phi}_B}\ .
\eea
Their scaling dimension is given by
\bea\label{llskjsl}
d=1-{1\over k}+{\beta^2\over k}\ .
\eea
For $0<\beta^2<1$, ${\boldsymbol V}_\pm$
are relevant boundary fields normalized as  in \eqref{uyttkalhj}, and  in what follows
we will  focus on   the  model described by the Hamiltonian
\bea\label{jksahsa}
{\boldsymbol H}_{\rm bsg}^{(k)}={\boldsymbol H}_{\rm free}-
\mu\, \Big(\,   \re^{\ri Vt}\ {\boldsymbol V}_++
\re^{-\ri Vt}\,  {\boldsymbol V}_-\, \Big)\ ,
\eea
where $\mu$ and $V$ are parameters.
The parameter $\mu$ carries the dimension
of $[\, length\,]^{d-1}$.
The boundary interaction contains an explicit time dependence through the factors $\re^{\pm \ri Vt}$.
Below we always assume that $V>0$. Notice that, although by means
of  the formal  change of the field variable,
${\boldsymbol \Phi}\to {\boldsymbol \Phi} -{\sqrt{k} \over\beta}\, t$,
the Hamiltonian \eqref{jksahsa} can be brought to the form\ \eqref{uyywtrsahsa},
effects of the external fields $V$ in \eqref{jksahsa} and $h$ in \eqref{uyywtrsahsa} are very
different. Whereas the system \eqref{uyywtrsahsa} possesses  the  thermal equilibrium state even for
non-zero values of $h$,
the system \eqref{jksahsa},
at a nonzero $V$ and a temperature $T$,  develops a
 nonequilibrium steady   state
which can be thought of as the result of an
infinite time evolution of the equilibrium state of the corresponding
``free'' system, with the interaction term
adiabatically switched on. We  denote by
$\langle\,{\boldsymbol  O}\, \rangle_{\rm bsg}^{(k)}$  the
expectation value of an observable ${\boldsymbol  O}$ over this
nonequilibrium steady  state.

The model\ \eqref{jksahsa}\ was already discussed in the literature.
In particular in the absence of the parafermionic sector ($k=1$)  it
coincides with the massless boundary sine-Gordon model \cite{gz,Warner}.
For this reason we shall call\ \eqref{jksahsa}
the  BP  sine-Gordon model.
Besides being an interesting model of Quantum Field Theory on its own,
the theory finds important  applications in condensed matter physics.
In Ref.~\cite{saleur} the model \eqref{jksahsa}  was suggested to describe
a point-like  impurity in the ``multichannel quantum wire''\footnote{The cases $k=1$ (without
the parafermionic sector)  and $k=2$ were previously studied
in Refs.\cite{Moon,FSLSN,Fen} and\ \cite{Kane,Simonetti}, respectively.}.
In this context
the integer $k$ is the number of spinless electron channels,
$\alpha_0=k\beta^2$ is the conductance (in  units $e^2/ h$)  of the
wire without impurity,
while $V$ is proportional to the
voltage drop across the impurity.
The quantities of special  interest
are  correlation functions of the Heisenberg  boundary operators
\bea\label{vpm}
{\boldsymbol  V}_{\pm}(t) = \re^{\pm \ri Vt}\
 {\boldsymbol  U}^{-1}(t)\, {\boldsymbol  V}_{\pm} \,  {\boldsymbol U}(t)\ ,
\eea
where ${\boldsymbol U}(t)$ is the time
evolution operator corresponding to the Hamiltonian\ \eqref{jksahsa}.
For instance, the energy  dissipation  rate    ${\dot E}$ for the steady state,
produced by
the time dependent  boundary perturbation
\eqref{jksahsa},
can be expressed
in terms of the expectation  values $\langle\, \boldsymbol V_{\pm}\, \rangle_{\rm bsg}^{(k)}$.
Indeed, as follows from \eqref{jksahsa}:
\bea\label{lsssl}
{\dot E}=- J_D V\ \ \ {\rm  with}\ \
J_D=\ri\, \mu\ \big(\, \langle\,{\boldsymbol  V}_{+}\, \rangle_{\rm bsg}^{(k)}-
\langle\,{\boldsymbol  V}_{-}\, \rangle_{\rm bsg}^{(k)}\, \big)
\ .
\eea
Notice that the quantity
\bea\label{alsksjl}
\alpha(V,T)=\alpha_0\ \big(1+\textstyle {2\pi \ {\beta^2\over k}}\
J_D/V\, \big)
\eea
is interpreted as a DC conductance of the multichannel
quantum wire \cite{saleur}.

\subsection
{Perturbative expansions for $\langle\,{\boldsymbol  V}_{\pm}\, \rangle_{\rm bsg}^{(k)}$}

If $V>0$  the system\ \eqref{jksahsa} evolves towards a steady state which is
characterized by the expectation values $\langle\,{\boldsymbol  V}_{\pm}\, \rangle_{\rm bsg}^{(k)}$.
This is   not a thermodynamic equilibrium state and no
simple explicit expression for its density matrix is known. Nevertheless,  one can
describe the definition of this state in terms of the real-time perturbation
theory\ \cite{Kel}.  It is useful to represent
the perturbative expansions for  $\langle\,{\boldsymbol  V}_{\pm}\, \rangle_{\rm bsg}^{(k)}$ in the
form
\bea\label{sssa}
\langle\,{\boldsymbol  V}_{\pm}\,
\rangle_{\rm bsg}^{(k)}={T\over 2}\, \partial_{\mu}\log A_{\pm}\  ,
\eea
where
$\log A_{\pm}$  are  formal   power series expansions,
\bea\label{ksasksa}
\log A_{\pm}=\sum_{n=1}^{\infty}a_{n}^{(\pm)}\  s^{n}\ \ \
{\rm with}\ \ s=\mu^2\, (2\pi T)^{2d-2}\ .
\eea
For $k=1$ the compact formula for the  perturbative coefficients
$a_{n}^{(\epsilon)}$
were obtained in  Refs.\cite{Wei,blz5}.
In essence
the derivation from \cite{blz5}  can be applied without  change
to the BP sine-Gordon model.
As a result, the following formula  emerges:
\bea\label{wf} a_{n}^{(\epsilon)} =\epsilon\ (-1)^{n+1}\
2^{2n}\ \ {\pi \over  n}
  \
 \sum_{\epsilon_1,\ldots,\epsilon_{2n-1}} \ \prod_{j=1}^{2n-1} \sin
(\pi d \eta_j)\ \
J_{\epsilon,\epsilon_1,\ldots,\epsilon_{2n-1}}\, ,
\eea
where the sum is taken over all arrangements of the ``charges''
$\epsilon_1,\ldots,\epsilon_{2n-1}=\pm 1$
with zero total charge $\epsilon+\sum_{s=1}^{2n-1}\epsilon_s=0$, and
\bea\label{cumcharges}
\eta_j=\epsilon+\sum_{s=1
}^{j-1} \epsilon_s\ .
\eea
The coefficients $J_{\epsilon,
\epsilon_1\ldots\epsilon_{2n-1}}$ in \eqref{wf} are expressed through the
$(2n-1)$-fold integrals over
the  real-time  correlation functions:
\bea\label{slsjsu}
&&J_{\epsilon,\epsilon_1\ldots\epsilon_{2j}}=\int_{-\infty}^0\rd t_1\int_{-\infty}^{t_1}
\rd t_2\cdots \int_{-\infty}^{t_{2n-2}}\rd t_{2n-1}\  \re^{\ri V\sum_{j=1}^{2n-1} \epsilon_j t_j}\times\\ &&
(2\pi T)^{2n(1-d)}\ \, \re^{-\ri\pi d\sum_{j=1}^{2n-1} \epsilon_j \eta_j}\
\big\langle\big\langle\,{\boldsymbol V}^{(\rm int)}_{\epsilon}(0)\,
{\boldsymbol V}^{(\rm int)}_{\epsilon_{1}}(t_{1}) \cdots
{\boldsymbol V}^{(\rm int)}_{\epsilon_{2n-1}}(t_{2n-1})\, \big\rangle\big\rangle_0\, .
\nonumber \eea
Here  ${\boldsymbol V}^{(\rm int)}_\pm(t)=\re^{\ri t {\boldsymbol H}_{\rm free} }\, {\boldsymbol V}_{\pm}\,
\re^{-\ri t {\boldsymbol H}_{\rm free} }$
 are  the boundary  operators in the interaction representation
and
\bea\label{khskskiw}
\big\langle\big\langle\,\cdots\, \big\rangle\big\rangle_0\equiv {\rm Tr}_{\cal H}\big[\,\cdots\,
\re^{ -{{\boldsymbol H}_{\rm free}\over T }}\big]/{\rm Tr}_{\cal H}\big[\re^{ -{{\boldsymbol H}_{\rm free}\over T }}
\big]
\eea
denotes an expectation value over
the  equilibrium  thermal state of the free system.

As  follows from the real-time OPE
\bea\label{slsre}
{\boldsymbol V}^{(\rm int)}_\pm (t){\boldsymbol V}^{(\rm int)}_\mp(t')\to \big(\, \ri\, (t-t')\,\big)^{-2d}
\, , \  \ \ \ \ 0<t-t'\to 0\ ,
\eea
the integrals\ \eqref{slsjsu} can be taken literally only if
the scaling dimension  $d<{1\over 2}$. For  ${1\over 2}<d<1$ we assume the ``analytic
regularization'' common in  conformal perturbation theory, i.e.
the integrals should be understood as analytically continued
from the convergence domain $\Re e\, d<{1\over 2}$.
For the sake of illustration we consider
explicitly the   first  coefficient $a_{1}^{(\pm)}$.
In this case the  correlators
appearing in the integrand of \eqref{slsjsu} are given by
\bea\label{isss}
\re^{\ri\pi d}\ \big\langle\big\langle\, {\boldsymbol V}^{(\rm int)}_{\pm}(0)\,
{\boldsymbol V}^{(\rm int)}_{\mp}(t_1)\, \big\rangle\big\rangle_0=
\bigg[\, {\pi T\over \sinh(-\pi T t_1)}\,\bigg]^{2d}\
\ \ \ \ (t_1<0)\ .
\eea
Calculating the elementary integral over $t_1$, one obtains
\bea\label{ksahsak}
a_{1}^{(\pm)}=4\pi\ \sin(\pi d)\  \ {\Gamma(1-2d)\,\Gamma(d\pm 2p)\over
\Gamma(1-d\pm 2p)}\ \ \ \ \ {\rm with}\ \ p=-\ri\ {V\over 4\pi T}\ .
\eea

\subsection{\label{secsixtre}
The system with $q$-oscillator}

Remarkably,  the formal power series \eqref{ksasksa}
can be interpreted as  high\,-\,temperature expansions
of the
{\it equilibrium}  free energy for  certain systems which differ from\ \eqref{jksahsa} in that
they involve  additional boundary degrees of freedom.
Originally this was demonstrated in Ref.\,\cite{blz5} for  the boundary
sine-Gordon model $(k=1)$, but
the same reasoning can be applied  to
the  generalized model\ \eqref{jksahsa}.

Let us
define the Hamiltonian
\bea\label{lsasal}
{\boldsymbol H}_{+}=
{\boldsymbol H}_{\rm free}-V{\boldsymbol h} -\mu\ \Big(\,   {\boldsymbol a}_-\ {\boldsymbol V}_++
{\boldsymbol a}_+\,  {\boldsymbol V}_-\, \Big)\ .
\eea
Here we use the same notations as in \eqref{jksahsa}.
Formula \eqref{lsasal} includes a new ingredient, namely, the
operators ${\boldsymbol h}$, ${\boldsymbol a}_+$, ${\boldsymbol a}_-$ which
commute with all the parafermionic and bosonic degrees of freedom.
They form among themselves the so-called
``$q$-oscillator algebra'',
\bea\label{ssmsas}
[\,{\boldsymbol h}\, ,{\boldsymbol a}_{\pm}\,]=\pm {\boldsymbol a}_{\pm}\, ,\ \ \ \
q\, {\boldsymbol a}_{+} {\boldsymbol a}_{-}-q^{-1}\, {\boldsymbol a}_{-} {\boldsymbol a}_{+}=
q-q^{-1}\ ,
\eea
with
\bea\label{ssljs}
q=\re^{\ri \pi d}\ .
\eea
Recall that the parameter $1-{1\over k}<d<1$ is  the scaling dimension of the boundary fields ${\boldsymbol V}_\pm$\ \eqref{llskjsl}.
Let $\rho_+$ be some representation of
\eqref{ssmsas}   such that the spectrum of $\rho_+({\boldsymbol h})$ is real and
bounded from {\it above}.  The Hamiltonian\ \eqref{lsasal}\ acts in the space
\bea\label{ksashks}
{\cal H}_+={\cal H}\otimes\rho_{+}\ ,
\eea
with ${\cal H}$ given by\ \eqref{llsajl}.
For $V>0$ this Hamiltonian is bounded from below. Then,  the system \eqref{lsasal}
possesses a thermal equilibrium state described by the standard density matrix, $Z^{-1}_+(\mu, V)\,
\re^{-{{\boldsymbol H}_+\over T}}$.

Strictly speaking the partition function
\bea\label{kjshsk}
Z_+(\mu, V)={\rm Tr}_{{\cal H}_+}\Big[\,  \re^{-{{\boldsymbol H}_+\over T}}
\, \Big]
\eea
is
ill-defined even for the non-interacting system. Indeed,
since  the space of states ${\cal H}_+$  has a structure of
tensor product (see Eqs.\,\eqref{llsajl},\,\eqref{ksashks}), the partition function
$Z_+(\mu, V)|_{\mu=0}$
factorizes into a product of three terms. The first one
coincides with the boundary degeneracy ${\rm g}_{\rm free}$
\eqref{aslss}. The  component $\rho_{+}$ in \eqref{ksashks} gives
rise to the factor ${\rm Tr}_{\rho_+}\big[\, \re^{V{\boldsymbol h}\over
T}\, \big]$ which is well-defined
if the spectrum of $\rho_+({\boldsymbol h})$ is bounded from  above.
The problem  comes from a  divergent  contribution of the   zero-mode 
of the uncompactified  Bose field subjected by the  free CBC.
To cancel  this formal divergence  we shall consider  below  the ratio
${Z_+(\mu,V)/ Z_+(0,V)}$.

Carrying out the finite temperature Matsubara procedure,
one obtains the  weak coupling expansion of the  form
\bea\label{lssaljsa}
{Z_+(\mu,V)\over Z_+(0,V)}=1+\sum_{n=1}^{\infty}\  A^{(+)}_n\ s^{n}\ \ \
{\rm with}\ \ s=\mu^2\, (2\pi T)^{2d-2}\ ,
\eea
where the expansion  coefficients are given by
\bea\label{lsajslasa}
A^{(+)}_n=\sum_{\epsilon_k=\pm\atop
\epsilon_1+\cdots\epsilon_{2n}=0}
T_{\epsilon_1\ldots\epsilon_{2n}}\ G_{\epsilon_1\ldots\epsilon_{2n}}\, .
\eea
Whereas  $T_{\epsilon_1\ldots\epsilon_{2n}}$ are expressed in terms of  traces over
the representation $\rho_+$,
\bea\label{kjsk}
T_{\epsilon_1\ldots\epsilon_{2n}}={
{\rm Tr}_{\rho_+}\big[\, \re^{V{\boldsymbol h}\over T}\, {\boldsymbol a}_{-\epsilon_1}\cdots
{\boldsymbol a}_{-\epsilon_{2n}}\, \big]\over
{\rm Tr}_{\rho_+}\big[\, \re^{V{\boldsymbol h}\over T}\, \big]}\ ,
\eea
the coefficients $G_{\epsilon_1\ldots\epsilon_{2n}}$ are multiple integrals:
\bea\label{slsjlaas}
G_{\epsilon_1\ldots\epsilon_{2n}}&=&  \int_{0}^{{1\over T}}\rd\tau_{1}
\int_{0}^{\tau_{1}}\rd\tau_{2}\cdots
\int_{0}^{\tau_{2n- 1}}\rd\tau_{2n}\ \re^{V \sum_{j=1}^{2n}\epsilon_j\tau_j}\times\\ &&
(2\pi T)^{2n(1-d)}
\ \  \langle\langle\, {\boldsymbol V}^{\rm (Mats)}_{\epsilon_{1}}(\tau_{1})\cdots\,
{\boldsymbol V}^{\rm (Mats)}_{\epsilon_{2n}}(\tau_{2n})\, \rangle\rangle_0\ .\nonumber
\eea
Here ${\boldsymbol V}^{\rm (Mats)}_\pm(\tau)=
\re^{\tau {\boldsymbol H}_{\rm free}}\,{\boldsymbol V}_\pm\,
\re^{-\tau {\boldsymbol  H}_{\rm free}}$ are unperturbed  Matsubara
operators associated with the  boundary operators
\eqref{alsksausy},
and $\langle\langle\,\cdots\, \rangle\rangle_0$  is defined in \eqref{khskskiw}.
As well as in formula \eqref{slsjsu},
we should  assume  the analytic regularization
of $G_{\epsilon_1\ldots\epsilon_{2n}}$ in the case    ${1\over 2}<d<1$.
One might also note that all the coefficients $T_{\epsilon_1\ldots\epsilon_{2j}}$
in \eqref{lsajslasa} are  determined unambiguously  via  the commutation relations  \eqref{ssmsas} and
the cyclic property of  trace. Thus,   they  do not depend on  a particular choice of
the representation
$\rho_+$ provided ${\rm Tr}_{\rho_+}\big[\,\re^{V{\boldsymbol h}\over T}\,\big]$ exist.

Now, following  along the lines of  Ref.\,\cite{blz5}, it is possible to  prove  that the
formal power series $A_+$
defined in  equation
\eqref{ksasksa}\ coincides with  \eqref{lssaljsa}, i.e.,
\bea\label{sajsahsa}
A_+={Z_+(\mu,V)\over Z_+(0,V)}\ .
\eea
Of course,  $A_-$\ \eqref{ksasksa} is also related  to
a certain
equilibrium-state   partition function.
Namely, let $\rho_{-}$ be any
representation of\ \eqref{ssmsas} such that the spectrum of
$\rho_{-}({\boldsymbol h})$ is bounded from {\it below}.
Consider the Hamiltonian
\bea\label{ssmmasal}
{\boldsymbol H}_{-}={\boldsymbol H}_{\rm free}+V{\boldsymbol h}
-\mu\ \Big(\,   {\boldsymbol a}_+\ {\boldsymbol V}_++
{\boldsymbol a}_-\,  {\boldsymbol V}_-\, \Big)
\eea
acting in ${\cal H}_-={\cal H}\otimes \rho_{-}$ and the associated partition function
\bea\label{uythsk}
Z_-(\mu, V)={\rm Tr}_{{\cal H}_-}\Big[\, \re^{-{{\boldsymbol H}_-\over T}}
\, \Big]\ .
\eea
Similarly to\ \eqref{sajsahsa}, one can prove the relation
\bea\label{lsashksa}
A_-={Z_-(\mu,V)\over Z_-(0,V)}\ .
\eea

\subsection{Characteristic properties of $A_{\pm}$}

Equation\,\eqref{ksasksa}  shows that
$\log A_{\pm}$ are formal power series   of the dimensionless variable
\bea\label{asalkls}
s=\mu^2\ (2\pi T)^{2d-2}\ ,
\eea
while the expansion coefficients $a_n^{(\pm)}$  depend on the dimensionless ratio  $V/T$.
In what follows, we  treat   $a_n^{(\pm)}$ as  functions
of the complex  variable
\bea\label{slksjsl}
p= -\ri\ {V\over 4\pi T}\ .
\eea
Here are the  main
properties of $A_{\pm}(s, p)$.

\begin{itemize}

\item
$\log A_{\pm}(s, p)$ are formal power series   of the form \eqref{ksasksa},
or, equivalently,
\bea\label{ssssjl}
A_{\pm}(s, p)=1+\sum_{n=1}^{\infty}A_n^{(\pm)}(p)\, s^{n}\ .
\eea
The coefficients $a_{n}^{(\pm)}(p)$ \eqref{ksasksa} and
$A_n^{(\pm)}(p)$\ \eqref{ssssjl}
are meromorphic functions in the whole complex plane
of
$p$ and
\bea\label{sjsjs}
A_{n}^{(-)}(p)=A_{n}^{(+)}(-p)\ ,\ \ \ \ \ \ a_{n}^{(-)}(p)=a_{n}^{(+)}(-p)\ .
\eea

\item
Both $a_{n}^{(+)}(p)$ and $A_n^{(\pm)}(p)$  are   analytic
in the right half-plane,
\bea\label{lsljsal}
\Re e\, p>p_n\ \ \ {\rm with}\ \ \ 2\,p_n=(1-d)\, n-1\ .
\eea
For $n=1,\,2,\,\ldots k$,  they  have a simple pole
located at $p=p_n$ of residue
\bea\label{ssaslsa}
{\rm Res}_{p=p_n}\big[\,A_n(p)\,\big]&=&{\rm Res}_{p=p_n}\big[\,a_n(p)\,\big]=\\ &&
\bigg[{2\pi\over \sqrt{k}\,\Gamma(d)}
\bigg]^{2n}\ {k!\, \Gamma(1-n(1-d))\over
2\, n!(k-n)!\,\Gamma(n(1-d))}\ .\nonumber
\eea
There is no singularity at  $p=p_n$ for $n>k$, therefore  the domain of analyticity
\eqref{lsljsal}
can be extended to the half plane $\Re e\, p>p_n-\delta_n$ with some finite
$\delta_n>0$.

\item
The following leading  power-low asymptotic holds in the domain of analyticity,
\bea\label{ljald}
a^{(+)}_{n}(p)\to C_n\ p^{1-2 n(1-d)}\ \ \ {\rm as}\ \ \ p\to\infty\ \ \ (\Re e\, p>p_n)\ ,
\eea
where $C_n$ is some constant.

\item
The formal power series $A_\pm(s, p)$ obey the so-called
``quantum Wronskian'' relation,
\bea\label{sjssjk}
\re^{2\pi\ri p} A_+(qs,p)A_-( q^{-1}s,p)-
\re^{-2\pi\ri  p} A_+( q^{-1}s,p)A_-( qs,p)=
2\ri\ \sin(2\pi p)\ ,
\eea
with $q=\re^{\ri\pi d}$.

\end{itemize}

\bigskip

\noindent
The proof of   functional relation  \eqref{sjssjk}
is based on  results of the work~\cite{blz3}.
It
is outlined  in Appendix B.
All   other  properties    readily follow
from formulas \eqref{wf} and \eqref{lssaljsa}.
For example   the  integrals\ \eqref{slsjlaas}
are entire functions  of the variable $p$\ \eqref{slksjsl}.
Thus,   singularities  of  the coefficients
 $A_n^{(+)}(p) $ are due to the traces over the
representation $\rho_+$.
A brief inspection of \eqref{kjsk} shows
that $A_n^{(+)}(p)$ possesses only simple poles.
They  may be   located only
 at  points  $2p=-( \, d\, (n+l)+m\,)$
where $l=0,\,1,\,\ldots n-1$ and $m=0,\,\pm 1,\pm 2,\,\ldots$.
Hence, $A_n^{(+)}$ are meromorphic functions  in the whole complex plane of $p$.
The same, of course,  holds true for
the coefficients
$a_{n}^{(+)}(p)$. Furthermore, since \eqref{sjsjs}\ holds trivially
for pure imaginary $p$, the claim   is  also true for any complex $p$.

\begin{figure}[ht]
\centering
\includegraphics[width=7cm]{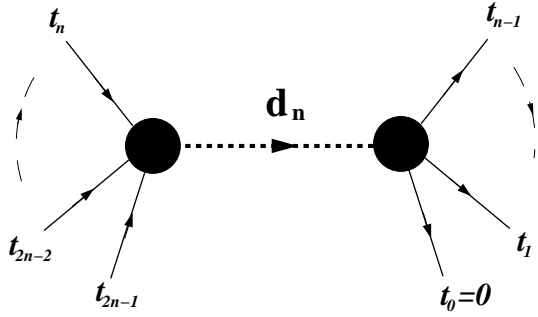}
\caption{The amplitude  defining the boundary $p_n$
 of the domain of  analyticity
$\Re e\, p>p_n$
for  $a^{(+)}_n(p)\ (n\leq k)$.
The incoming and outgoing  legs correspond  to the
insertions of  boundary fields ${\boldsymbol V}^{(\rm int)}_-(t)$
and  ${\boldsymbol V}^{(\rm int)}_+(t)$, respectively .
For $n\leq k$ the intermediate
state carries  the scaling dimension \eqref{lsslbqfg}    and
$p_n=-{d_n\over 2n}$.}
\label{fig-scat}
\end{figure}

The analyticity  of  $a^{(+)}_n(p)$ in  the right half-plane $\Re e\, p>p_n$ for
some $p_n$ and   the asymptotic formula\ \eqref{ljald} follow
immediately from Eqs.\,\eqref{wf},\,\eqref{slsjsu}. To determine  $p_n$ and
calculate the  residue
one needs to examine these equations in more details.
It is possible to show that the boundary of the domain of analyticity
is
defined by the integral $J_{+,\epsilon_1\ldots \epsilon_{2n-1}}$\ \eqref{slsjsu} whose
``cumulative charges'' $\eta_j$
\eqref{cumcharges} have  maximum admissible
values\footnote{Note that because of the factor
$\prod_{j=1}^{2n-1} \sin
(\pi d \eta_j)$ the sum \eqref{wf} enjoys
a  remarkable property, namely, it
contains only the terms where none of the ``cumulative charges'' $\eta_j$
 $(j= 1, \ldots, 2n-1)$ vanish.}, i.e.
\bea
\epsilon_1=\ldots=\epsilon_{n-1}=+1\, ,\ \ \ \
\epsilon_n=\ldots=\epsilon_{2n-1}=-1\ .
\eea
Moreover, as $p\to p_n$, the dominant contribution to  this  multiple  integral
comes from the region of integration
where the  operators ${\boldsymbol V}^{(\rm int)}_{\epsilon_j}$
with the same ``charges'' $\epsilon_j$ combine into   two well-separated  clusters:
$|t_1| T\sim\cdots\sim |t_{n-1}| T\sim  1$
and $|t_{n}| T\sim \cdots\sim |t_{2n-1}|T \gg 1$. Therefore,
to determine the singular behavior  as   $p\to p_n$
we invoke
the cluster property and
replace
the correlation function
in the integrand\ \eqref{slsjsu} by
\bea\label{eqwwyslslsa}
&&\re^{\ri\pi n d}\, \big\langle\big\langle\,
{\boldsymbol V}^{(\rm int)}_{+}(t_{0}) \cdots {\boldsymbol V}^{(\rm int)}_{+}(t_{n-1})\,
{\boldsymbol V}^{(\rm int)}_{-}(t_{n}) \cdots
{\boldsymbol V}^{(\rm int)}_{-}(t_{2n-1})
\, \big\rangle\big\rangle_0 \to\nonumber \\ &&\ \ \ \ \ \ \  F(t_0,\ldots,t_{n-1})\
\re^{2\pi T d_n(t_{n}-t_{0})}\ F^*(t_n,\ldots,t_{2n-1})
\  ,
\eea
where $t_0=0$ and\footnote{Here it is  assumed that $n\leq k$.}
\bea\label{lsslbqfg}
 d_n=n\ (\, n(d-1)+1)
\eea
is  the scaling dimension  of the
boundary operator
${\boldsymbol \Psi}_{n}\,
\re^{\ri {n\beta\over\sqrt{k}} {\boldsymbol \Phi}_B}$
(see Fig.\,\ref{fig-scat}).
The latter is a product of the bosonic exponential field and the
boundary   field ${\boldsymbol \Psi}_{n}$ \eqref{saskisys}.
The correlation function appearing  in  \eqref{eqwwyslslsa},
\bea\label{sksahksa}
F(t_0,\ldots,t_{n-1})&=&(2\pi T)^{-d_n}\times\\ &&
\big\langle\big\langle\,  {\boldsymbol V}^{(\rm int)}_{-}(t_{0})\ldots
{\boldsymbol V}^{(\rm int)}_{-}(t_{n-1})\
{\boldsymbol \Psi}^{(n)}_B\, \re^{\ri {n\beta\over\sqrt{k}} {\boldsymbol \Phi}_B}(-\infty)
\,\big\rangle\big\rangle_0\ , \nonumber
\eea
has a very  simple form in the case $n\leq k$. Namely, for $t_1>\ldots> t_n$
\bea\label{ahksa}
F(t_1,\ldots,t_n)=(2\pi T)^{nd}\ \prod_{m=1}^{n-1}C_{1m}\ \ \prod_{m<j}\Big(\,
2\ri \sinh \big(\pi T(t_m-t_j)\big)
\, \Big)^{2d-2}\ .
\eea
Here $C_{1m}=\sqrt{(k-m)(m+1)\over k}$
are the structure constants  \eqref{alssjlsa} with $j=1$.
Combining \eqref{eqwwyslslsa} with \eqref{wf}
one finds that the  coefficients
$a_n(p)$  $(n=1,\ldots k)$ possess
a simple pole singularity  at $p=p_n=-{d_n\over 2n}$ and
\bea\label{sshkss}
{\rm Res}_{p=p_n}\big[\,a^{(+)}_n(p)\,\big]=(-1)^{n+1}\,
{2^{2n-1}\pi k!n!\over  n^2k^n (k-n)!}\ \sin(\pi n d)\, S^2\
\prod_{j=1}^{n-1}\sin^2(\pi j d)\, ,
\eea
where $S$ is the Selberg integral\ \cite{Selberg,Dot},
\bea\label{kjshksah}
S&=&\int_{-\infty}^0\rd u_1\int_{-\infty}^{u_1}\rd u_2\ldots \int_{-\infty}^{u_{2n-2}}
\rd u_{2n-1}
    \prod_{l<j}\bigg[\,
2 \sinh \Big({u_l-u_j\over 2}\Big)\,\bigg]^{2 d-2}=\nonumber\\  &&(-1)^{1+{n(n+1)\over 2}}\
\bigg[{\pi\over \Gamma(d)}\bigg]^n\
{\Gamma\big(1-n(1-d)\big) \over \pi (n-1)!\prod_{j=1}^{n-1} \sin(\pi j d)}\ .
\eea
Plugging \eqref{kjshksah} into\ \eqref{sshkss}
one arrives at equation \eqref{ssaslsa}.

Finally it is easy to see
that for $n>k$  there is no pole at $p=p_n$   and
$a^{(+)}_n(p)$ is analytic  in some half plane $\Re e\, p>p_n-\delta_n$ where $\delta_n$ is
a finite positive constant.

\subsection{\label{Aps}Uniqueness of the solution of the   Riemann-Hilbert problem}

The above-described properties  of
the formal power series expansions
$A_{\pm}(s,p)$   constitute
a certain Riemann-Hilbert problem defining  these series uniquely.
For $k=1$ this was shown  in the Appendix of the work \cite{blz5}.
Here we  generalize  the result of \cite{blz5} to the case $k>1$.

Let us  substitute the formal power series \eqref{ksasksa}
into the quantum Wronskian relation\ \eqref{sjssjk}.
This  yields  an infinite set of relations for the
meromorphic  functions $a^{(+)}_{n}(p)$
of the following  form
\bea\label{jsdgtr}
\sin(\pi n d+2\pi p)\,
a^{(+)}_n(p)-\sin(\pi n d-2\pi p)\,
a^{(+)}_n(-p)=R_n(p)\, ,
\eea
where   $R_n(p)$
are expressed through $a^{(+)}_l(p)$ with $l=1,\ldots n-1$ only.
For example,
\bea\label{skjdyt}
R_1(p)&=&0\, , \\
R_2(p)&=&\big( \, q\, a^{(+)}_1(p)+q^{-1}\, a^{(+)}_1(-p)\, \big)^2\
\re^{4\pi \ri p}\ \sin(2\pi p)/2\ .\nonumber
\eea
Introduce the  function $r_n(p)$ such that
\bea\label{ksjhskhs}
r_n(p)= {\Gamma(n-1-n d+2 p)\over \Gamma(2-n+n d+2 p)}\ a^{(+)}_n(p)\ ,
\eea
then \eqref{jsdgtr} takes the form
\bea\label{slssksuyoi}
r_n(p)-r_n(-p)&=&{(-1)^{n-1}\over\pi}\times\\
&& \Gamma(n-1-n d+2 p)\,
\Gamma(n-1-n d -2 p)\,  R_n(p)\, .\nonumber
\eea

In the case $n=1$, the RHS in \eqref{slssksuyoi} vanishes and
this equation supplemented
by the analyticity condition\ \eqref{lsljsal}
implies that
$r_1(p)$ is an even entire function of the variable $p$. Moreover,
in view of  the asymptotic condition \eqref{ljald}, one should
conclude that $r_1=const$. The  constant
is defined by the residue condition  \eqref{ssaslsa}.
This yields formula \eqref{ksahsak}.

We shall  prove the  uniqueness  of the solution of \eqref{jsdgtr} for $n>1$   by using induction.
Let us assume all the functions $a^{(+)}_m(p)$ for $m=1,\ldots n-1$ are already determined, so
the LHS's of equations \eqref{jsdgtr} and \eqref{slssksuyoi} are given.
Suppose also for the moment that
the complex  parameter $d$ belongs to the strip ${n-1\over n}-\delta< \Re e\, d<{n-1\over n}$,
where $\delta>0$ is
some small  number.
If $\Re e\big({n-1\over n}-d\big)$ is  sufficiently small,
then
$r_n(p)$ is analytic in the half plane $\Re e\, p\geq 0$ except
for a simple pole at
$p=p_n\equiv{1\over 2}\,( \, n(1-d) -1\,) $.
The  residue of $r_n(p)$ at this pole  is fixed by the condition \eqref{ssaslsa}. Also, it is
easy to see that
 $r_n(p)\to 0$ as $p\to \infty$ within  $\Re e\, p>0$.
Hence, relation \eqref{slssksuyoi} supplemented by the above
quoted  requirements  for $r_n(p)$
constitutes
a simple factorization  problem.
The uniqueness of this Riemann-Hilbert   problem  follows immediately from the Liouville theorem.
To complete the induction step
we   note that the function $r_n(p)$   within   $\Re  e\, d<1$
can be obtained
via the  analytic continuation in  the variable $d$
from the strip  ${n-1\over n}-\delta< \Re e\, d<{n-1\over n}$.

\bigskip
Thus we see that\ \eqref{jsdgtr} provides a recursion for the evaluation of $a^{(+)}_n(p)$
which
allows one to express   $a^{(+)}_n(p)$ in terms of an $(n-1)$-fold integral.
Explicit formulas for $a^{(+)}_2$ and $a^{(+)}_3$ in the case $k=1$  can be found
in the Appendix of  \cite{blz5}. Those  formulas can be easily generalized to $k>1$.
It is worth noting  that such integral representations prove to be     convenient
for  numerical calculations because the  integrals converge very fast
at infinity.

\subsection{Exact expressions for $A_{\pm}$}

It was observed some years ago in the work~\cite{toteo},
that $A_{\pm}(s,p)$
for $k=1$ and some particular values of $p$ can be related exactly to
the eigenvalue problem of a certain ordinary differential operator.
This remarkable
observation was  generalized and proven  for  any   $p$ in Ref.~\cite{blzz}.
Here we    extend the result of~\cite{blzz} to  all integers $k>1$.

Let us consider the  Schr${\ddot {\rm o}}$dinger equation
\bea\label{OsDER}
\Big\{\, -\partial_u^2+
\kappa^2\,  \Big[\, \exp\big({\textstyle{\beta^2\, u\over 1-\beta^2}}\,\big)+
\exp\big({\textstyle{u\over 1-\beta^2}}\big)
\, \Big]^k-\xi^2
\, \Big\}\, \Theta(u) = 0\, .
\eea
For $0<\beta^2<1$ the potential 
term decays at $u\to-\infty$ and therefore for real $\xi>0$
there is a  Jost solution  which 
is asymptotically plane wave as $u\to-\infty$:
\bea\label{ksjksksj}
\Theta_-(u,\xi)\to \re^{-\ri \xi u}\ .
\eea
As a matter of fact,
this particular  solution is
a meromorphic function in the whole complex plane  of the  parameter  $\xi$,
so
it solves  \eqref{OsDER} for all values $\xi$ except
points where $\Theta_-(u,\xi)$
has simple poles  (see, e.g., \cite{Calogero}). Notice that  $\Theta_-(u,-\xi)$ is
another, linearly independent (for $\xi\not=0$) solution of~\eqref{OsDER}.

Since the potential in  \eqref{OsDER} grows rapidly as $u\to+\infty$,
the  Schr${\ddot {\rm o}}$dinger equation admits also
a  solution which decays at large positive $u$. We denote this solution
by $\Theta_+(u,\xi)$
and fix its normalization by the condition
\bea\label{slksajsla}
\Theta_+(u,\xi)\to
 (2\kappa)^{-{1\over 2}}\ \re^{F( \ri\beta\,|\,u)}
\ \ \ \ \ \ \ \ {\rm as}\ \ \ \ \ \ \  u\to+\infty\ ,
\eea
where $F( b\, |\, u)$ is defined in  Eq.\,\eqref{nahggt}.
$\Theta_+(u,\xi)$ is an entire function of  $\xi^2$ and solves
differential  equation \eqref{OsDER} for all values of this complex parameter.

The main objects of our interest  are properly normalized Wronskians of the
above-introduced solutions, namely,
\bea\label{uslssl}
D_{\pm}(\kappa,\xi)=
{\sqrt{2\pi\, (1-d)}
 \over  \Gamma(1\mp 2\ri\xi (1-d) )}\ \big(\,\kappa (1-d)\,\big)^{\mp 2\ri\xi (1-d)}\ \
W[\, \Theta_+(u,\xi)\,,\, \Theta_-(u,\pm \xi)\,]\ .
\eea
Here  $d$ is given by \eqref{llskjsl} and
the overall factor     provides
the normalization condition
\bea\label{lsulaas}
\lim_{\kappa\to 0}D_\pm(\kappa,\xi )=1\ .
\eea
In what follows we will show that for 
$0<\beta^2<1$ and all values of $\kappa$ and $\xi$
\bea\label{lksjlks}
A_{\pm}(s,p)=D_\pm(\kappa,\xi)\ ,
\eea
provided  the variables  are identified according  to the relations:
\bea\label{lsalsuy}
\xi={ \ri\,  p\over 1-d}\ ,\ \ \ \ \ \ \ \ \ \ \ \ \ \ \ \ \
\kappa = {1\over 1-d}\ \, \bigg[{2\pi \sqrt{s}\over \sqrt{k}\Gamma(d)}
\bigg]^{1\over 1-d}\ .
\eea
More specifically,  we  will check  that $D_\pm$ considered as functions
of the variables $s$ and $p$, obey  the same set of
conditions \eqref{ssssjl}--\eqref{ljald}   as  $A_{\pm}(s,p)$.
Thus \eqref{lksjlks} will  follow from   the  uniqueness
of the solution of the   Riemann-Hilbert problem.

\begin{itemize}

\item
In order to  examine
properties of   $D_{\pm}$
it is useful to make a change of  the variable in \eqref{OsDER},
\bea\label{kjsssk}
z={{u\over 2-2d}}+\log\big(\, 2\kappa\, (1-d)\, \big)\ ,
\eea
and rewrite the differential equation
using
the notations $s,\, p$ \eqref{lsalsuy} and $d$\ \eqref{llskjsl}:
\bea\label{ulsaslsa}
\big(-\partial_{z}^2+\re^{2z}+4p^2+\delta U(z)\, \big)\ {\tilde \Theta}(z)=0\ ,
\eea
with
\bea\label{lksjs}
\delta U(z)=\sum_{m=1}^{k}{k!\over m! (k-m)!}\ \bigg( {2^{2-d}\pi\over
\sqrt{k}\Gamma(d)}\bigg)^{2m}\ \
s^m\ \re^{2 z(1-  (1-d)\, m\,)}\ .
\eea
The  solutions $\Theta_\pm$\ \eqref{ksjksksj},\,\eqref{slksajsla}
should be  considered now as functions of  the complex parameters $s,\, p$
provided $1-{1\over k}<d<1$ and  $z$ is real.
In fact, it is convenient for the present discussion to change slightly
their normalizations:
\bea\label{lkanksa}
{\tilde \Theta}_-(z, p)&=& \big(\,2\kappa\,(1-d)\,\big)^{\mp 2\ri\xi (1-d)}\
\, \Theta_-(u,\xi)\ ,\nonumber\\
{\tilde \Theta}_+(z, p)&=&{1\over \sqrt{1-d}}\ \ \   \Theta_+(u,\xi)\ .
\eea
Then formula \eqref{uslssl} takes the form:
\bea\label{slsls}
D_{\pm}(s,p)=\sqrt{\pi\over 2}\ \
{ 2^{\mp 2p}\over
\Gamma(1\pm 2p)}\ \ W[\, {\tilde \Theta}_+(z,p)\,,\, {\tilde \Theta}_-(z,\pm p)\,]\ .
\eea

Both ${\tilde \Theta}_+$ and ${\tilde \Theta}_-$
are entire functions of $s$. In addition,  ${\tilde \Theta}_-$
is a meromorphic function of $p$, while
${\tilde \Theta}_+$ is an even entire function of
this complex parameter.
Hence $D_{\pm}(s,p)$  can be expanded in a power series of $s$ similar
to\ \eqref{ssssjl}, and  the expansion  coefficients,
$D^{(\pm)}_n(p)\,:\,D^{(-)}_n(p)=D^{(+)}_n(-p)$,
are meromorphic functions of $p$.

\item

To proceed further we will need the following conventional formula
for $D_+(s,p)$, valid   over  the domain $\Re e\, p \geq 0$,
\bea\label{ssajgsks}
{1\over D_+(s,p)}=1-
\int_{-\infty}^{\infty}\rd z\ I_{2p}(\re^{z})\ \delta U(z)\ \chi(z)\ .
\eea
Here $I_{p}(x)$ is the
modified Bessel function and  $\chi(z)$
solves
the Lipman-Schwinger    equation,
\bea\label{hjagg}
\chi(z)=K_{2p}(\re^z)-\int_{-\infty}^{\infty}\rd z'\ G(z,z')\ \delta U(z')\ \chi(z')\ ,
\eea
where $G(z,z')$ is the Green function of \eqref{ulsaslsa}  with $\delta U=0$ subject
of the asymptotic condition
$\lim_{z\to\infty}G(z,z')=0$.
Notice that  $\chi(z)$ differs from the solution ${\tilde \Theta}_+(z,p)$ by some
overall $z$-independent factor only.
The coefficients $D_{n}^{(+)}(p)$ can be calculated within 
standard perturbation theory developed
with respect to $\delta U$. In fact, to determine $D_{n}^{(+)}(p)$
for given $n$, one needs to perform   $n-1$-perturbative iterations
in the Lipman-Schwinger equation.
In particular,
\bea\label{lssmnbh}
D_{+}=1+\int_{-\infty}^{\infty}\rd z\ I_{2p}(\re^{z})\ K_{2p}(\re^{z})
\ \delta U(z)+O\big((\delta U)^2\big)\ .
\eea
Making use  of \eqref{lksjs}
and
the table integral
\bea\label{lssal}
\int_{0}^{\infty}\rd x\ x^{2\alpha-1}\ I_{2p}(x)  K_{2p}(x)=
{\Gamma({1\over 2}-\alpha)\ \Gamma(\alpha)\Gamma(\alpha+2p)\over
4\sqrt{\pi}\ \Gamma(1-\alpha+2p)}\ ,
\eea
one finds for  $n\leq k$:
\bea\label{kshsk}
&&D^{(+)}_n(p)=
\bigg( {2^{2-d}\pi\over
\sqrt{k}\Gamma(d)}\bigg)^{2n}\ \Gamma(-{\textstyle{1\over 2}}+n(1-d))\times\\ &&
\ \ \  {k!\over n! (k-n)!}\ {\Gamma(1-n (1-d))\, \Gamma(1-n (1-d)+2p)\over
4\sqrt{\pi} \Gamma(n (1-d) +2p)}+\ldots\ .\nonumber
\eea
Here the dots   mean  the terms corresponding  to the
higher-order perturbative in $\delta V$  contributions.
Further, one may argue that the first-order contribution\ \eqref{kshsk}
is  analytic in the half-plane
$\Im m (p)>p_n={1\over 2} (n (1-d)-1)$, whereas
the omitted terms  have a wider domain of analyticity.
Hence,   $D^{(+)}_n(p)$ with $n\leq k$ is an analytic function   for
$\Im m (p)>p_n$ and has a simple pole located at $p=p_n$
with the same  residue as in \eqref{ssaslsa}.
Similarly,  it is possible to  show that
$D^{(+)}_n(p)$ with $n>k$  is  analytic in the half-plane
$\Im m (p)>p_n-\delta_n$ with  some  $\delta_n>0$.

\item

The   large $p$ behavior of the expansion  coefficients $d_n^{(+)}(p)$
for the series $\log D_n^{(+)}(s,p)$ can be
explored by means of  standard semiclassical methods.  A simple  WKB
analysis of the Schr${\ddot {\rm o}}$dinger
equation\ \eqref{ulsaslsa} shows that
$d_n^{(+)}(p)\to C_n\ p^{1-2n(1-d)}$ as $p\to\infty$ and
$\Im m\, p>p_n$, with  the  constants $C_n$ given by
\bea\label{lksla}
C_n&=& (-1)^{n+1}\  k \ \bigg[{2\pi\over
\sqrt{k}\Gamma(d)}\bigg]^{2n}\ \times \\
&&{\Gamma(n-(1-d)\, k\, n\,)\,\Gamma(1-(1-d)\, n\,)\,\Gamma(-{1\over 2}+(1-d)\,n\,)\over
2\sqrt{\pi}\, n!\ \Gamma(1-(1-d)\, k\, n\,)}\ .\nonumber
\eea

\item

Finally, let us prove  that $D_{\pm}(s,p)$\ \eqref{slsls} obey  the  quantum Wronskian
relation.
We start with an observation that the following transformations
of the variable $z$ and the parameters $s$ and $p$,
\bea\label{muynmsjs}
{\hat \Lambda}&:&\ z\to z\, ,\ \ \ \ \ \ \ \ \ \ s\to  s\,,\
\, \ \ \ \ \ \  p\to -p\ ,\nonumber\\
{\hat \Omega}&:&\ z\to z+\ri\pi\, ,\ \ \ \, s\to q^{-2}\,s\,,
\ \ \  p\to p\,, \eea
leave the differential equation \eqref{ulsaslsa}
unchanged while acting nontrivially on its solutions.
It has been mentioned already, that
the transformation ${\hat \Lambda}$ applied to the solution ${\tilde \Theta}_-(z,p)$\
\eqref{lkanksa} yields another solution, and the pair 
\bea\label{lksjssauj} {\tilde \Theta}_-={\tilde \Theta}_-(z,p)\, ,\ \ \
{\hat \Lambda}{\tilde \Theta}_-={\tilde \Theta}_-(z,-p)\ ,
\eea
forms a basis in the space of solutions of\ \eqref{ulsaslsa}.
Indeed,
\bea\label{mnasajsla} W\big[{\tilde \Theta}_-,{\hat
\Lambda}{\tilde \Theta}_-\big]=-4p\ , \eea
therefore the solutions\ \eqref{lksjssauj}\ are  linearly independent
provided $p\not=0$.
The solution ${\tilde \Theta}_+$\ \eqref{lkanksa} can be always  expanded
in the basis \eqref{lksjssauj},
\bea\label{askss} {\tilde \Theta}_+=c_1\ {\tilde \Theta}_-+c_2\
{\hat \Lambda}{\tilde \Theta}_-\ . \eea
Making use of \eqref{mnasajsla} we have
\bea\label{lakjsla} c_2={1\over 4p}\
W(s,p)\ , \eea
with
$W(s,p)=W\big[{\tilde \Theta}_+,{\tilde \Theta}_-\big]$.
The solution ${\tilde \Theta}_+$ is an even function of $p$, therefore
\bea\label{uddylakjsla}
c_1=-{1\over 4 p}\ W(s,-p)\ .
\eea
Let us apply now the transformation
${\hat \Omega}$ \eqref{muynmsjs} to both sides of
\eqref{askss}. 
Since

${\hat \Omega}{\tilde \Theta}_-(z, \pm p)
=
\re^{\pm 2\pi\ri  p}\,{\tilde \Theta}_-(z, \pm p)$, so
\bea\label{lsajsa}
{\hat \Omega}{\tilde \Theta}_+={1\over 4 p}\ \big(
\ \re^{- 2\pi\ri  p}\, W(s q^{-2},p)\ 
{\hat \Lambda}{\tilde \Theta}_--\re^{2\pi\ri  p}
\, W(s q^{-2},-p)\ {\tilde \Theta}_- \big)\ .
\eea
Now, it is not difficult to check that
\bea\label{muyasajsla}
W_z\big[{\tilde \Theta}_+,{\hat \Omega}{\tilde \Theta}_+\big]=-2\, \ri \,  \ .
\eea
Combining formulae\,\eqref{askss},\,\eqref{lsajsa} and\ \eqref{muyasajsla}   one obtains
the relation
\bea\label{yjsla}
\re^{2\pi\ri  p}\, W(s,p)\, W(q^{-2}s,-p)-
\re^{-2\pi\ri  p}\, W(q^{-2}s,p)\, W(s,-p)=8\ri\,p  \ .
\eea
Hence $D_{\pm}(s,p)$\ \eqref{slsls}
satisfy the  quantum Wronskian
relation \eqref{sjssjk} with  $A_\pm(s,p)$
replaced by $D_{\pm}(s,p)$.

\end{itemize}

\bigskip
\noindent
Notice that once the  relation\ \eqref{lksjlks} is established
we may assert that series  \eqref{lssaljsa},\,\eqref{ssssjl}
have   an infinite  radius of convergence.

\subsection{Formula for the DC conductance}

Since the parafermionic boundary sine-Gordon model\ \eqref{jksahsa} has been of
a certain interest   in   the    quantum impurity problem\ \cite{saleur},
we present
here an  exact  formula for the DC conductance\ \eqref{alsksjl}
which follows immediately from the above consideration.

Using equation \eqref{sssa}, the conductance can be written in the form
\bea\label{hssgfd}
\alpha(V,T)/\alpha_0=
1+\beta^2\,  \ {s\over 2pk}\ \, \partial_{s}\log\Big(\,  {A_+\over A_-}\Big)\ .
\eea
This formula has a nice interpretation in terms of
the Schr${\ddot {\rm o}}$dinger problem\ \eqref{OsDER}.
Indeed, as $v\to -\infty$,
the solution $\Theta_+(v,\xi)$\ \eqref{slksajsla} develops the
asymptotic behavior:
\bea\label{lklslks}
\Theta_+(u,\xi)
\to const\ \big(\,\re^{\ri\xi u}+S(\xi, \kappa)\ \re^{-\ri\xi u}\, \big)\ ,
\eea
where
\bea\label{kslsasj}
S(\xi, \kappa)=-{ W[\, \Theta_+(u,\xi)\, ,\, \Theta_-(u,-\xi)\,]\over
W[\,\Theta_+(u,\xi)\,,\, \Theta_-(u,\xi)\,]}
\eea
is the reflection scattering amplitude  for the
Schr${\ddot {\rm o}}$dinger equation\ \eqref{OsDER}.
Therefore, in view of
\eqref{uslssl} and  \eqref{lksjlks},  
the DC conductance is expressed in terms of $S(\xi,\kappa)$
as follows,
\bea\label{lssajssy}
\alpha(V,T)
/\alpha_0=
1+{ k\over 4\ri\, 
\xi\, (\beta^{-1}-\beta)^2}\ \ \kappa\,\partial_{\kappa}\,\log S(\xi, \kappa)\ ,
\ \ \ \
V/T={4\pi\xi\over k}\ (1-\beta^2)\ .
\eea
Recall   that $\beta^2=\alpha_0/k$ is the
dimensionless conductance per channel   of the wire without impurity,
while $\kappa$
in \eqref{lssajssy} is the dimensionless inverse
temperature
measured in  unit  of the Kondo scale:
$\kappa=E_*/T,$\ \  $E_*\propto \mu^{k-\alpha_0\over k^2}$.

Formula \eqref{lssajssy} extends the  result of \cite{blzz} to the case  $k>1$.
We note also that as  $T\to 0$, the
dimensionless variable $p=-\ri  V/(4 \pi T)\to\infty$.
In this limit,  the coefficients $a_{n}^{(+)}(p)$  develop
the simple asymptotic behavior\ \eqref{ljald} where the $C_n$ are given
by \eqref{lksla}. This immediately yields
the large-$V$ expansion of the  conductance at $T=0$
proposed in Ref.\cite{saleur}:
\bea\label{skshks}
{\alpha(V,0)\over \alpha_0}
=1+{ {\sqrt{\pi}\over 2}}\sum_{n=1}^{\infty}
(-1)^n\, {
\Gamma(n+1-kn\gamma)\,\Gamma(1-n\gamma)
\over n!\, \Gamma(1-kn\gamma)\, \Gamma({3\over 2}-n\gamma
)}\,
\bigg[{4\pi E_*\gamma\over V}\bigg]^{2n\gamma},
\eea
where $\gamma={k-\alpha_0\over k^2}$.

\section{
Appendix : Quantum Wronskian relation}

The purpose of this appendix is twofold. First, it is intended to
sketch how the
quantum Wronskian relation\ \eqref{sjssjk}  can be proven for the
perturbative series $A_{\pm}$\ \eqref{sajsahsa},\,\eqref{lsashksa}.
Second, it provides a short review of some  well-known issues  of  CFT
which are useful for understanding the models  under consideration.

Our  previous discussion has always dealt with
the so-called
open string picture (channel). In this  Hamiltonian  picture
the variable $x\leq 0$ plays the role of
the space coordinate, while the coordinate  $\tau$ is treated as the
compact Euclidean time.
We now want to switch over to the ``closed string'' channel. In this channel
the world-sheet coordinate $x$ is viewed as  the Euclidean time, so that
the  CFT lives in the finite volume $1/T$.
The closed string channel  Hilbert space   is embedded in  a tensor product of the
chiral spaces  of  states  of  right  and   left
movers.
Below  the following notational    convention is applied: If the
open string  Hilbert space was denoted by ${\cal H}$, then the
right  and left movers'   spaces of  states
in the closed string channel   will be
denoted (with  some abuse of notations) as
${\tilde {\cal H}}$ and ${\overline{ {\tilde  {\cal H}}}}$ respectively.
Of course, the spaces ${\tilde {\cal H}}$ and ${\overline{ {\tilde  {\cal H}}}}$  are
isomorphic  to each other.
We will focus mainly on the  ``right''   component ${\tilde {\cal H}}$.

In the closed string channel
an effect of the boundary at $x=0$ can be  described in terms of the boundary
state which incorporates all
information about boundary conditions\ \cite{nappi,Ishibashi,Card,gz}.
In the case of the ``free''  CFT\ \eqref{lssaasj} the corresponding
conformal boundary state $|\, B\,\rangle_{\rm free}$ takes the   form,
\bea\label{kshks}
|\, B\,\rangle_{\rm free}=
|\, B\,\rangle^{(\Phi)}_{\rm free}\otimes|\, B\,\rangle^{(k)}_{\rm free}\ ,
\eea
where
\bea\label{ytrslksslk}
|\, B\,\rangle^{(\Phi)}_{\rm free}\subset
{\tilde {\cal H}}^{(\Phi)}\otimes {\overline{ {\tilde  {\cal H}}}{}^{(\Phi)}}
\eea
is the boundary state of the Gaussian CFT subject to
the von Neumann boundary condition, while
\bea\label{slkslssa}
|\, B\,\rangle^{(k)}_{\rm free}\subset
{\tilde {\cal H}}^{(k)}\otimes {\overline{ {\tilde  {\cal H}}}{}^{(k)}}
\eea
is the boundary state  of the minimal parafermionic model with free boundary condition.

As an important  step toward  the derivation of \eqref{sjssjk}, we will show
that the power series $A_{\pm}(s,p)$ are vacuum eigenvalues of certain   operators
acting in the chiral Hilbert space
${\tilde {\cal H}}^{(\Phi)}\otimes{\tilde {\cal H}}^{(k)}$.
With this aim in mind
we need to    describe explicitly
${\tilde {\cal H}}^{(\Phi)}$ and  ${\tilde {\cal H}}^{(k)}$,
as well as the boundary states\ \eqref{ytrslksslk}
and \eqref{slkslssa}.

\subsection{\label{huysgt}
Chiral space ${\tilde {\cal H}}^{(\Phi)}$ and
boundary state $|\, B\,\rangle^{(\Phi)}_{\rm free}$  }

A general solution to  the bulk equation of motion
$\triangle{\boldsymbol \Phi}=0$ has
the form:
\bea\label{ansgs}
{\boldsymbol  \Phi}(\tau,x)
={\boldsymbol \phi}(\tau-\ri x)+{\bar {\boldsymbol \phi}}(\tau+\ri x)\ ,
\eea
where ${\boldsymbol \phi}(v)$ and ${\boldsymbol \phi}({\bar v})$ are right  and
left  chiral  fields respectively.
In the case of  the uncompactified  Bose field ${\boldsymbol  \Phi}$:\
${\boldsymbol  \Phi}(\tau+1/T,x)=
{\boldsymbol  \Phi}(\tau,x)$,
the  chiral fields    admit the  Fourier  mode expansions of the form:
\bea\label{vars}
{\boldsymbol \phi}(v) = {\textstyle {1\over 2}}\,  {\boldsymbol  \Phi}_0+
\pi v T\ {\boldsymbol  \Pi}_0 +\ri \sum_{m\neq 0}
{{ {\boldsymbol  \phi}_{m}}\over m}\
\re^{-2\pi \ri m v T}\ \ \ \ \ (v=\tau-\ri\, x)\ ,
\eea
and
\bea\label{varsiu}
{\bar {\boldsymbol \phi}}(v) = {\textstyle {1\over 2}}\,  {\boldsymbol  \Phi}_0 -
\pi   {\bar v} T\  {\boldsymbol  \Pi}_0+\ri\, \sum_{m\neq 0}
{{ {\bar {\boldsymbol  \phi}}_{m}}\over m}\
\re^{2\pi \ri m {\bar v} T}\ \ \ \ \ ({\bar v}=\tau+\ri\, x)\ .
\eea
The canonical quantization procedure
leads to the following   nontrivial commutation relations:
\bea\label{osc}
[\, {\boldsymbol \Phi}_0\,,\, {\boldsymbol \Pi}_0\,] =\ri\  ;
\qquad [\, {\boldsymbol \phi}_n\, ,\,  {\boldsymbol \phi}_m\,] =
[\, {\bar {\boldsymbol \phi}}_n\, ,\,  {\bar {\boldsymbol \phi}}_m\,]=
{\textstyle {n\over 2}}\  \delta_{n+m, 0}\ .
\eea
Let ${\cal F}_P$  $({\bar {\cal F}}_P)$ be the Fock space, i.e. the space
generated by a free action
of the operators ${\boldsymbol \phi}_n$  $({\bar {\boldsymbol \phi}}_n)$
with $n < 0$ on the ``vacuum''
vector $|\, P\, \rangle$  which satisfies
\bea\label{kuyhg}
&& {\boldsymbol \phi}_n\,   |\, P\, \rangle  =
{\bar {\boldsymbol \phi}}_n\,   |\, P\, \rangle= 0\ , \quad {\rm  for} \quad n > 0\ ; \nonumber\\
&& {\boldsymbol \Pi}_0 \, |\, P\, \rangle = P\, |\, P\, \rangle\, , \ \ \ \  \ \ \ \ \ \
{\boldsymbol \Pi}_0 \, {\overline {|\, P\, \rangle}} = P\, {\overline {|\, P\, \rangle}}\ .
\eea
Then the closed string  space of states coincides with the direct integral
$\int_p {\cal F}_P\otimes {\bar {\cal F}}_P$, and the corresponding
chiral Hilbert space is given by
\bea\label{lssalsa}
{\tilde {\cal H}}^{(\Phi)}=\int_p {\cal F}_P\ .
\eea

The boundary state \eqref{ytrslksslk} associated with
the von Neumann boundary condition
obeys   the equation,
\bea\label{ylksslk}
\big(\, \partial_{x}{\boldsymbol \Phi}(\tau,x)\, \big)\big|_{x=0}\,
|\, B\,\rangle^{(\Phi)}_{\rm free}=0\ .
\eea
It reads explicitly as follows~\cite{nappi,Ishibashi}
\bea\label{lssalsaj}
|\, B\,\rangle^{(\Phi)}_{\rm free}=
{\rm g}_D\ \exp\Big(\sum_{n=1}^{\infty}{\textstyle {2\over n}}\, (-1)^n\
{\boldsymbol \phi}_{-n}{\bar {\boldsymbol \phi}}_{-n}\Big)\, |\, {\rm vac}\, \rangle^{(\Phi)}\ ,
\eea
where $|\, {\rm vac}\, \rangle^{(\Phi)}=|\, 0\, \rangle\otimes {\overline {|\, 0\, \rangle}}$ is
the ground state of the Gaussian theory in the closed string channel.
Now, in view of \eqref{lssalsaj}
it is easy to see that
for $1/T>\tau_1>\cdots>\tau_{n}>0$ the following relation,
involving  the correlators of the boundary exponential fields, holds:
\bea\label{strjsaj}
&&{}_{\rm free}^{(\Phi)}\langle\, B\, |\,
\re^{\ri a_1 {\boldsymbol \Phi}_B}(\tau_1)\, \cdots\,
\re^{\ri a_n {\boldsymbol \Phi}_B}(\tau_{n})\, |\, {\rm vac }\, \rangle^{(\Phi)}=
\\ &&\ \ \ \ \ \ \ \ \ \ \ \ \ \ \ \ \ \ \ \ \ \
\re^{-2\pi \ri  a P\sum_{j=1}^n\tau_j}\
\langle\, P\, |\,
\re^{2\ri a_1 {\boldsymbol \phi}}(\tau_1)\, \cdots\,
\re^{2\ri a_n {\boldsymbol \phi}}(\tau_{n})\, |\, P\,\rangle
\, .\nonumber
\nonumber
\eea
Here
\bea\label{lss}
\re^{2\ri a{\boldsymbol\phi} }(\tau)&=&(2\ri\pi T)^{a^2\over 2}\ \exp\Big(2a \sum_{m> 0}
{{ {\boldsymbol  \phi}_{-m}}\over m}\
\re^{2\pi \ri m \tau T}\Big)\ \exp\Big(
\ri a\,  {\boldsymbol  \Phi}_0 +2\ri \pi a\,
 \tau T\ {\boldsymbol  \Pi}_0\Big) \times \nonumber \\ &&
\exp\Big(-2a \sum_{m> 0}
{{ {\boldsymbol  \phi}_{m}}\over m}\
\re^{-2\pi \ri m \tau T}\Big)
\eea
is an  intertwining  operator for
the Fock spaces:
\bea\label{lusdyss}
\re^{2\ri a{\boldsymbol \phi}}(\tau)\ :\ {\cal F}_P\to {\cal F}_{P+a}\ .
\eea

\subsection{\label{nshgsgt}
Chiral space ${\tilde {\cal H}}^{(k)}$ and
boundary state $|\, B\,\rangle^{(k)}_{\rm free}$  }

\subsubsection{Parafermionic and ${\rm W}_k$  current algebras}

Here we review shortly
some well-known facts about the chiral Hilbert space of the minimal
parafermionic  models. The details can be found in Refs.\,\cite{ZamFat,FATZ,FL}.

Again, let  the variable $v$ be a complex coordinate, $v=\tau-\ri x$, on
the 2D cylinder.
The chiral  parafermionic  algebra  contains a set of
currents
$\big\{{\boldsymbol \psi}_{n}(v)\big\}_{n=1}^{k-1}$
of conformal dimensions\ \eqref{lsjs}.
This  algebra possesses a global symmetry with respect to
the ${\mathbb Z}_k$ transformations
${\boldsymbol \psi}_{n}
\to \omega^{na }\ {\boldsymbol \psi}_{n}$ with $a=0,\,\ldots\, k-1$,
and its ${\mathbb Z}_k$  invariant component
can be  generated by the  currents  with the lowest
conformal dimension, i.e., $ {\boldsymbol \psi}_+
\equiv  {\boldsymbol \psi}_{1}$
and $  {\boldsymbol \psi}_-\equiv  {\boldsymbol \psi}_{k-1}$, through
the OPE:
\bea\label{lkss}
{\boldsymbol \psi}_+(u) {\boldsymbol \psi}_-(v)&=&(u-v)^{2-
{2\over k}}\ \Big\{ {\boldsymbol I}+\textstyle{{k+2\over 2\,k}}\,  (u-v)^2\, \big(\, {\boldsymbol W}_2(u)+
{\boldsymbol W}_2(v)\, \big)+\nonumber \\ &&
\textstyle{{1\over 2\, k^{3\over 2}}}\, (u-v)^3\ \big(\, {\boldsymbol
W}_3(u)+{\boldsymbol W}_3(v)\, \big)+
\ldots\,
\Big\}\ .
\eea
The spin-2 current ${\boldsymbol W}_2$  \eqref{lkss}, in turn,
generates the Virasoro algebra with the central charge
\bea\label{saisjs}
c_k={2(k-1)\over k+2}\ .
\eea
The higher-order terms involve
higher-spin currents, $ {\boldsymbol W}_3,
\, {\boldsymbol W}_4\ldots$ forming the
${\rm W}_k$ algebra introduced   in
\cite{FATZ,FL} with
the special value\ \eqref{saisjs} of the central charge.
Notice that
the higher   currents can also be produced recursively from
singular parts of OPE's of the lower currents.
In this sense the chiral  ${\rm W}_k$  algebra is generated
by two basic currents ${\boldsymbol W}_2$ and
${\boldsymbol W}_3$.

The  parafermionic  algebra   has  a set of irreducible representations (irreps),
\bea\label{lksajsal}
\big\{\, {\cal M}_J\, \big|\ J=0,\,{\textstyle {1\over 2}},\,\ldots\, {\textstyle { k\over 2}}\,
\big\}\ ,
\eea
obtained through the action of the  ${\boldsymbol \psi}_+$ and  ${\boldsymbol \psi}_-$ modes
on the highest vectors (see Ref.\cite{ZamFat} for details).
The irrep ${\cal M}_J$ decomposes into a direct sum of
subspaces specified by the ${\mathbb Z}_k$  charge $\ell $
(number of ${\boldsymbol \psi}_+$ modes minus number of ${\boldsymbol \psi}_-$ modes  modulo $k$):
\bea\label{slsjsa}
{\cal M}_J=\oplus_{\ell=0}^{k-1}\ {\cal M}_{J}^{(\ell)}\ .
\eea
Each   component
${\cal M}_{J}^{(\ell)}$ in \eqref{slsjsa} is an
irrep  of the ${\rm W}_k$ algebra.

Recall that an irrep of the ${\rm W}_k$ algebra
is  obtained by factorizing a  highest weight  modules
over submodules of ``null-vectors''\ \cite{FATZ,FL}.
In the case under consideration, the  highest weight can be thought of as a pair
of
eigenvalues $(\Delta,\, w)$ of  mutually commuting
operators
\bea\label{uytriu}
{c_k\over 24}+(2\pi T)^{-1}\, \int_0^{1/T}{\rd v\over 2\pi}\ {\boldsymbol W}_2(v)
\, ,\ \ \ \ \ \ (2\pi T)^{-2}\ \int_0^{1/T} {\rd v\over 2\pi}\ {\boldsymbol W}_3(v)\ ,
\eea
corresponding to the highest weight vector.
In fact, it is  most convenient to label
the highest weight vectors
by  means of  another   pair of numbers   $(J,\, M)$ such that
\bea\label{yttrlsiy}
\Delta_{(J,M)}&=&{J(J+1)\over k+2}-{M^2\over k}\, ,\ \ \nonumber \\
w_{(J,M)}&=&{M\over 3\sqrt{k}}\ \Big(\, { {2(3k+4)\over k}}\ M^2-6\ J(J+1)+k\, \Big)\ .
\eea
If  ${\cal W}_{(J,M)}$    denotes
the
irrep of
highest weight  $(\Delta_{(J,M)}, w_{(J,M)})$\ \eqref{yttrlsiy},
then the ${\rm W}_k$  structure of the parafermionic irreps
\eqref{slsjsa} is  described as follows:
\bea\label{ksdsdk}
{\cal M}_{0}^{(0)}&\simeq& {\cal W}_{(0, 0)}\, ,\ \ \ \ \
{\cal M}_{0}^{(j)}\simeq {\cal W}_{({k\over 2},
{k\over 2}-j)}\ \ \ \
(j=1,\ldots k-1)\ , \\
{\cal M}_{J}^{(\ell)}&\simeq& {\cal W}_{(J, J-\ell)}\ \ \ \ {\rm for}\ \ \ \ell< 2J\  \ (J>0)\ ,
\nonumber
\eea
and
\bea\label{lsslss}
{\cal M}_{J}^{(\ell)}\simeq  {\cal M}_{{k\over 2}-J}^{(\ell')}\ \ \ \ \
{\rm for}\ \ \ \ell-\ell'=2J\ ({\rm mod }\  k)\ .
\eea

Let  $\Sigma$ be a set containing ${k(k+1)\over 2}$ number  of pairs  $(J,M)$ of the form:
\bea\label{lssnmdndh}
\Sigma=
\big\{\, (J,M)\,  &\big| & J=0,\, {\textstyle {1\over 2}},\,\ldots {\textstyle { k\over 2}}
\, ;\  M=-J, -J+1,\, \ldots\, ,\,  J\,;\nonumber \\ && (J,-J)\equiv
\big({\textstyle {k\over 2}}-J, {\textstyle {k\over 2}}-J\big)\ \big\}\ .
\eea
As implied above, the chiral parafermionic  algebra can be thought of
as an algebra of intertwiners  in the
 set
\bea\label{iuyssslsks}
\big\{ {\cal W}_{(J,M)}\big\}_{(J,M)\in\Sigma}
\eea
of
``integrable'' irreps of  ${\rm W}_k$.
In particular,
\bea\label{kasak}
{\boldsymbol \psi}_{\pm}(v)
\, :\, {\cal W}_{(J,M)} \longrightarrow {\cal W}_{(J,M\pm 1)}\ .
\eea
Notice that
the pairs $(J,-J)$ and
$({k\over 2}-J, {k\over 2}-J)$ are subject to the equivalence relation  in \eqref{lssnmdndh} since
the corresponding  highest weights $(\Delta,\,w)$\ \eqref{yttrlsiy} are the same, i.e.,
${\cal W}_{(J,-J)}\simeq {\cal W}_{({k\over 2}-J, {k\over 2}-J)}$.
Finally, the closed string Hilbert space in the case of minimal parafermionic models
is given by the direct sum
$\oplus_{(J,M)\in\Sigma}\ {\cal W}_{(J,M)}\otimes {\overline {{\cal W}}}_{(J,M)}$,
while the corresponding chiral Hilbert space can be
described  as follows:
\bea\label{uikjshsk}
{\tilde {\cal H}}^{(k)}=\oplus_{(J,M)\in\Sigma}\ {\cal W}_{(J,M)}\,  .
\eea

\subsubsection{${\rm W}_k$  invariant boundary states}

The ${\rm W}_k$   algebra posses an automorphism of the form
\bea\label{sjsklaslk}
{\mathbb C}\ :\ \
{\boldsymbol W}_s\to  (-1)^s\ {\boldsymbol W}_s\ ,\ \ \ s=2,\, 3,\,\ldots\ .
\eea
Thus  \cite{Ishibashi,Card}, there are two types of     ${\rm W}_k$  invariant CBC's.
In the string theory
they are referred to as  A and B-branes \cite{Moor}.
The corresponding  boundary states obey the equations
\bea\label{slsksal}
{\rm A}-{\rm brane} \ :\ \ \
\big[\, {\boldsymbol  W}_{s}(\tau)-(-1)^s\ {\bar
{\boldsymbol  W}}_{s}(\tau)\, \big]_{x=0}\, |\, B\, \rangle=0\ ,
\eea
and
\bea\label{ssdkdjddjh}
{\rm B} - {\rm brane} \ :\ \ \
\big[\, {\boldsymbol   W}_{s}(\tau)-
{\bar {\boldsymbol  W}}_{s}(\tau)\,  \big]_{x=0}\, |\, B\, \rangle=0\ .
\eea

Let $|\, {\cal I}_{(J,M)}\, \rangle_{\rm A}$ be the A-type
Ishibashi state\ \cite{Ishibashi,Card,Moor}
corresponding to the integrable irreps ${\cal W}_{(J,M)}$ from the set\ \eqref{iuyssslsks}.
In other words, it is a  solution of~\eqref{slsksal} in
the space
${\cal W}_{(J,M)}\otimes {\overline {\cal W}_{(J,M)}}$,
that is unique, provided
\bea\label{weslsjsl}
\big\{\,
\langle\, (J,M)
\,|\otimes {\overline {\langle\, (J,M)\,|}}\,\big\}\,\cdot |\, {\cal I}_{(J,M)}
\, \rangle_{\rm A}=1\ ,
\eea
where $|\,  (J,M)\,\rangle$ is the highest weight vector of ${\cal W}_{(J,M)}$.
Now, following  the  formalism of Ref.~\cite{Card}, we can construct a  set of  states
satisfying  both the Cardy consistency condition and
\eqref{slsksal}:
\bea\label{skshfd}
|\, B_{(j,m)}\, \rangle=\sum_{(J,M)\in\Sigma}\ {S^{(J,M)}_{(j,m)}\over \sqrt{
S_{(0,0)}^{(J,M)}}}\  |\, {\cal I}_{(J,M)}\, \rangle_{\rm A}\ , \ \ \ \ \ (j,m)\in\Sigma\ .
\eea
Here  $S^{(J,M)}_{(j,m)}$ is  a modular matrix of the characters of the integrable
irreps.\footnote{In the case under consideration the modular matrix is given by
\cite{Gepner,Kac},
$$S^{(J,M)}_{(j,m)}={2\, \omega^{2Mm}
\over \sqrt{k(k+2)}}\ \sin\Big({{\pi (2j+1)(2J+1)\over k+2}}\Big)\ .$$
}
The common wisdom then is that the ${k(k+1)\over 2}$ amount of  Cardy
states~\eqref{skshfd}
are the boundary states  corresponding to
some local,  ${\rm W}_k$  invariant boundary conditions.
In Section~\ref{thirdssec} these CBC's
have been  referred to as ${\cal B}_{l,n}$ $(0\leq n\leq l\leq k-1)$,
where the
integers    $l,\, n$ are  related to the pair  $(j,m)$ in~\eqref{skshfd}
as follows: $l=k-j-m$ and $n=j-m$.
In particular, the $k$ amount of the boundary  states
$\big\{\, |\, B_{({k\over 2}, {k\over 2}-s)}\,
\rangle\, \big\}_{s=0}^{k-1}$
correspond
to the $k$ possible types of the fixed CBC's,
while  $ \big\{\,
|\, B_{({k-1\over 2}, {k-1\over 2}-s) }\,
\rangle\, \big\}_{s=0}^{k-1}$
are the boundary states associated with  the non-trivial fixed points of the
boundary flow ${\cal R}^{(k)}_\theta$.

Let us consider now equation~\eqref{ssdkdjddjh}.
Again, to construct nontrivial boundary states of the B-type one
must draw on    the B-type Ishibashi states $|\, {\cal I}_{(J,M)}\,\rangle_{\rm B}$,  i.e.,
the unique normalized solutions  of~\eqref{ssdkdjddjh} in the space
${\cal W}_{(J,M)}\otimes {\overline {\cal W}_{(J,M)}}$.
Obviously $|\, {\cal I}_{(J,M)}\,\rangle_{\rm B}$ exists only if $J$ and $M$
satisfy the condition
\bea\label{slssklyt}
w_{(J,M)}=0\ ,
\eea
where the polynomial   $w_{(J,M)}$
is given by~\eqref{yttrlsiy}.
The following solutions of the Cardy
consistency condition\ \cite{Card} can be easily verified
\cite{Card,SalB,AOSR,Moor}:\footnote{
Equation~\eqref{soshfd} requires  special consideration for even $k$ and $j={k\over 4}$
 (see Ref.~\cite{Moor} for details)}
\bea\label{soshfd}
|\, B_{j}\, \rangle=\sqrt{k}\ 
\sum_{J=0}^{[{k\over 2}]}\ {S^{(J,0)}_{(j,0)}\over \sqrt{
S_{(0,0)}^{(J,0)} }}\  |\, {\cal I}_{(J,0)}\, \rangle_{\rm B}\  \ \ \
\big(\,j=0,\, 1,\, 2,\,\ldots \big[{\textstyle {k\over 2}}\big]\, \big)\ .
\eea

There are strong reasons to believe\ \cite{SalB, Card,AOSR} that
the solution $|\, B_j\, \rangle$ with $j=0$
corresponds to the free CBC:
\bea\label{shfd}
|\, B\, \rangle_{\rm free}^{(k)}= |\, B_0\, \rangle\ .
\eea
By virtue of  the  above described structure of this  boundary state, one
can show that the following relations are satisfied
when  $\sum_{s=1}^{n}\epsilon_s=0\ ({\rm mod}\, k)$:
\bea\label{ulssqwear}
{}_{\rm free}^{(k)}\langle\, B\, |\, {\boldsymbol \psi}_{\epsilon_1}(v_1)\, \ldots\,
 {\boldsymbol \psi}_{\epsilon_n}(v_n)\,
|\, {\rm vac}\, \rangle^{(k)}= {\rm g}_{\rm free}\
 \langle\, (0,0)\, |\, {\boldsymbol \psi}_{\epsilon_1}(v_1)\, \ldots\,
 {\boldsymbol \psi}_{\epsilon_n}(v_n)\,
|\, (0,0)\, \rangle\, ,
\eea
and
\bea\label{lssqwear}
{}_{\rm free}^{(k)}\langle\, B\, |\, {\bar {\boldsymbol \psi}}_{\epsilon_1}({\bar v}_1)\,
\ldots\,  {\bar {\boldsymbol \psi}}_{\epsilon_n}({\bar v}_n)\,
 |\, {\rm vac}\, \rangle^{(k)}={\rm g}_{\rm free}\
\ {\overline {\langle\, (0,0)\, |}}\, {\bar {\boldsymbol \psi}}_{\epsilon_1}({\bar v}_1)\,
\ldots\, {\boldsymbol \psi}_{\epsilon_n}({\bar v}_n)\,
{\overline {|\, (0,0)\, \rangle}}\, .
\eea
Here we use the notation
$|\, {\rm vac}\, \rangle^{(k)}\equiv
|\, (0,0)\, \rangle\,\otimes\, {\overline {|\, (0,0)\, \rangle}}$
for the closed string  ground state of the minimal  parafermionic  model.

It is essential to point out
that correlators \eqref{ulssqwear} and \eqref{lssqwear} are
not single-valued functions of their variables. To resolve the phase ambiguity
in \eqref{ulssqwear} we  require that,  (a)
when the arguments  are real, $v_s=\tau_s$,
then  \eqref{ulssqwear} is a real function within  the domain
$1/T>\tau_1>\ldots >\tau_n>0$, and  (b) the short distance
expansions of \eqref{ulssqwear}  must be consistent with
 the OPE's\ \eqref{KJaskjs} provided ${\boldsymbol \Psi}_n(\tau)
\equiv {\boldsymbol \psi}_n(\tau-\ri x)|_{x=0}$.
Making use of  the relation between  parafermions
and  ${\widehat {SU_k(2)}}$ WZW currents\ \cite{ZamFat}, it is possible to
show that both conditions can be indeed satisfied.
Similarly, one can  resolve the phase ambiguity of \eqref{lssqwear}.
As a result, the boundary values  of the
 correlation functions \eqref{ulssqwear} and
\eqref{lssqwear}  at real $v_s={\bar v}_s=\tau_s$   coincide within the
domain $1/T>\tau_1>\ldots >\tau_n>0$.

\subsection{Baxter's operators}

In the boundary state formalism,
thermal expectation values
\eqref{khskskiw} are expressed in terms of    correlators of the form:
\bea\label{lajska}
\big\langle\big\langle\, \cdots\,
\big\rangle
\big\rangle_0
={{}_{\rm free}\langle\, B\, |\,\cdots\, |\, {\rm vac}\,\rangle\over
{}_{\rm free}\langle\, B\, |\, {\rm vac}\,\rangle}\, ,\ \ \ \ |\, {\rm vac}\,\rangle=
|\, {\rm vac}\,\rangle^{(\Phi)}\otimes |\, {\rm vac}\,\rangle^{(k)}
\, ,\eea
with $|\, B\, \rangle_{\rm free}$ defined in~\eqref{kshks}.
Our previous  analysis implies, therefore,   that
the integrand
in~\eqref{slsjlaas} can be written as an expectation value over
the highest vectors $|\, P, (J, M)\, \rangle=
|\, P\,\rangle \otimes |\, (J,M)\, \rangle$
with $J=M=0$:
\bea\label{ytrklsjssha}
&&\re^{V \sum_{j=1}^{2n}\epsilon_j\tau_j}\
\langle\langle\, {\boldsymbol V}^{\rm (Mats)}_{\epsilon_{1}}(\tau_{1})\cdots\,
{\boldsymbol V}^{\rm (Mats)}_{\epsilon_{2n}}(\tau_{2n})\, \rangle\rangle_0=
\\ &&\ \ \ \ \ \ \ \ \ \ \ \ \ \ \ \ \ \ \ \
\langle\,  (0,0), P\,|\, {\bf U}_{\epsilon_{1}}(\tau_{1})\cdots\,
{\bf U}_{\epsilon_{n}}(\tau_{n})\, |\, P, (0, 0)\, \rangle \ .
\nonumber
\eea
Here
the parameter $V$  is related to the zero-mode momentum 
$P$:
\bea\label{slsaksa}
V=2\pi\ri\ T\ \  {\beta P\over \sqrt{k}}\ ,
\eea
and
\bea\label{saslslk}
{\bf U}_{\pm }(\tau)=
{\boldsymbol \psi_\pm}\ \re^{\pm \ri\ {2\beta\over\sqrt{k}}\, {\boldsymbol
\phi}}(\tau)\ .
\eea
From the mathematical
point of view,
${\bf U}_{\pm }(v)$ are  intertwining operators (chiral vertex operators)
of the conformal dimension $d$\ \eqref{llskjsl}. They
act  in the chiral Hilbert space
\bea\label{lksajhsk}
{\tilde{\cal H}}_{\rm free}=\int_P\ \oplus_{(J,M)}\, {\cal V}_{P,(J,M)}\,  \ \ \ \ {\rm with}\ \ \
\ \  \  {\cal V}_{P,\, (J,M)}\equiv {\cal F}_P\otimes{\cal W}_{(J,M)}
\eea
in accordance with the formula
\bea\label{ssaksa}
{\bf U}_{\pm }(\tau)\ :\ {\cal V}_{P,\, (J,M)}\to {\cal V}_{P\pm {\beta\over \sqrt{k}},
(J,M\pm 1)}\ .
\eea
Equation~\eqref{ytrklsjssha}, in turn, implies that
the power series $A_{\pm}(s,p)$\ \eqref{lsajslasa},\,\eqref{lsashksa}
can be thought of as expectation values over the state
$|\, P, (0, 0)\, \rangle$ with $P=2\sqrt{k}\, p/\beta$ of the operators
\bea\label{slisajsl}
&&\ \ \ \ \ \ \ \ \ \ \ \ \ 
{\mathbb A}_{\pm}(s) = \bigg(\, {\rm Tr}_{\rho_\pm}
\Big[\, \omega^{\mp 2{\boldsymbol M} {\boldsymbol h}}\,
\re^{ \pm 2\pi \ri {\beta\over\sqrt{k}} {\boldsymbol \Pi}_0 {\boldsymbol h}}
\, \Big]\,\bigg)^{-1}\times \\ &&
{\rm Tr}_{\rho_\pm}\bigg[\, \omega^{\mp 2{\boldsymbol M} {\boldsymbol h}}\
\re^{ \pm 2\pi \ri {\beta\over\sqrt{k}}\,
{\boldsymbol \Pi}_0 {\boldsymbol h} }\
\  {\cal T}\exp\Big\{\mu
\int_{0}^{1/T}\rd\tau\  {\bf K}_\pm(\tau)
\, \Big\}\,\bigg]\, .\nonumber
\eea
Here  ${\bf K}_\pm(\tau)= {\boldsymbol a}_\mp {\bf U}_+(\tau) +
{\boldsymbol a}_\pm {\bf U}_\mp(\tau)$
and $s=\mu^2\, (2\pi T)^{2d-2}$.

Some explanations of formula \eqref{slisajsl} are in order at this point.
First,
it contains the ordered exponential (the symbol ${\cal
T}$ denotes the path ordering) which is defined in terms of the
power series in $\mu$ as follows,
\bea\label{troi}
&&{\cal T}\exp\bigg\{\mu
\int_{0}^{1/T}\rd\tau \ {\bf K}_\pm(\tau)\,\bigg\}=1+
\sum_{n=1}^{\infty}
\mu^n\times \\ && \ \ \ \ \ \ \ \ \
 \int_{1/T> \tau_1 > \tau_2 >\ldots
>\tau_n> 0} \rd\tau_1\ldots\rd\tau_n\
{\bf K}_\pm(\tau_1)\, {\bf K}_\pm(\tau_2)\,\ldots\, {\bf K}_\pm(\tau_n)\ .\nonumber
\eea
In fact, only even powers of $\mu$  survive in the definition~\eqref{slisajsl},
so  ${\mathbb A}_{\pm}$ are
formal series in $s\propto \m^2$
whose expansion
coefficients are  operators acting in the chiral Hilbert space \eqref{lksajhsk}.
Second, the factor  $ \omega^{\mp 2{\boldsymbol M}{\boldsymbol h}} $ appearing in
\eqref{slisajsl}
was undefined yet.
As  follows from the commutation relations\ \eqref{ssmsas},
only integer powers of the operator  $\omega^{ 2{\boldsymbol M}}$
are  involved in \eqref{slisajsl}.
The  action  of $\omega^{ 2{\boldsymbol M}}$ in the chiral
Hilbert space\ \eqref{lksajhsk}
 is defined by the formula
\bea\label{slssl}
\omega^{ 2{\boldsymbol M}}\ {\cal V}_{P,\, (J,M)}=\omega^{2 M}\ {\cal V}_{P,\,(J,M)}\ .
\eea
Notice that, because of  the isomorphism
${\cal W}_{(J,-J)}\simeq {\cal W}_{({k\over 2}-J, {k\over 2}-J)}$,
the operator ${\boldsymbol M}$ itself  is ill-defined.

In spite of the fact that
the factor $ \omega^{\mp 2{\boldsymbol M}{\boldsymbol h}} $ does not contribute
to the matrix element
$\langle\, P, (0,0)\,|\,{\mathbb A}_\pm\,|\, P, (0, 0)\, \rangle$,
it plays an important
role in the   definition of  ${\mathbb A}_\pm$. Due to this phase
factor the operators ${\mathbb A}_\pm$ commutes with the operator
\bea\label{ssiyudl}
{\mathbb I}_{1}=\int_0^{1/T}{\rd v\over 2\pi}\ \Big(\, (\partial
{\boldsymbol \phi})^2+
{\boldsymbol W}_2\, \Big)\ .
\eea
A proof of the commutation relation
\bea\label{lkssj}
[\, {\mathbb I}_{1}\, ,\, {\mathbb A}_\pm(s)\, ]=0\ ,
\eea
is based on  the quasiperiodic properties of the intertwining operators
${\boldsymbol\psi}_{\pm}(u)$:
\bea\label{slsjls}
{\boldsymbol \psi}_{\pm}(u+1/T)={\boldsymbol
\psi}_{\pm}(u)\ \omega^{1\mp 2{\boldsymbol M}}\ .
\eea
It follows along  the  line 
of  Appendix C of Ref.\,\cite{blz3}.
Notice that the operators ${\mathbb I}_1+{\bar {\mathbb I}}_1$ and
${\mathbb I}_1-{\bar {\mathbb I}}_1$ coincide, respectively, with
Hamiltonian and the  momentum operator in the closed string channel. Hence,
commutation relation \eqref{lkssj} manifests the invariance of
${\mathbb A}_\pm(s)$ with respect to translations along the $\tau$-direction.

It is worth to note that the
operators ${\mathbb A}_{\pm}(s)$\ \eqref{slisajsl}
act invariantly in  ${\cal V}_{P, (J, M)}$. Furthermore, since
each  space ${\cal V}_{P, (J, M)}$ naturally splits into the direct sum
of finite dimensional ``level subspaces''
\bea\label{alksls}
{\cal V}_{P, (J, M)}=\oplus_{\ell=0}^{\infty}\ {\cal V}_{P, (J, M)}^{(\ell)}\ ;\ \ \
{\mathbb L}\ {\cal V}_{P, (J, M)}^{(\ell)}=\ell\ {\cal V}_{P, (J, M)}^{(\ell)}\ ,
\eea
and the grading operator ${\mathbb L}$ essentially coincides with  ${\mathbb I}_1$,
then ${\mathbb A}_{\pm}(s)$ act invariantly
in all the level subspaces ${\cal V}_{P, (J, M)}^{(\ell)}$.
In particular,  the highest vectors  $|\, P, (J,M)\,\rangle$ are
eigenvectors for ${\mathbb A}_{\pm}(s)$ and
\bea\label{kljshskh}
{\mathbb A}_{\pm}(s)\, |\, P, (0,0)\,\rangle=A_\pm(s,p)\  |\, P, (0,0)\,\rangle\ \ \ \ \ \
{\rm with}\ \ \ p={\beta P\over 2\sqrt{k}}\ .
\eea

Following the route of the work\ \cite{blz3}
it is possible to prove that ${\mathbb A}_{\pm}(s)$  commute
among themselves for any values of $s$, 
\bea\label{lksasal}
[\,{\mathbb A}_\pm(s)\, ,\, {\mathbb A}_\pm(s')\, ]=
[\,{\mathbb A}_+(s)\, ,\, {\mathbb A}_-(s')\, ]=0\ ,
\eea
and
satisfy the operator-valued identity
\bea\label{ueytetsjssjk}
\re^{\ri {2\pi \beta\over\sqrt{k}} {\boldsymbol \Pi}_0 }\,
{\mathbb A}_+(qs)\,{\mathbb  A}_-( q^{-1}s)-
\omega^{2\boldsymbol M}\,
{\mathbb  A}_+( q^{-1}s)\,{\mathbb A}_-( qs)=
\re^{\ri {2\pi \beta\over\sqrt{k}}  {\boldsymbol \Pi}_0 }
 -\omega^{2\boldsymbol M}\, ,
\eea
with $q=\re^{\ri\pi d}$. Again,
formulae \eqref{lksasal} and \eqref{ueytetsjssjk} should be understood   as
an infinite set of formal operator-valued   relations for the
expansion coefficients of ${\mathbb  A}_\pm$
without
any reference  to convergence issues.
The quantum Wronskian relation \eqref{sjssjk} follows immediately from
\eqref{kljshskh} and \eqref{ueytetsjssjk}.

A key idea of the proof of \eqref{lksasal} and \eqref{ueytetsjssjk}
is based on the observation that
the formal integrals
\bea\label{kssksha}
{\boldsymbol x}_0={1\over q-q^{-1}}\ \int_0^{1/T}\rd u\ {\boldsymbol U}_-(u)\, ,\ \ \ \ \
{\boldsymbol x}_1={1\over q-q^{-1}}\ \int_0^{1/T}\rd u\ {\boldsymbol U}_+(u)
\eea
satisfy the Serre relations for the quantum 
affine algebra $U_q\big({\widehat {sl(2)}}
\big) $
and can be identified  with the generators of its Borel subalgebra.
With this observation
the construction   elaborated  in \cite{blz3}  can be applied
without any  changes. Finally,  we note that
the operators ${\mathbb  A}_\pm$, acting in the CFT chiral Hilbert space,
play a role similar to that of
the Q-operators of
Baxter's lattice theory~\cite{Baxter}.

\section{
Appendix : Partition function of  the  BP  sinh-Gordon model}

If the parameter $\beta^2=-b^2$ is negative
the Schr${\ddot {\rm o}}$dinger equation \eqref{OsDER}
takes the form \eqref{OsDEra}.
It is quite natural to expect that  this differential equation
within the parametric domain $b^2>0$
is somehow related
to the BP sinh-Gordon model\ \eqref{uyywtrsahsa}.
The precise relation is proposed in  Eq.\,\eqref{ksssk}.
Unfortunately,  a rigorous derivation
of this formula
does not currently exist.
Here we    briefly  argue  in  favour of \eqref{ksssk}
based on the  high- and low-temperature expansions
of the partition function $Z^{(k)}_{\rm bshg}(h)$.

\subsection{High-temperature expansion}

\subsubsection{The BP Liouville model}

First of all, let us discuss  the leading high-temperature behavior
of \eqref{slssjshy} based on
properties of the  Hamiltonian\ \eqref{uyywtrsahsa}.
Qualitatively, one may expect that if $\Re  e\, h$ is not too small,
the limit $T\to\infty$ of\ \eqref{slssjshy} is controlled by either
the boundary operator ${\boldsymbol V}_+\equiv{\boldsymbol \Psi}_+ \re^{{b\over\sqrt{k}} {\boldsymbol \Phi}_B}$ 
or ${\boldsymbol V}_-\equiv
{\boldsymbol \Psi}_-\re^{-{b\over\sqrt{k}} {\boldsymbol \Phi}_B}$ in\ \eqref{uyywtrsahsa}
depending on the sign $\Re  e\, h$, with some crossover at small
$\Re  e\, h$.

More precisely, let us assume that  $\Re  e\, h<0$. Then
we can treat the term $\mu {\boldsymbol V}_-$
in the Hamiltonian\ \eqref{uyywtrsahsa}   as a perturbation, and in leading
approximation one has
\bea\label{lkjlsajsa}
Z_{\rm bshg}^{(k)}(h)|_{T\to\infty}\to Z_{\rm bl}^{(k)}(h)\  \ \ \ \ \ \ (\Re  e\, h<0)\ ,
\eea
where $ Z_{\rm bl}^{(k)}(h)$ is  the partition function
corresponding to the Hamiltonian of the ``BP Liouville'' model:
\bea\label{wtrsahsa}
{\boldsymbol H}_{\rm bl}^{(k)}={\boldsymbol H}_{\rm free}+ h\ {\boldsymbol \Phi}_B-
\mu\  {\boldsymbol \Psi}_+\ \re^{{b\over\sqrt{k}} {\boldsymbol \Phi}_B}\ .
\eea
An important feature of the QFT \eqref{wtrsahsa}
is   the conformal invariance.
To be precise, the theory can be made scale and conformally invariant by an
appropriate redefinition of the RG transformation, namely by supplementing it by
a formal  field redefinition ${\boldsymbol \Phi}\to {\boldsymbol \Phi}+
{Q\over\sqrt{k}}\, \delta t_{\rm RG}$,
where the RG ``time'' $t_{\rm RG}\sim \log(1/T)$ and $Q=b+b^{-1}$.
In the nomenclature of string theory this corresponds to introducing a linear
dilaton which  modifies  slightly
the   stress-energy tensor of the model:
\bea\label{skjhskjs}
{\boldsymbol T}^{(2)}&=&-(\partial {\boldsymbol \Phi})^2+
{\textstyle{Q\over \sqrt{k}}}\, \partial^2{\boldsymbol\Phi} +
{\boldsymbol W}_2 \ ,\\
{\bar {\boldsymbol T}}^{(2)}&=&-({\bar \partial} {\boldsymbol \Phi})^2+
{\textstyle{Q\over \sqrt{k}}}\,
{\bar \partial}^2{\boldsymbol \Phi} +
{\bar {\boldsymbol W}}_2\ .\nonumber
\eea
The first two terms in each  equation\ \eqref{skjhskjs}
constitute  the corresponding chiral  components of the modified stress-energy tensor of the Gaussian
theory, 
while ${\boldsymbol W}_2$ and ${\bar {\boldsymbol W}}_2$ are chiral components  of
the stress-energy tensor of  the minimal parafermionic model 
(see the OPE \eqref{lkss}).
The scaling dimension of the boundary operator ${\boldsymbol \Psi}_+ \re^{{b\over\sqrt{k}} {\boldsymbol \Phi}_B}$
with respect to the modified stress-energy tensor \eqref{skjhskjs} equals one and, hence,
the temperature and $\mu$-dependences of the partition function $Z_{\rm bl}^{(k)}(h)$ readily follows
from the  dimensional analysis:
\bea\label{slkusysls}
Z_{\rm bl}^{(k)}(h)=
(2\pi T)^{\ri { P Q\over\sqrt{k}}}\   \mu^{-\ri { P\sqrt{k}\over b}}\  G^{(k)}
(P\, |\, b)\ ,
\eea
with
\bea\label{qwelksals}
 P=\ri\ {h\over T}\ .
\eea
Here  $G^{(k)}$ is some function of the dimensionless parameters $P$, $b$ and $k$.

In a view of  the standard  heuristic arguments of Ref.\,\cite{Goulian},
it is  expected that  $G^{(k)}(P\, |\,b)$
is  an  analytical function of the complex variable $P$
in the lower half plane $\Im  m P<0$.
Furthermore,
following\ \cite{Goulian}  in
the ``perturbative'' calculation of the
partition function $Z_{\rm bl}^{(k)}$,
one can first integrate out the constant mode of the Bose field 
${\boldsymbol \Phi}$. This integration produces simple poles in the variable $P$
at the points $P=\ri\ n\,b\sqrt{k}$ $(n=0,\,1\ldots)$, and
the corresponding residues 
\bea\label{slksls}
G^{(k)}_n= {\rm Res}_{P=\ri\, b\sqrt{k}\ n }\big[\, G^{(k)}(P\, |\,b) \,\big]\
\ \ \ \ \ \ (n=0,\, 1,\,2\ldots\,)
\eea
are expressed through the integrals of the $k\times n $-points
Matsubara correlation functions
\bea\label{slsslj}
G^{(k)}_n&=&{{\rm g}_D\, {\rm g}_{\rm free}\over 2\pi\ri}\ (2\pi T)^{n-( kn-1)n b^2}\times\\
&&\int_0^{1\over T}\rd \tau_1\ldots\int_0^{\tau_{kn-1}}\rd\tau_{k n}\
\langle\langle\, {\boldsymbol V}^{\rm (Mats)}_+(\tau_1)\ldots
{\boldsymbol V}^{\rm (Mats)}_+(\tau_{k n})\,
\rangle\rangle'_0\ .\nonumber
\eea
Here  ${\rm g}_{\rm free}$ is the
boundary degeneracy of the free CBC in the
minimal parafermionic model (see Eq.\,\eqref{aslss}),
and the prime  means that
the constant mode  contribution is
excluded from the
thermal averaging
$\langle\langle\,\cdots\rangle\rangle_0$ \eqref{khskskiw}.
The free-field correlators in \eqref{slsslj} have an especially simple
form  for   $n=1$. In this
case the integral   \eqref{slsslj} 
can be brought to the form  of the Selberg integral\ \cite{Selberg,Dot}. This yields
the explicit   formula
\bea\label{sslssa}
G^{(k)}_1&=&
{{\rm g}_D\, {\rm g}_{\rm free}\over 2\pi\ri}\,  {k!\over k^{k\over 2}}\,
\int_0^{2\pi}\rd v_1\ldots \int_0^{v_{k-1}}\rd v_{k}\,
\prod_{i<j}\Big[2\, \sin\big({\textstyle{v_i-v_j\over 2}}\big)\Big]^{-{2(1+b^2)\over k} }=
\nonumber\\
&& {{\rm g}_D\, {\rm g}_{\rm free}\over 2\pi\ri}\ \ \Gamma(-b^2 )\ \bigg[{2\pi\over \sqrt{k}
\Gamma(1-{1+b^2\over k})}\bigg]^k\ .
\eea

Another highly nontrivial property of the
BP Liouville  partition function can be guessed  by  virtue
of the form of   stress-energy tensor\ \eqref{skjhskjs}. Namely, it is
invariant with respect  to the   transformation $b\to b^{-1}$.
In the case of the conventional Liouville model a similar
phenomenon
manifests  a remarkable non-perturbative  duality   of the theory.
If one admits     that the same symmetry  occurs
in  the  BP  Liouville model,
then, the function $G^{(k)}(P\,|\, b)$ should also possess
an additional   series of  ``dual'' poles at $ P=\ri\, b^{-1}\sqrt{k}\ n$.

It should be mentioned that 
the model \eqref{wtrsahsa}   has been already studied
in the works  \cite{FZZ} and \cite{Ahn}   for
the cases  $k=1$ and
$k=2$, respectively. 
In both cases the function
$G^{(k)}$ \eqref{slkusysls} looks as follows 
\bea\label{jiuysssysslssl}
G^{(k)}(P|b)=
{ {\rm g}_D{\rm g}_{\rm free} \over 2\pi \ri P}\ \Gamma\Big(1+{\ri bP\over \sqrt{k}}\Big)\,
\Gamma\Big(1+{\ri P \over b\sqrt{k}}\Big)\,
\bigg[
{2\pi\ b^{-{2\over k}}\over \sqrt{k}\Gamma(1-{1+b^2\over k})}\bigg]^{-{\ri P\sqrt{k} \over b}}.
\eea
Notice  that 
\eqref{jiuysssysslssl} satisfies   
the all above-mentioned conditions
even thought  $k>2$.

\subsubsection{\label{alsjakju}General structure of the high-temperature expansion }

Above we have discussed 
the leading
high-temperature  behavior of  $Z_{\rm bshg}^{(k)}$ in the case
$\Re  e\, h<0$. Obviously for $\Re  e\, h>0$, $Z_{\rm bl}^{(k)}(h)$
in  asymptotic formula
\eqref{lkjlsajsa} should be replaced by $Z_{\rm bl}^{(k)}(-h)$.
Since $Z_{\rm bl}^{(k)}(\pm h)$ vanishes in the limit $T\to\infty$ if $h$ is
taken in the ``wrong'' half plane
(note the factor $T^{\ri { P Q\over\sqrt{k}}}$ in \eqref{slkusysls}),
this in turn suggests that the overall $T\to\infty$ asymptotics of the
partition function is correctly
expressed by the sum
\bea\label{ssysls}
Z_{\rm bshg}^{(k)}|_{T\to\infty}
\to Z_{\rm bl}^{(k)}(h)+Z_{\rm bl}^{(k)}(-h)\ .
\eea

What can be said about corrections to the leading asymptotic?
Again we assume that $\Re  e\, h<0$ and consider the perturbative
effect of the   term  
$\mu\,{\boldsymbol V}_-$  in\ \eqref{uyywtrsahsa} to the
partition function\ \eqref{slkusysls}.
In the unperturbed BP Liouville model
theory   the parameter $\mu$ is a dimensionless  constant. Let
us   eliminate $\mu$  from the  Hamiltonian\ \eqref{wtrsahsa} by   shifting
the field ${\boldsymbol \Phi}$. Then the coupling $\mu$ 
in front of the ${\boldsymbol V}_-$
in Eq.\,\eqref{uyywtrsahsa} is replaced
by $\mu^2$. Since the  anomalous dimension of the boundary
operator ${\boldsymbol V}_-$ with respect to the BP Liouville
stress-energy tensor \eqref{skjhskjs} is given by $1-{2bQ\over k}$, then
$\mu^2\sim\ 
E_*^{{2bQ\over k}}$, where $E_*$ is the physical energy scale of the
BP sinh-Gordon model.
Hence we deduce that the perturbative
corrections to the leading asymptotics\ \eqref{lkjlsajsa}\ should be
in a form of power series expansion of the dimensionless parameter
$\kappa^{{2bQ\over k}}$, with  $\kappa=E_*/T$.
The
case $\Re  e\, h>0$ can be analyzed similarly and one comes to the
same conclusion about the form of perturbative corrections to \eqref{ssysls}.
As was mentioned above, it is quite natural 
to  expect that the leading asymptotics \eqref{ssysls}
is invariant with respect to the duality  transformation $b\to b^{-1}$.
If one assumes that   the BP sinh-Gordon partition function
possesses this property of self-duality as well, then
the high-temperature expansion of $Z_{\rm bshg}^{(k)}$ 
should be of the form
\bea\label{slskls}
Z_{\rm bshg}^{(k)}=  Z_{\rm bl}^{(k)}(h) \ M(\kappa,h)+Z_{\rm bl}^{(k)}(-h)\ M(\kappa,-h)\ .
\eea
Here $M(\kappa,h)$ is a formal double power series in integer powers of $\kappa^{{2bQ\over k}}$ and
$\kappa^{{2Q\over bk}}$:
\bea\label{slsaksal}
M(\kappa,h)\simeq 1+\sum_{n,m}M_{n,m}(h)\ \kappa^{{2Q\over k}\, ( n b+ m b^{-1})}\ .
\eea

\subsubsection{Small $\kappa$ expansion of the spectral determinant}

In order to justify  the  
proposed  formula for  $Z_{\rm bshg}^{(k)}$ \eqref{ksssk},
we should  observe  the structure  \eqref{slskls} in
the
small-$\kappa$ expansion of
the spectral determinant $D(\kappa,\xi)$\ \eqref{qqzsksls}.
When $\kappa$ goes to zero
and
$u\gg 1$ the potential in the Schr${\ddot {\rm o}}$dinger problem
\eqref{OsDEra}
can be approximated by $\kappa^2\
\re^{{ku\over bQ} }$. To 
understand the quality of this approximation one can make
a change the variable 
\eqref{kjsssk} with $d=1-b Q/k$  and bring 
the differential equation \eqref{OsDEra} to the form
\eqref{ulsaslsa}.
Therefore for $u\gg 1$
\bea\label{sjuyy}
\Theta_+(u)\to {\textstyle \sqrt{2Qb\over \pi k}}\ K_{ \ri { b P\over\sqrt{k}}}
\big({\textstyle {2Qb\kappa\over k}}\, \re^{{ku\over 2bQ}}\big)\ .
\eea
Here and below, in order to write formulae in the most instructive form,
we use the notation $P={2Q\over \sqrt{k}}\, \xi$
rather than $\xi$ (see relations \eqref{qwelksals} and
\eqref{lkasnlas}).
The overall normalization in \eqref{sjuyy} is 
chosen to ensure the asymptotic
condition \eqref{psiassminus}. Within  the domain,
\bea\label{sksunag}
-{\textstyle {Q\over kb}}\ \log\big({\textstyle{1\over \kappa^{2}}}\big) \ll u
\ll {\textstyle{Qb\over k}}\
\log\big({\textstyle{1\over \kappa^{2}}}\big)\, ,
\eea
the potential in \eqref{ulsaslsa} develops a wide plateau and
the solution
$\Theta_{+}$\ \eqref{sjuyy}\
becomes a combination
of the  two plane waves,
\bea\label{waves}
\Theta_{+}(u) &=&\sqrt{Qb\over 2\pi k}\ \Gamma\Big(   -\ri\,{bP\over \sqrt{k}}\Big)\
\Big({Qb\kappa\over
k}\Big)^{-{\rm i}\,
 {bP\over \sqrt{k}}}\ A(\kappa,-P\,|\, b)\ \re^{{\rm i} {\sqrt{k}P\over 2Q}\,u}+\nonumber\\ &&
\sqrt{Qb\over 2\pi k}\ \Gamma\Big(   \ri\,{bP\over \sqrt{k}}\Big)\
\Big({Qb\kappa\over
k}\Big)^{{\rm i} {bP\over \sqrt{k}}}\
A(\kappa, P\,|\, b)\ \re^{-{\rm i} {\sqrt{k}P\over 2Q}\,u}\ ,
\eea
with
\bea\label{sjdsudu}
A( \kappa, P\, |\, b)=1+O\big(\kappa^{2Qb\over k}\big)\ .
\eea
When $\kappa$ is small but finite, corrections to \eqref{sjdsudu} can be obtained
using the perturbation theory. In view of 
equation \eqref{lksjs}, these
corrections have
the form of a power series in
$\kappa^{2Qb\over k}$:
\bea\label{sksahkasj}
A( \kappa,P\, |\, b)
\simeq 1+\sum_{n=1}^{\infty}\ A_n(P\,|\, b)\ \kappa^{2 Q b n\over k}\ .
\eea
The formal power series $A( \kappa,P\, |\,b)$ amounts to
$A_{+}(s,p)$ \eqref{lksjlks}
taken  at $\beta^2=-b^2<0$ with
the variable  $s$ related to $\kappa$ as
in \eqref{lsalsuy} and   $p=\ri {b P\over 2\sqrt{k}}$.
We have  no reason to
expect  the convergence of the power series \eqref{sksahkasj} for $b^2>0$,  i.e., the series should
be understood as a formal asymptotic expansion.

To find  the form of the solution $\Theta_-$ in the domain\ \eqref{sksunag}, we note
that the transformation $u\to -u$ and $b\to b^{-1}$  leaves the
Schr${\ddot {\rm o}}$dinger equation\ \eqref{OsDEra} unchanged 
while  interchanging
the solutions $\Theta_+$ and 
$\Theta_-$\ \eqref{psiassminus}.\footnote{For this reason the
partition function\ \eqref{ksssk}
is invariant with respect to
the  duality transformation $b\to b^{-1}$.} Hence,
$\Theta_-$ can be obtained 
by means of this   transformation
from \eqref{waves} 
within the domain\ \eqref{sksunag}.
Now   the Wronskian\ \eqref{qqzsksls} can be   calculated
at any point from\ \eqref{sksunag}, where both 
$\Theta_\pm$ are combinations ofplane waves.
It yields the following  
form of the small-$\kappa$ expansion of the spectral determinant:
\bea\label{sskpsa}
D(\kappa,\xi)\simeq R(P)\  A(\kappa,P\,|\, b)\  A(\kappa,P\, |\, b^{-1})+
R(-P)\  A(\kappa,-P\, |\,b )\  A(\kappa,-P\, |\, b^{-1})\ ,
\nonumber\eea
with
\bea\label{lkslsj}
R(P)={\ri\, P\over 2\pi\sqrt{k}}\ \Gamma\Big( {\ri bP\over\sqrt{k}}\Big)\,
\Gamma\Big( {\ri P\over b\sqrt{k}}\Big)\ b^{\ri {P\over\sqrt{k}}\, (b^{-1}-b)}\
\Big({\kappa Q\over k}\Big)^{-\ri {QP\over\sqrt{k}}}\ ,
\eea
and $P={2Q\over \sqrt{k}}\, \xi$.

The  high-temperature expansion\ \eqref{sskpsa} 
is in   agreement with \eqref{slskls}.
Furthermore,
it implies  that   formula \eqref{jiuysssysslssl}
holds true 
for any $k\geq 1$ .

\subsection{Low-temperature expansion}

Here we elucidate the low-temperature behavior of $Z_{\rm bshg}(h)$ using the
boundary state formalism described in Appendix B.

In the case of the BP sinh-Gordon model
the boundary state $|\, B\,\rangle_{\rm bshg}$
is some particular vector in the closed string space of state
\bea\label{sksjsjsk}
|\, B\,\rangle_{\rm bshg}\in
\int_P\ \oplus_{(J,M)}\, {\cal V}_{P,(J,M)}\otimes{\bar {\cal V}}_{P,(J,M)}\ ,
\eea
where we use the same notations  as in formula  \eqref{lksajhsk}.
Unlike  the conformal boundary states
discussed in  Appendix B, $|\, B\,\rangle_{\rm bshg}$ essentially depends
on the temperature  via the dimensionless
parameter $\kappa=E_*/T$.
In the presence of the external field  $h$ \eqref{uyywtrsahsa},
the partition function
$Z_{\rm bshg}(h)$  can be   expressed  in terms of
the vacuum overlap  of the boundary state  \eqref{sksjsjsk}
analytically continued
to pure imaginary values of the variable $P$ \eqref{qwelksals}:
\bea\label{usslkssl}
Z_{\rm bshg}(h)={}_{\rm bshg}\langle\, B\, |\cdot \Big\{\, | \, P,\, (0,0)\,\rangle\otimes
{\overline{| \, P,\, (0,0)\,\rangle}}\, \Big\}\Big|_{P=\ri\ {h\over T}}\ .
\eea
The motivation behind  \eqref{usslkssl} can be found  in Ref.\,\cite{SLAZ}
where a similar relation  was exploited  in
the context
of the circular brane model.

The simplest  idea about
the infrared fixed point   of  the BP sinh-Gordon boundary
flow is that it   corresponds  to the fixed  CBC
for  both bosonic and parafermionic sectors of the theory.
This  scenario implies that
\bea\label{slssa}
|\, B\,\rangle_{\rm bshg}\to
\re^{-{E_{\rm bshg}\over T}}\
|\, B\,\rangle^{(\Phi)}_{\rm D}\otimes |\, B_{(0,m)}\, \rangle\ \ \ \ \ \ {\rm as}\ \ \
\kappa\to\infty\ .
\eea
Here $E_{\rm bshg}\propto E_*$ is the ground state energy,
$|\, B_{(0,m)}\, \rangle$ is  the boundary state
corresponding to the fixed CBC of the minimal
parafermionic model~\eqref{skshfd},\footnote{Recall,
in order to fix the value of  integer $m=0,\,1,\ldots k-1$ in \eqref{slssa}, one needs to specify
the  normalization of the boundary fields ${\boldsymbol \Psi}_\pm$ unambiguously
(see  Section~\ref{firstssec}).} and
\bea\label{uyalsaj}
|\, B\,\rangle^{(\Phi)}_{\rm D}=
{\rm g}_D\ \int_{P} \exp\Big(\sum_{n=1}^{\infty}\,{\textstyle {2\over n}}\
{\boldsymbol \phi}_{-n}{\bar {\boldsymbol \phi}}_{-n}\Big)\,
|\, P\, \rangle\otimes {\overline {|\, P\, \rangle}}\
\eea
is the boundary state associated with the Dirichlet CBC~\cite{nappi,Ishibashi}.
The assumption~\eqref{slssa}   leads,  in turn,
to  the  leading  low-temperature
asymptotics   for  the  BP sinh-Gordon  partition function:
\bea\label{aksasak}
\log Z_{\rm bshg}(h )=-{E_{\rm bshg}\over T}+ \log\big({\rm g}_D\, {\rm g}_{\rm fixed} \big)+
O\big(\, T\,\big)\ .
\eea

To make  more definite   predictions about the  low-temperature
expansion of $ Z_{\rm bshg}(h )$, one should take into account
the  integrability of the model.
In a forthcoming paper~\cite{Pillow} we intend to present  a  comprehensive   discussion of
a relevant set of local  Integral of Motions (IM's)
in the context of  more general QFT model with  boundary
interaction.
Skipping details of the computational complexity we  mention
here that
the boundary state\ \eqref{sksjsjsk}
is expected to satisfy an infinite set of
conditions
\bea\label{alksla} (\,
{\mathbb I}_{2l-1}- {\bar {\mathbb I}}_{2l-1}\,)\ |\, B\, \rangle_{\rm
bshg}=0\ \ \ \ \ \ \ \ \ (\,l=1,\,2\ldots\,)
\eea
for some operators
\bea\label{alkslakj}
{\mathbb I}_{2l-1}(x)=\int_0^{1/T}{\rd\tau\over 2\pi}\ {\boldsymbol T}_{2l}(\tau-\ri x)\ , \ \ \
\ \ \ \ {\bar {\mathbb I}}_{2l-1}(x)= \int_0^{1/T}{\rd\tau\over 2\pi}\
{\bar {\boldsymbol T}}_{2l}(\tau+\ri x)\ ,
\eea
where
${\boldsymbol T}_{2 l} $ and ${\bar {\boldsymbol T}}_{2 l}$ are
chiral currents of the Lorentz spin $2l$ and $(-2l)$, respectively.
As has been explained  in Appendix C, the local  chiral currents
of the parafermionic minimal model
constitute
${\rm W}_k$ algebra.   Therefore
the local densities ${\boldsymbol T}_{2 l} $
are, in fact,  appropriately regularized polynomial in
$\partial  {\boldsymbol \phi}$ and
${\rm W}_{k}$ currents as well as  their derivatives.
Similarly, ${\bar {\boldsymbol T}}_{2 l} $ are constructed using
${\bar \partial}   {\boldsymbol \phi}$ and   ${\bar {\rm  W}}_k$ currents.
Notice that any  operators in the form\ \eqref{alkslakj}
are conserving charges in the sense that
\bea\label{slkssla}
{\rd \over \rd x}\ {\mathbb I}_{2l-1}(x)={\rd \over \rd x}\ {\bar {\mathbb I}}_{2l-1}(x)=0\ ,
\eea
and
the meaning of \eqref{alksla} is that the boundary neither emits
nor absorbs any amount of the combined charges ${\bar {\mathbb
I}}_{2l-1}- {\mathbb I}_{2 l-1}$~\cite{gz}.
For this reason ${\mathbb I}_{2 l-1}$ and   ${\bar {\mathbb I}}_{2 l-1}$
are referred to  as
local  IM's.

The local IM's  ${\mathbb I}_{1}$ and ${\mathbb I}_{3}$ are
known in the closed form. Namely,
${\mathbb I}_{1}$ is given by \eqref{ssiyudl}  and
an explicit bosonized form of the first nontrivial local density ${\boldsymbol T}_{4}$
is presented  in Ref.~\cite{fateq}.
There are strong indications  that
the commutativity conditions
\bea\label{sjquyqy}
[\, {\mathbb I}_{2l-1}\,,\,
{\mathbb I}_{1}\,]=[\,{\mathbb I}_{2l-1}\,,\,
{\mathbb I}_{2}\,]=0
\eea
(a) fix all the operators  ${\mathbb I}_{2l-1}$ for $l>2$
uniquely up to the normalization, and (b) such defined
operators   ${\mathbb I}_{2l-1}$   mutually commute:
\bea\label{treslslsa}
\big[\, {\mathbb I}_{2l-1}\, ,\, {\mathbb I}_{2m-1}\, \big]=0\ .
\eea
In order to specify ${\mathbb I}_{2l-1}$ unambiguously,
it is convenient to normalize the corresponding  local density
in such a way that
the monomial $(\partial\phi)^{2l}$ appears  in ${\boldsymbol T}_{2l}$
with the coefficient equals one, i.e.
\bea\label{uytslsjsl}
{\mathbb I}_{2l-1}= \int_0^{1/T}{\rd\tau\over 2\pi}\ \big(\,
(\partial\phi)^{2l}+\ldots\, \big)\ .
\eea

We can now  take advantage of  the  general prediction  from the
Introduction of Ref.~\cite{LVZ} about the form
of  the low-temperature expansion in the case of an
RG boundary flow which   possesses
an infinite set of mutually commuting local IM's, and 
terminated at  the  ``trivial''   infrared
fixed point (the one which in Cardy's classification\ \cite{Card} corresponds to the
identity primary state). In the case under consideration  the   prediction 
implies that the low-temperature expansion of $\log Z_{\rm bshg}(h ) $ looks as follows
\bea\label{irexpan}
\log Z_{\rm bshg}(h ) \simeq  -f_{0}\
{E_*\over T}+ \log\big({\rm g}_D\, {\rm g}_{\rm fixed} \big)
-\sum_{l=1}^{\infty}\,
f_{l}\ \Big({T\over E^*}\Big)^{2l-1}\
\  I_{2l-1}\Big(\ri\, {h\over T} \Big)\ ,
\eea
where  $f_{l}$ are some  coefficients   and
$I_{2l-1}(P)$ are vacuum   eigenvalues of the operators ${\mathbb I}_{2l-1}$:
\bea\label{lksls}
{\mathbb I}_{2l-1}\, |\, P,\, (0,0)\, \rangle=
(2\pi T)^{2l-1}\ I_{2l-1}(P)\ |\, P,\, (0,0)\, \rangle\ .
\eea
Let us emphasize that the $I_{2l-1}(P)$  are unique
polynomials of order $l$ of the variable $P^2$, and, as  follows from the
normalization condition\ \eqref{uytslsjsl}
\bea\label{salsaaslk}
I_{2l-1}(P)=\Big({P\over 2}\Big)^{2l}+\ldots\ ,
\eea
where the omitted terms contain lower powers of $P^2$.
In particular,
in view of the
explicit formulae for
${\mathbb I}_{1}$ \eqref{ssiyudl}
and ${\mathbb I}_{3}$~\cite{fateq} it is possible to show that
\bea\label{asasla}
I_{1}(P)&=&\Big({P\over 2}\Big)^2-{ k\over 8\, (2+k)}\ ,\\
I_{3}(P)&=&\Big({P\over 2}\Big)^4-{5\, k\over 4\,  (2 + 3 k)}\ \Big({P\over 2}\Big)^2+
{ k\, (\, 9k+4\,Q^2\,) \over
64\,  (2 + k)\, (2 + 3 k)}\ ,
\nonumber
\eea
with $Q=b+b^{-1}$.

Applying the standard WKB iterational scheme~\cite{Landau}  to
the  spectral determinant~\eqref{qqzsksls}, 
one can  reproduce exactly  the structure of the
low-temperature expansion~\eqref{irexpan} as well as
the eigenvalues\ \eqref{asasla}. En passant,
the WKB calculation provides  an explicit form
of the  the expansion coefficients:
\bea\label{lssuqy}
f_l={\Gamma( l-{1\over 2})\over \Gamma(kl-{k\over 2})}\ \
{\Gamma\big(\, {(2l-1)k\over 2Q}\, b\, \big)\,
\Gamma\big(\, {(2l-1)k\over 2Q}\, b^{-1}\,\big)\over
2\,\sqrt{\pi}\ l!}\ \bigg({{\sqrt{k}\over Q}}\bigg)^{2l}\ .
\eea

\section{ Appendix :
Low-temperature expansion
of ${\bar Z}^{(k)}_{\theta}$}

At low temperature the regularized  partition function \eqref{dsgfasta}
can be studied using the WKB expansion.
The leading WKB approximation yields an explicit expression for
the  regularized ground state energy\ \eqref{lisso}:
\bea\label{lsksls}
{{\bar E}^{(k)}_\theta\over  E_*}=-2 \big(1-{\textstyle{1\over k}}\big)
+
\sum_{m=2}^{\infty}{{\Gamma(m-{1\over 2})\over
2\sqrt{\pi} m!}}\Big[\psi\big(m -{\textstyle{
1\over 2}}\big)-\psi\big( mk-{\textstyle{k\over 2}}\big)+
\psi\big( {\textstyle{k\over 2}}\big)-\psi\big( {\textstyle{1\over 2}}\big)\Big]
\sin^{2m}(\theta)\,.\nonumber
\\
\eea
It is useful to bear in mind that
\eqref{lsksls} is applicable for
$0\leq \theta<{\pi\over 2}$ only.
It can be analytically continued
within  the domain ${\pi\over 2}\leq \theta<\pi$
by means of the   relation:
\bea\label{sssslsa}
&&{\bar E}^{(k)}_{\theta}/E_*=-{\bar E}^{(k)}_{\pi-\theta}/E_*+
{\sqrt{\pi}\over k}\  \sum_{l=1\atop l\not={k\over 2}}^{k-1}\,
{\Gamma(-{l\over k}-{1\over 2})\over
\Gamma(1-{l\over k})}\ \big[\, \sin(\theta)\, \big]^{1+{2l\over k}}
{}_2F_1\big({\textstyle{l\over k}}\,,\,
1\, ;\,
 {\textstyle{3\over 2}}+{\textstyle{l\over k}} \,
|\, \sin^2(\theta)\,\big)+\nonumber \\ &&
{\textstyle {4\over k}}\, \sin^2\big({\textstyle{\pi (k-1)\over 2}}\big)\, \Big[\,
\big(1-\cos(\theta) \big)\log\big[\sin\big(
{\textstyle{ \theta\over 2}}\big) \big]+
\big(1+\cos(\theta) \big)\log\big[\cos\big(
{\textstyle{ \theta\over 2}}\big) \big]\,  \Big]\, .
\eea
The following formula  for the  regularized ground state energy
holds  within   $0\leq \theta<\pi$
\bea\label{wejsre}
&& {\bar E}^{(k)}_\theta/ E_*=C^{(k)}_0-\big(\,\gamma_E+2\log 2+
\psi\big({\textstyle{k\over 2}}\big)-\log k\, \big)\, \cos(\theta)+ \\ &&
{\textstyle{\sqrt{\pi}\over k}}\ \cos^2(\theta)\ \sum_{l=1}^{k-1}\, {\Gamma({l\over k}+{1\over 2})\over
\Gamma({l\over k})}\ {}_3F_2\big({\textstyle{l\over k}}+{\textstyle{ 1\over 2}}\,,\,
{\textstyle{ 1\over 2}}\, ,\, 1\, ;\, 2
\,  ,\, {\textstyle{3\over 2}}\, |\, \cos^2(\theta)\,\big)-\nonumber\\
&&{\textstyle{2\over 3 k}}\ \cos^3(\theta)\  \sum_{l=1}^{k-1}\, {\textstyle {l\over k}}\ \,
 {}_3F_2\big({\textstyle{l\over k}}+1\,,\, 1\, ,\, 1\, ;\, 2
\,  ,\, {\textstyle{5\over 2}}\, |\, \cos^2(\theta)\, \big)\ ,\nonumber
\eea
where
\bea\label{saassalk}
C^{(k)}_0=
-{\textstyle{\sqrt{\pi}\over k}}\ \sum_{l=1\atop l\not={k\over 2}}^{k-1}\, {\Gamma({l\over k}-{1\over 2})\over
\Gamma({l\over k})}-
{\textstyle{2\log 2\over k}}\ \sin^2\big({\textstyle{\pi (k-1)\over 2}}\big)\ .
\eea

The  systematic WKB expansion of  \eqref{dsgfasta}
leads to an  asymptotic series  of the form
\bea\label{slksasls}
\log {\bar Z}^{(k)}_\theta\simeq
-{{\bar E}^{(k)}_\theta\over  T}+\log({\rm g}_{\rm fixed})+\sum_{l=1}^{\infty}
F^{(k)}_{l}(\theta)\ \Big({T\over E_*}\Big)^{ 2 l-1}\ ,
\eea
where the coefficient $F^{(k)}_1$ reads explicitly as
\bea\label{sshssg}
&&F^{(k)}_{1}(\theta)=
{k-1\over 12 (k+2)}\ {}_2F_1\big({\textstyle{1\over 2}}, {\textstyle{1\over 2}}+
{\textstyle{1\over k}};
{\textstyle{3\over 2}}+{\textstyle{1\over k}}\,|\,\sin^2(\theta)\, \big)=\\
&&-
{\sqrt{\pi} \, k\Gamma({1\over 2}+{1\over k})\over
24\, \Gamma({1\over k}-1)}\, \big[\,\sin(\theta)\,\big]^{-1-{2\over k}}-
{k-1\over  12 k}\, \cos(\theta)\,
 {}_2F_1\big(1+{\textstyle{1\over k}}, 1;
{\textstyle{3\over 2}}\,\,\cos^2(\theta)\big)\, .\nonumber
\eea


\begin{thebibliography}{99}


\bibitem
{SLAZ} S.L. Lukyanov and A.B. Zamolodchikov, J.  Stat. Mech.:
Theor. Exp. P05003, (2004).

\bibitem
{AES} V. Ambegaokar, U. Eckern and G. Sch\"on, Phys. Rev. Lett.
{\bf 48}, 1745 (1982).

\bibitem
{SWar} S.L. Lukyanov and P. Werner, 
Universal scaling behavior of the single electron box in the strong tunneling 
limit, Preprint RUNHETC-06-02 (arXiv:
cond-mat/0606).

\bibitem{saleur}
P. Fendley and H. Saleur, Phys. Rev. {\bf B60}, 11432 (1999).

\bibitem
{ZamFat}
A.B. Zamolodchikov and V.A. Fateev, Sov. Phys. JETP {\bf
62}, 215 (1985).


\bibitem{Wu}
F.Y. Wu and Y.K. Wang, J. Math. Phys. {\bf 17}, 439 (1976).


\bibitem{Elit}
S. Elitsur, R. Pearson and J. Shigemitau, Phys. Rev. {\bf 19D},
3698 (1979).

\bibitem{Frad}
E. Fradkin and L. Kadanoff, Nucl. Phys. {\bf B170} (FS1), 1,
(1980).

\bibitem{FZaue}
V.A. Fateev and A.B. Zamolodchikov, Phys. Lett. {\bf A92}, 37
(1982).

\bibitem{Gepner}
D. Gepner and Z. Qiu, Nucl. Phys. {\bf B285}, 423, (1987).



\bibitem{Carda}
J.L. Cardy, Nucl. Phys. {\bf B275} [FS12], 200, (1986).


\bibitem{SalB}
H. Saleur and M. Bauer, Nucl. Phys. {\bf B320}, 591, (1989).

\bibitem{Card}
J.L. Cardy, Nucl. Phys. {\bf B324}, 581, (1989).

\bibitem{AOSR}
I. Affleck, M. Oshikawa, H. Saleur, Nucl. Phys. {\bf B594}, 535,
(2001).

\bibitem{Moor}
J. Maldacena, G. Moore and  N. Seiberg,
JHEP {\bf 0107}, 046 (2001).

\bibitem{AfLud}
I. Affleck and A.W.W. Ludwig, Phys. Rev. Lett.  {\bf 67}, 161 (1991).

\bibitem{FRied}
D. Friedan and A. Konechny,
Phys. Rev. Lett. {\bf 93}, 030402 (2004).

\bibitem{CZ}
R. Chatterjee and  A. Zamolodchikov,
Nucl. Phys. {\bf B432}, 427 (1994).


\bibitem{AOS}
I. Affleck, M. Oshikawa, H. Saleur,
J. Phys. {\bf A31}, 5827, (1998).


\bibitem{Schom}
S. Fredenhagen, V. Schomerus,
JHEP  {\bf 0202}, 005 (2002).


\bibitem
{Frolov} V.A. Fateev, I.V. Frolov and A.S. Shwarz, Nucl. Phys.
{\bf B154}, 1 (1979).

\bibitem{Luscher} M. L\"uscher, Nucl. Phys.
B  {\bf 200} [FS4], 61 (1982).

\bibitem
{AlZ}
Al.B. Zamolodchikov, unpublished notes (2001).


\bibitem{BLZ}
V.V. Bazhanov, S.L. Lukyanov,  and A.B. Zamolodchikov, Comm. Math.
Phys. {\bf 177}, 381 (1996).


\bibitem{blz2}
V.V. Bazhanov, S.L. Lukyanov and A.B. Zamolodchikov, Comm. Math.
Phys. {\bf 190}, 247 (1997).

\bibitem{blz3}
V.V. Bazhanov, S.L. Lukyanov and A.B. Zamolodchikov, Comm. Math.
Phys. {\bf 200}, 297 (1999).

\bibitem{blz5}
V.V. Bazhanov, S.L. Lukyanov and A.B. Zamolodchikov, Nucl. Phys.
{\bf B549 } [FS], 529 (1999).

\bibitem
{blzz} V.V. Bazhanov, S.L. Lukyanov and A.B. Zamolodchikov, Jour.
Stat. Phys. {\bf 102}, 567 (2001).



\bibitem{Kac}
V.G. Kac, Infinite dimensional Lie algebras, Birkh${\ddot{\rm a}}$user, Boston (1983).

\bibitem{FatZam}
V.A. Fateev and Al.B. Zamolodchikov, Phys. Lett. {\bf B271}, 91
(1991).

\bibitem{LVZ}
S.L. Lukyanov, E.S. Vitchev, and A.B. Zamolodchikov, Nucl. Phys.
{\bf B683}, 423 (2004).


\bibitem{Witten}
E. Witten, Phys. Rev. {\bf D47}, 3405 (1993).

\bibitem{Landau}
L.D. Landau and E.M. Lifshitz, Quantum Mechanics:
Non-Relativistic Theory, Vol.3, Third Edition
Pergamon Press, Oxford
(1997).

\bibitem{LTZ}
S.L. Lukyanov, A.M. Tsvelik and A.B. Zamolodchikov, Nucl. Phys.
{\bf B719}, 103 (2005).

\bibitem{SasZ}
A.B. Zamolodchikov, Teor. Mat. Fiz. {\bf 65}, 347 (1985).

\bibitem{FATZ}
V.A. Fateev and A.B. Zamolodchikov, Nucl. Phys. {\bf B280}, 644,
1987.


\bibitem
{FL} V. Fateev and S. Lukyanov, Int. J. Mod. Phys. {\bf A3}, 507
(1988).

\bibitem{Kosterlitz}
J.M. Kosterlitz, Phys. Rev. Lett. {\bf 37}, 1577 (1976).

\bibitem
{Zwerger} W. Hofstetter and W. Zwerger, Phys. Rev. Lett. {\bf
78}, 3737 (1997).


\bibitem
{Leigh} R.G. Leigh, Mod. Phys. Lett. {\bf A4}, 2767 (1989).

\bibitem
{Rychkov} V.S. Rychkov, JHEP {\bf 212}, 68  (2002).

\bibitem
{Wang} X. Wang  and  H. Grabert, Phys. Rev. {\bf B53}, 12621
(1996).

\bibitem
{APolyakov} A.M. Polyakov, Phys. Lett. {\bf B59}, 79 (1975).


\bibitem
{Belavin} A.A. Belavin  and A.M. Polyakov, Pis'ma Zh. Eksp. Teor.
Fiz. {\bf 22}, 503 (1975) [JETP Lett. {\bf 22}, 245 (1975)].




\bibitem
{Korshunov} S.E. Korshunov, Pis'ma Zh. Eksp. Theor. Phys. {\bf
45}, 342 (1987) [JETP Lett. {\bf 45}, 434 (1987)].

\bibitem
{Nazarov} Y.V. Nazarov, Phys. Rev. Lett. {\bf 82}, 1245 (1999).











\bibitem
{Feigelman} M.V. Feigelman, A. Kamenev, A.I. Larkin and M.A.
Skvortsov, Phys. Rev. {\bf B66}, 054502 (2002).






\bibitem
{Calan} C.G. Callan and L. Thorlacius, Nucl. Phys. {\bf B329},
117 (1990).

\bibitem
{Legett} A.O. Caldeira and A.J. Leggett, Phys. Rev. Lett. {\bf
46}, 211 (1981).

\bibitem
{Zaikin} G. Sch$\ddot{\rm o}$n and A.D. Zaikin, Phys. Rep. {\bf
198}, 237 (1990).

\bibitem
{Larkin}
I.S. Beloborodov, A.V. Andreev and A.I. Larkin, Phys. Rev. {\bf B68}, 024204 (2003).


\bibitem
{Averin}
D.V. Averin and K.K. Likharev,
in {\it Mesoscopic Phenomena in Solids},
eds. B.L. Altshuler  {\it et al.}
(Elsevier, Amsterdam, 1991), p.173.



\bibitem
{Matveev} K.A. Matveev, Phys. Rev. {\bf B51}, 1743 (1995).




\bibitem{Kel}
L.V. Keldysh, Sov. Phys. JETP {\bf 20}, 1018 (1965).

\bibitem
{gz} S. Ghoshal and A. Zamolodchikov, Int. J. Mod. Phys. {\bf
A9}, 3841 (1994).

\bibitem
{Warner} P. Fendley, H. Saleur and N.P. Warner, Nucl. Phys. {\bf
B430}, 577 (1994).



\bibitem{Moon}
K. Moon, H. Yi, C.L. Kane, S.M. Girvin and M.P.A. Fisher, Phys.
Rev. Lett. {\bf 27}, 4381 (1993).


\bibitem{FSLSN}
P. Fendley,  A.W.W. Ludwig and  H. Saleur, Phys. Rev. Lett. {\bf
74}, 3005 (1995).

\bibitem{Fen}
P. Fendley,  A.W.W. Ludwig and  H. Saleur, Phys. Rev. {\bf B52},
8934 (1995).

\bibitem{Kane}
C.L. Kane and M.P.A. Fisher, Phys. Rev. {\bf B46}, 15233 (1992).

\bibitem{Simonetti}
F. Lesage, H. Saleur, P. Simonetti, Phys. Rev. {\bf B56}, 7598
(1997).

\bibitem{Wei}
U. Weiss, Sol. St. Comm. {\bf 100}, 281 (1996).

\bibitem{Selberg}
A. Selberg,  Norsk. Mat. Tidsskr. {\bf 26}, 71 (1944).

\bibitem{Dot}
V. Dotsenko and V. Fateev, Nucl. Phys. {\bf B251}, 691 (1985).

\bibitem{toteo}
P. Dorey and R. Tateo, J. Phys.  {\bf A32}, L419 (1999).

\bibitem{Calogero}
F. Calogero and A. Degasperis, Spectral Transform and Solitons,
North Holland, Amsterdam (1982).


\bibitem
{nappi} C.G. Callan, C. Lovelace, C.R. Nappi and S.A. Yost, Nucl.
Phys. {\bf B293}, 83 (1987).

\bibitem
{Ishibashi} N. Ishibashi, Mod. Phys. Lett. {\bf A4}, 251 (1989).


\bibitem{Baxter}
R.J. Baxter, Exactly Solved Models in Statistical Mechanics, Academic Press,
Academic Press (1982).

\bibitem{FZZ}
V. Fateev, A. Zamolodchikov and Al. Zamolodchikov,
Boundary Liouville field theory. 1. Boundary state and boundary
two point function.
Preprint RUNHETC-2000-01
(arXiv: hep-th/0001012).

\bibitem{Ahn}
 C. Ahn, C. Rim and  M. Stanishkov,
Nucl. Phys. {\bf B636}, 497 (2002)

\bibitem{Goulian}
M. Goulian and M. Li, Phys. Rev. Lett. {\bf 66}, 2051 (1991).

\bibitem{Pillow}
S.L. Lukyanov and A.B. Zamolodchikov, to appear.

\bibitem{fateq} V.A. Fateev,
Nucl. Phys. {\bf B473}, 509 (1996).


\end{thebibliography}
\end{document}